\documentclass[11pt]{article}
\usepackage{amssymb}
\usepackage{mathtools}
\usepackage{amsmath}
\usepackage{amstext}
\usepackage{graphicx,epsfig}
\usepackage{epsfig}
\usepackage{verbatim} 
\usepackage{caption}
\usepackage{fancybox}
\usepackage{jcapmod}
\usepackage{slashed}
\usepackage{color}
\usepackage{ulem}
\usepackage{enumitem}
\usepackage{subfigure}
\usepackage{bbm}	
\usepackage{parskip}
\usepackage{dsfont}
\usepackage{tabu}
\usepackage[numbers,sort&compress]{natbib}
\usepackage{xparse}

\allowdisplaybreaks

\usepackage{tikz}
\usepackage{tikzlines}

\usepackage{tkz-euclide}
\usetkzobj{all}

\NewDocumentCommand\Perp{O{0.5}mmmO{1cm}}{%
  \coordinate (#4) at
    ($ ($ #2!#1!#3 $) ! {sin(90)} ! 90:#3 $) {};
  \draw[thick,densely dotted,->] ($ #2!#1!#3 $) -- ($ ($ #2!#1!#3 $) ! #5 ! (#4)$);
  \coordinate (#4) at ($ ($ #2!#1!#3 $) ! #5 ! (#4)$);  
}

\newcommand{\Comment}[1]{{}}
\definecolor{darkblue}{rgb}{0.15,0.35,0.55}
\definecolor{reddish}{rgb}{0.65, 0.2, 0.2}
\usepackage[linktocpage=true]{hyperref}
\hypersetup{
colorlinks=true,
citecolor=darkblue,
linkcolor=reddish,
urlcolor=darkblue,
pdfauthor={},
pdftitle={},
pdfsubject={}
}

\newcommand{\nn}{ \nonumber\\}

\definecolor{greyish}{rgb}{.90,.90,.90}
\definecolor{greyish2}{rgb}{.96,.96,.96}
\usepackage{xcolor,colortbl}
\usepackage{tcolorbox}

\newcommand{\rd}{{\rm d}}

\newcommand{\MP}{M_{\rm Pl}}
\newcommand{\im}{\, {\rm Im} \, }
\newcommand{\re}{\, {\rm Re} \, }
\newcommand{\Tr}{\, {\rm Tr} \, }
\newcommand{\bfk}{{\bf  k}}
\newcommand{\bfp}{{\bf  p}}
\newcommand{\bfu}{{\bf  u}}
\newcommand{\bfq}{{\bf  q}}
\newcommand{\bfx}{{\bf  x}}

\newcommand{\bfy}{{\bf  y}}
\newcommand{\bfz}{{\bf  z}}

\newcommand{\ts}{\tau_{\star}}
\def\be{\begin{equation}}
\def\ee{\end{equation}}

\DeclareRobustCommand\gravitonline{\tikz[line width=1.2 pt,baseline=-0.5ex] \draw[graviton] (-1,0) -- (1,0);}
\DeclareRobustCommand\pmline{\tikz[line width=1.2 pt,baseline=-0.7ex] \draw[pmfield] (-1,0) -- (1,0);}
\DeclareRobustCommand\mixing{\begin{tikzpicture}[baseline=-0.7ex] \node at (0,0) [circle,draw=black,inner sep=.1cm,line width= .2mm](mixT) {};	\node at (0,0) [cross,line width=1pt,minimum size=.2cm](mixTcross) {};\end{tikzpicture}}

\title{}
\author{}

\numberwithin{equation}{section}

\begin{document}

~
%
%
\renewcommand{\thefootnote}{\fnsymbol{footnote}}
\begin{center}
{\huge \bf{Shapes of gravity: Tensor non-Gaussianity\\and massive spin-2 fields}}
\end{center} 

\vspace{1truecm}
\thispagestyle{empty}
\centerline{\Large Garrett Goon$,{}^{\rm a,b,}$\footnote{\href{mailto:gg399@cam.ac.uk}{\tt gg399@cam.ac.uk}} Kurt Hinterbichler$,{}^{\rm c,}$\footnote{\href{mailto:kurt.hinterbichler@case.edu}{\tt kurt.hinterbichler@case.edu}} Austin Joyce$,{}^{\rm d,}$\footnote{\href{mailto:austin.joyce@columbia.edu}{\tt austin.joyce@columbia.edu}} and Mark Trodden$,{}^{\rm e,}$\footnote{\href{mailto:trodden@physics.upenn.edu}{\tt trodden@physics.upenn.edu}}}
\vspace{.7cm}

\centerline{\it ${}^{\rm a}$Institute for Theoretical Physics and Center for Extreme Matter and Emergent Phenomena,}
\centerline{\it Utrecht University, Leuvenlaan 4, 3584 CE Utrecht, The Netherlands}

\vspace{.3cm}

\centerline{\it ${}^{\rm b}$Department of Applied Mathematics and Theoretical Physics}
\centerline{\it Cambridge University, Cambridge, CB3 0WA, UK}

\vspace{.3cm}

\centerline{\it ${}^{\rm c}$CERCA, Department of Physics, Case Western Reserve University, }
\centerline{\it 10900 Euclid Ave, Cleveland, OH 44106, USA}

\vspace{.3cm}

\centerline{\it ${}^{\rm d}$Center for Theoretical Physics, Department of Physics, }
\centerline{\it Columbia University, New York, NY 10027}

\vspace{.3cm}

\centerline{\it ${}^{\rm e}$Department of Physics and Astronomy, Center for Particle Cosmology,}
\centerline{\it University of Pennsylvania, 209 S. 33rd St., Philadelphia, PA 19104, USA}

\vspace{.5cm}
\begin{abstract}
\noindent\noindent
If the graviton is the only high spin particle present during inflation, then the form of the observable tensor three-point function is fixed by de Sitter symmetry at leading order in slow-roll, regardless of the theory, to be a linear combination of two possible shapes. This is because there are only a fixed number of possible on-shell cubic structures through which the graviton can self-interact. If additional massive spin-2 degrees of freedom are present, more cubic interaction structures are possible, including those containing interactions between the new fields and the graviton, and self-interactions of the new fields.  We study, in a model-independent way, how these interactions can lead to new shapes for the tensor bispectrum. In general, these shapes cannot be computed analytically, but for the case where the only new field is a partially massless spin-2 field we give simple expressions. It is possible for the contribution from additional spin-2 fields to be larger than the intrinsic Einstein gravity bispectrum and provides a mechanism for enhancing the size of the graviton bispectrum relative to the graviton power spectrum.
\end{abstract}

\newpage

\tableofcontents

\newpage
\renewcommand*{\thefootnote}{\arabic{footnote}}
\setcounter{footnote}{0}

\section{Introduction\label{sec:Introduction}}

A generic prediction of inflation is 
a nearly Gaussian spectrum of primordial tensor perturbations, arising from fluctuations of the graviton about the nearly de Sitter inflationary background. 
In addition to these two-point correlations, 
tensor non-Gaussianities are generically produced, for instance three-point correlations between three helicity-2 gravitons.  
Recently there has been some effort to understand the form of these tensor non-Gaussianities \cite{Maldacena:2002vr,Maldacena:2011nz,Gao:2011vs,Soda:2011am,Shiraishi:2011st,Dimastrogiovanni:2015pla,Biagetti:2017viz,Dimastrogiovanni:2018uqy,Dimastrogiovanni:2018gkl} and upcoming experiments, such as LISA \cite{2017arXiv170200786A}, will be able to bound the level of tensor non-Gaussianity in the universe \cite{Adshead:2009bz,Bartolo:2018qqn}.

If the only field of spin $\geq 2$ active during inflation is the graviton itself---and if its leading interactions are described by General Relativity---then the form of tensor non-Gaussianity is fixed by the cubic graviton self-interaction vertex present in the Einstein--Hilbert action \cite{Maldacena:2002vr,Maldacena:2011nz}.  Allowing for higher-derivative corrections, there is another possible shape corresponding to the six-derivative, cubic vertex present in a (Weyl)$^3$ interaction. These are the only possible 3-point shapes for a massless spin-2 field, regardless of other possible higher-derivative terms in the Lagrangian~\cite{Maldacena:2011nz}, up to slow roll corrections parametrizing departures from pure de Sitter.\footnote{There is a possible parity-violating cubic interaction $\mathcal{L}\sim \epsilon WWW$, which would naively generate another shape, but this term only generates slow-roll suppressed non-Gaussianities \cite{Soda:2011am,Shiraishi:2011st}. We do not consider possible parity-violating interactions in this paper. Throughout the introduction we specialize to four spacetime dimensions.}  This fact follows from de Sitter invariance, which restricts the possible on-shell, cubic vertices (cubic terms in the Lagrangian modulo field re-definitions and total derivatives) to be one of two structures, and the fact that the non-Gaussianity is determined solely by the on-shell cubic vertex.  This is equivalent to the statement that conformal symmetry on the boundary of de Sitter space fixes the form of the 3-point correlators up to a finite number of constants.  

It is possible that other fields besides the inflaton and graviton are present during inflation with masses of order the Hubble scale.  If any of these fields have spin $\geq 2$, or if the graviton itself has a small mass, then there will be new possible on-shell three-point vertices which are not possible for the massless graviton alone, and correspondingly new possible shapes of the 3-point function if the new degrees of freedom mix with the graviton tensor modes.  For example, a single massive spin-2 field has four possible parity-invariant self-interaction structures, as opposed to the 2 possible for a massless spin-2 field (in four spacetime dimensions).  These extra structures arise because the interactions for a massive particle are not required to satisfy the gauge invariance constraints that a massless particle must.  If these additional massive particles mix with or otherwise transmit their fluctuations to the graviton, then these new structures can imprint themselves on graviton non-Gaussianities and could be detected by futuristic gravitational wave observations.  This would be evidence for the presence of new higher-spin degrees of freedom.

In this paper, we perform a model-independent study of the possible new shapes of tensor non-Gaussianity that can arise due to the presence of additional heavy spin-2 fields.  We do this by classifying the possible on-shell cubic vertices through which a massive spin-2 field can interact with itself and with the graviton.  Any theory of gravity coupled to additional massive spin-2 particles (for example, those derived from ghost-free massive gravity \cite{deRham:2010kj,Hinterbichler:2011tt,deRham:2014zqa}, bi-gravity \cite{Hassan:2011zd,Schmidt-May:2015vnx} and multi-gravity \cite{Hinterbichler:2012cn} theories) must, on-shell at cubic order, reduce to a linear combination of these vertices. The tensor non-Gaussianity depends only on these vertices and can be computed from them, and so the results apply to any model with these degrees of freedom. A powerful feature of the 3-point function in de Sitter space is that its form is completely fixed by the isometries to be a linear combination of a finite number of shapes. This is true non-perturbatively, and therefore it is not even necessary for the additional spin-2 field to be fundamental: even if it is composite or a resonance, its 3-point interactions and correlators will be fixed to by de Sitter symmetry.\footnote{This is analogous to 3-point scattering amplitudes in flat space, whose exact structure is fixed by Lorentz invariance.}

The approach we take to evaluating late-time 3-point correlation functions is to compute the cosmological wavefunction as the on-shell action evaluated on a classical solution. This approach makes it manifest that correlation functions are sensitive only to on-shell vertices. In order to perform this computation, we derive the bulk-to-boundary propagator for a general mass spin-2 field, along with the generic on-shell vertices between arbitrary admixtures of massless, partially massless, and massive spin-2 fields, which may be of independent interest for AdS/CFT applications. 

For generic spin-2 masses, the integrals required to evaluate the on-shell action and compute the late-time non-Gaussianity cannot be evaluated in closed form.  In these cases, the best that can be done is a numerical evaluation of the general expression for particular masses of interest.  There are, however, specific mass values for which the non-Gaussianity can be evaluated analytically. One of these is the so-called {\it partially massless} value where $m^2=2H^2$.  In these cases which only involve massless and partially massless spin-2 fields, we will give explicit expressions for all possible 3-point correlation functions in exact de Sitter space.

The partially massless point is special for several reasons.  It saturates the Higuchi bound $m^2\geq 2H^2$ \cite{Higuchi:1986py}, which gives a minimum mass for stable spin-2 fields on de Sitter space (the massless graviton at $m^2=0$ is the only exception).  At the partially massless point, a new scalar gauge symmetry emerges, which removes the longitudinal degree of freedom of the massive spin-2, leaving a field with only tensor and vector modes.  There are studies and no-go theorems that would seem to forbid consistent theories of a single {interacting} partially massless spin-2 particle~\cite{Zinoviev:2006im,Hassan:2012gz,Hassan:2012rq,deRham:2012kf,Hassan:2013pca,Deser:2013uy,deRham:2013wv,Zinoviev:2014zka,Garcia-Saenz:2014cwa,Hinterbichler:2014xga,Joung:2014aba,Alexandrov:2014oda,Hassan:2015tba,Hinterbichler:2015nua,Cherney:2015jxp,Gwak:2015vfb,Gwak:2015jdo,Garcia-Saenz:2015mqi,Hinterbichler:2016fgl,Bonifacio:2016blz,Apolo:2016ort,Apolo:2016vkn,Bernard:2017tcg,Boulanger:2018dau}.  However, it may still be possible to have theories of partially massless spin-2 particles interacting with other fields, possibly an infinite number of them (for instance there are known examples of Vasiliev-like higher spin theories with infinite towers of partially massless fields \cite{Bekaert:2013zya,Basile:2014wua,Alkalaev:2014nsa,Joung:2015jza,Brust:2016zns}). Questions about the non-linear completion of the theory will not affect the systematics of our computations in this paper because to compute bi-spectra we only need consistency on-shell up to cubic order, which amounts to finding the vertices that are invariant on-shell under the linearized partially massless symmetry. These vertices are known \cite{Joung:2012rv,Joung:2012hz}, and we reproduce the partially massless ones as a byproduct of our analysis. In situations where the partially massless symmetry fails to exist beyond cubic order there could be additional contributions to correlation functions beyond those that we consider coming from vertices that do not have the partially massless gauge symmetry.

Once we have computed the wavefunction involving mixtures of massless and partially massless fields, we turn to phenomenology.  Wavefunction coefficients involving PM fields on external legs can produce observable signatures in multiple ways. For instance, they can be seen directly if the partially massless spin-2 couples directly to matter. Alternatively, if there exists a linear mixing between the PM spin-2 and the graviton, then these coefficients imprint themselves on the graviton bispectrum.
Here we explore a mechanism involving a linear mixing between the partially massless field and the graviton. Such a mixing is not possible in exact de Sitter space, and therefore carries a slow-roll suppression. Mixing of this type has two effects: it can transmit non-Gaussianity into the graviton sector at the end of inflation, so that the wavefunctional computed in exact de Sitter space accurately captures the non-Gaussianities in the tensor sector. Another effect is that the PM spin-2 field can mix into the graviton during inflation. We estimate this latter process, but this requires a more-involved computation to treat fully. Aside from 3-point correlation functions, it is of course possible for additional spin-2 fields to affect higher-point graviton correlation functions, and we comment briefly on possible signatures in the four-point function. Our analysis in this case is somewhat preliminary, but there are some intriguing features.

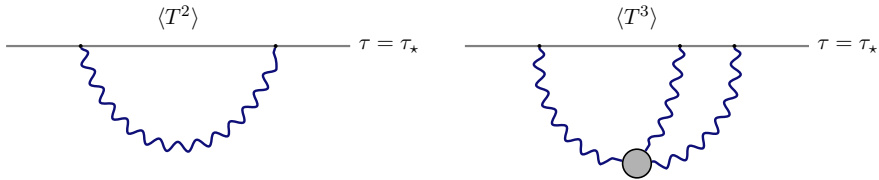
\begin{figure}
\begin{center}
\resizebox{12cm}{!}{
\begin{tikzpicture}
\node at (0,.5) {$\langle T^{2}\rangle$};
\draw[gray, line width=1.1pt] (-3,0) -- (3,0) node [right,black] {$\tau=\ts$};
\node at (-1.7,0) [smalldot,minimum size=.05cm](dotL) {};
\node at (1.7,0) [smalldot,minimum size=.05cm](dotR) {};
\coordinate (C) at (0,-1.5);
\draw[graviton, line width=1.2 pt] (dotL) .. controls (-1,-2.22) and (1,-2.22) .. (dotR);

\begin{scope}[xshift=8cm]
\node at (0,.5) {$\langle T^{3}\rangle$};
\draw[gray, line width=1.1pt] (-3,0) -- (3,0) node [right,black] {$\tau=\ts$};
\node at (0,-2.05) [vertex,line width=.8pt,minimum size=.5cm](vertex) {};
\node at (-1.7,0) [smalldot,minimum size=.05cm](dotL) {};
\node at (1.7,0) [smalldot,minimum size=.05cm](dotR) {};
\node at (.75,0) [smalldot,minimum size=.05cm](dotC) {};
\path[]
	(dotL) edge[bend right=40,graviton, line width=1.2 pt] node  {} (vertex)
	(dotC) edge[bend left=14,graviton, line width=1.2 pt] node  {} (vertex)
	(dotR) edge[bend left=43,graviton, line width=1.2 pt] node  {} (vertex);
\end{scope}
\end{tikzpicture}
}
\end{center}
\caption{Diagrams giving rise to the $\langle T^{2}\rangle$ and $\langle T^{3}\rangle$ wavefunction coefficients, which determine $\langle \gamma^{2}\rangle$ and $\langle \gamma^{3}\rangle$ to leading order.  Wavy lines correspond to graviton bulk-to-boundary propagators. It should be noted that other references (such as \cite{Arkani-Hamed:2015bza,Lee:2016vti}) use similar diagrams to denote the entire in-in correlator, while for us the above diagrams only correspond to wavefunction coefficients.  The three-point interaction vertex in the right diagram can arise from the Einstein--Hilbert term or a $W_{\mu\nu\rho\sigma}^{3}$ higher-derivative interaction (other diffeomorphism invariant interactions are redundant with these \cite{Maldacena:2011nz}).}
\label{fig:Graviton2And3PointCoefficients}
\end{figure}

Our results are also of more formal theoretical interest. There has been great progress recently both in systematizing the computation of correlation functions in cosmology using de Sitter symmetry~\cite{Maldacena:2011nz,Creminelli:2011mw,Antoniadis:2011ib,Mata:2012bx,Ghosh:2014kba,Kundu:2014gxa,Arkani-Hamed:2015bza,Pajer:2016ieg,Arkani-Hamed:2018kmz} and in the related problem of studying conformal field theories in momentum space~\cite{Bzowski:2013sza,Bzowski:2015pba,Bzowski:2018fql,Isono:2018rrb}. To this point, investigations involving external fields with spin have focused on the massless cases, where gauge invariance (or current conservation) provides numerous simplifications. The results provided here provide the first steps toward an understanding of correlation functions in de Sitter space with more general external states. The building blocks we provide may enable many new computations besides the ones we present here and provide additional data to help further develop our systematic understanding of perturbative field theory in cosmological spacetimes.

Many of our results are rather technical, so to orient the reader we first give a brief overview of the computation we perform. Additionally, we provide a rough bound on the size of non-Gaussianity which can be induced by additional spin-2 fields. For concreteness, we focus on the imprint that partially massless fields can leave on the graviton bispectrum $\langle\gamma^{3}\rangle$.

\subsection{Estimating the size of non-Gaussianity: A sketch\label{sec:SketchOfCalculation}}

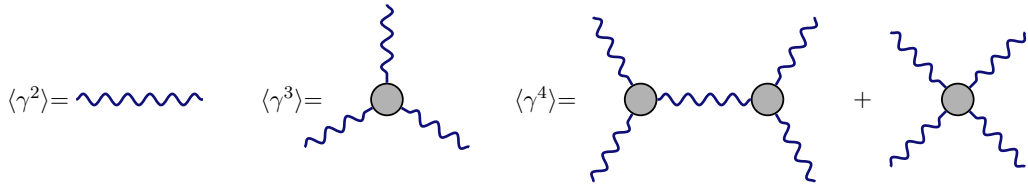
\begin{figure}
\begin{center}
\resizebox{14cm}{!}{
\begin{tikzpicture}
\node at (0,0) {$\langle \gamma^{2}\rangle$=};
\draw[graviton, line width=1.2 pt] (.6,0)--(2.6,0);

	\begin{scope}[xshift=4cm]

	\node at (0,0) {$\langle \gamma^{3}\rangle$=};

	\node at (1.5,0) [vertex,line width=.8pt,minimum size=.5cm](vertexL) {};
	
	\coordinate (dotLeftL) at ([shift=({210:1.5})]vertexL) {};

\coordinate (dotRightL) at ([shift=({-30:1.5})]vertexL) {};

\coordinate (dotTopL) at ([shift=({90:1.5})]vertexL) {};

\path[]
	(dotTopL) edge[graviton,line width=1.2 pt] node  {} (vertexL)
	(dotLeftL) edge[graviton,line width=1.2 pt] node  {} (vertexL)
	(dotRightL) edge[graviton,line width=1.2 pt] node  {} (vertexL);
	
	\end{scope}

	\begin{scope}[xshift=8cm]

	\node at (0,0) {$\langle \gamma^{4}\rangle$=};

	\node at (1.5,0) [vertex,line width=.8pt,minimum size=.5cm](vertexL) {};
	\node at (3.5,0) [vertex,line width=.8pt,minimum size=.5cm](vertexR) {};

\coordinate (dotUpL) at ([shift=({120:1.5})]vertexL) {};

\coordinate (dotDownL) at ([shift=({240:1.5})]vertexL) {};

\coordinate (dotUpR) at ([shift=({60:1.5})]vertexR) {};

\coordinate (dotDownR) at ([shift=({-60:1.5})]vertexR) {};

\path[]
	(dotUpL) edge[graviton,line width=1.2 pt] node  {} (vertexL)
	(dotDownL) edge[graviton,line width=1.2 pt] node  {} (vertexL)
	(dotUpR) edge[graviton,line width=1.2 pt] node  {} (vertexR)
	(dotDownR) edge[graviton,line width=1.2 pt] node  {} (vertexR)
	(vertexR) edge[graviton,line width=1.2 pt] node  {} (vertexL);
	
	\node at (5,0) {+};
	
	\node at (6.5,0) [vertex,line width=.8pt,minimum size=.5cm](vertexRR) {};
	
	\coordinate (dot1RR) at ([shift=({45:1.5})]vertexRR) {};

	\coordinate (dot2RR) at ([shift=({135:1.5})]vertexRR) {};

	\coordinate (dot3RR) at ([shift=({225:1.5})]vertexRR) {};

	\coordinate (dot4RR) at ([shift=({315:1.5})]vertexRR) {};

	\path[]
	(dot1RR) edge[graviton,line width=1.2 pt] node  {} (vertexRR)
	(dot2RR) edge[graviton,line width=1.2 pt] node  {} (vertexRR)
	(dot3RR) edge[graviton,line width=1.2 pt] node  {} (vertexRR)
	(dot4RR) edge[graviton,line width=1.2 pt] node  {} (vertexRR);
	
	\end{scope}

\end{tikzpicture}
}
\end{center}
\caption{Diagrammatic expressions for the correlation functions $\langle \gamma^{2}\rangle$, $\langle \gamma^{3}\rangle$, and $\langle\gamma^{4}\rangle$.  Wavy graviton lines correspond to factors of $\re \langle T^{2}\rangle^{-1}$, while three- and four-point vertices correspond to factors of $\re\langle T^{3}\rangle$ and $\re\langle T^{4}\rangle$, respectively.  Note that these diagrams have a fundamentally different meaning than the wavefunction diagrams of Fig.~\ref{fig:Graviton2And3PointCoefficients}, despite the similar notation. These instead represent equal-time correlation functions on the time slice $\tau=\ts$.}
\label{fig:SchematicGravitonCorrelatorCalculation}
\end{figure}

The approach we follow is to compute graviton non-Gaussianities via the wavefunction of the universe, $\Psi[\bar{\varphi}_{\bfk},\ts]$, which when squared gives a probability distribution for fields, collectively denoted by $\varphi$, to take on a given profile, $\bar{\varphi}_{\bfk}$, at time $\tau=\ts$.  We are interested in momentum space correlators, hence our boundary conditions are phrased in momentum space, as indicated.  Equal-time expectation values are computed via the usual quantum mechanics formula:
\be
\langle \varphi_{\bfk_{1}}(\ts)\ldots \varphi_{\bfk_{n}}(\ts)\rangle=\int\mathcal{D}\bar{\varphi}\, |\Psi[\bar{\varphi}_{\bfk},\ts]|^{2}\bar{\varphi}_{\bfk_{1}}\ldots\bar{\varphi}_{\bfk_{n}}\, .\label{QMFormula}
\ee

Much of our effort will be devoted to computing the wavefunctional itself. This can be done 
via the path integral which, at leading order, is approximated by the action evaluated on the classical solution, $\varphi_{\rm cl}$, which takes on the prescribed boundary values $\bar{\varphi}_{\bfk}$ at $\tau=\ts$, integrated up to time $\ts$
\be
\Psi[\bar{\varphi}_{\bfk},\ts]= \int^{\varphi(\tau_\star)=\bar{\varphi}}\mathcal{D}\varphi\, \exp[iS]\approx \exp\left (iS_{\rm cl}[\varphi_{\rm cl}]\right )\ .
\ee
This can be expanded in powers of $\bar{\varphi}$, schematically as 
\be
\Psi[\bar{\varphi}_{\bfk},\ts]\approx \exp\left (-\frac{1}{2!}\langle \mathcal{O}^{2}\rangle\bar{\varphi}^{2}-\frac{1}{3!}\langle \mathcal{O}^{3}\rangle\bar{\varphi}^{3}-\frac{1}{4!}\langle \mathcal{O}^{4}\rangle\bar{\varphi}^{4}+\cdots\right )\, ,\label{SchematicWavefunctionCoefficients}
\ee
for some functions $\langle\mathcal{O}^{n}\rangle$. Using the wavefunctional \eqref{SchematicWavefunctionCoefficients}, the correlators \eqref{QMFormula} can then be computed perturbatively via standard Gaussian integral formulae.  A more detailed discussion of the wavefunction formalism may be found in Appendix \ref{app:TheWavefunction}.

The wavefunction coefficients in~\eqref{SchematicWavefunctionCoefficients} have a convenient diagrammatic representation.  For instance, in the case of graviton correlators we use the notation $\varphi\to\gamma_{ij}$ and $\mathcal{O}\to T_{ij}$.  Considering only self-interactions, the $\langle T^{2}\rangle$ and $\langle T^{3}\rangle$ coefficients (indices suppressed) would arise from the diagrams\footnote{The pure GR calculation corresponding to Fig.~\ref{fig:Graviton2And3PointCoefficients} and Fig.~\ref{fig:SchematicGravitonCorrelatorCalculation} can be found in \cite{Maldacena:2002vr} while the effects of higher derivative curvature terms were considered in \cite{Maldacena:2011nz}.} in Fig.~\ref{fig:Graviton2And3PointCoefficients}.  Here and throughout, \gravitonline~is used to represent graviton lines.    The left diagram corresponds to evaluating the quadratic action on the linear classical solution, $\gamma_{\rm cl}\propto\bar{\gamma}_{\bfk}$, while in the right diagram the same solution is inserted into the cubic interaction.  The actions are integrated over all of spacetime up to the $\tau=\ts$ surface where correlators are to be computed.  Such calculations are familiar to AdS/CFT practitioners: diagrams such as Fig.~\ref{fig:Graviton2And3PointCoefficients} are the dS version of Witten diagrams and lines correspond to bulk-to-boundary propagators (or bulk-to-bulk propagators in diagrams involving internal lines).

\begin{figure}
\begin{center}
\resizebox{16cm}{!}{
\begin{tikzpicture}

\begin{scope}[xshift=0cm]
\node[font=\Large] at (0,.5) {$\langle \Sigma T^{2}\rangle$};
\draw[gray, line width=1.1pt] (-3,0) -- (3,0) node [right,black] {$\tau=\ts$};
\node at (0,-2.05) [vertex,line width=.8pt,minimum size=.5cm](vertex) {};
\node at (-1.7,0) [smalldot,minimum size=.05cm](dotL) {};
\node at (1.7,0) [smalldot,minimum size=.05cm](dotR) {};
\node at (.75,0) [smalldot,minimum size=.05cm](dotC) {};
\path[]
	(dotL) edge[bend right=40,pmfield, line width=1.2pt] node  {} (vertex)
	(dotC) edge[bend left=14,graviton, line width=1.2pt] node  {} (vertex)
	(dotR) edge[bend left=43,graviton, line width=1.2pt] node  {} (vertex);
\end{scope}

\begin{scope}[xshift=8cm]
\node[font=\Large] at (0,.5) {$\langle T \Sigma^{2}\rangle$};
\draw[gray, line width=1.1pt] (-3,0) -- (3,0) node [right,black] {$\tau=\ts$};
\node at (0,-2.05) [vertex,line width=.8pt,minimum size=.5cm](vertex) {};
\node at (-1.7,0) [smalldot,minimum size=.05cm](dotL) {};
\node at (1.7,0) [smalldot,minimum size=.05cm](dotR) {};
\node at (.75,0) [smalldot,minimum size=.05cm](dotC) {};
\path[]
	(dotL) edge[bend right=40,pmfield, line width=1.2pt] node  {} (vertex)
	(dotC) edge[bend left=14,pmfield, line width=1.2pt] node  {} (vertex)
	(dotR) edge[bend left=43,graviton, line width=1.2pt] node  {} (vertex);
\end{scope}

\begin{scope}[xshift=16cm]
\node[font=\Large] at (0,.5) {$\langle \Sigma^{3}\rangle$};
\draw[gray, line width=1.1pt] (-3,0) -- (3,0) node [right,black] {$\tau=\ts$};
\node at (0,-2.05) [vertex,line width=.8pt,minimum size=.5cm](vertex) {};
\node at (-1.7,0) [smalldot,minimum size=.05cm](dotL) {};
\node at (1.7,0) [smalldot,minimum size=.05cm](dotR) {};
\node at (.75,0) [smalldot,minimum size=.05cm](dotC) {};
\path[]
	(dotL) edge[bend right=40,pmfield, line width=1.2pt] node  {} (vertex)
	(dotC) edge[bend left=14,pmfield, line width=1.2pt] node  {} (vertex)
	(dotR) edge[bend left=43,pmfield, line width=1.2pt] node  {} (vertex);
\end{scope}
\end{tikzpicture}
}
\end{center}
\caption{Wavefunction diagrams arising from interactions between the graviton and a partially massless spin-2 field.  When combined with a mixing term $\langle T\Sigma\rangle$, these terms allow the partially massless field to imprint itself on $\langle \gamma^{3}\rangle$.}
\label{fig:GravitonPM3PointCoefficients}
\end{figure}
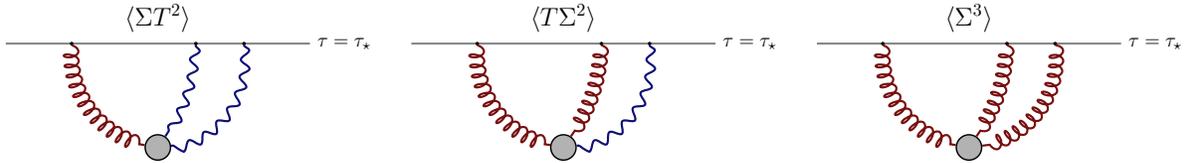

The graviton power spectrum and bispectrum are related to the wavefunction coefficients $\langle T^{2}\rangle$ and $\langle T^{3}\rangle$ via relations of the form
\begin{align}
\langle \gamma^{2}\rangle \sim \frac{1}{\re \langle T^{2}\rangle} \ , \qquad\qquad \langle \gamma^{3}\rangle \sim \frac{\re\langle T^{3}\rangle}{\re\langle T^{2}\rangle^{3}} \ ,\label{SchematicPowerBispectrumGraviton}
\end{align}
which follow from~\eqref{QMFormula} and~\eqref{SchematicWavefunctionCoefficients}.
Since the relations \eqref{SchematicPowerBispectrumGraviton} arise from Gaussian integrals, these final equal-time correlators also have their own diagrammatic expansion, using the wavefunctional coefficients as effective vertices. For a schematic of this, see Fig.~\ref{fig:SchematicGravitonCorrelatorCalculation}.\footnote{Though we use the same lines and vertices to depict both wavefunctional and actual correlator diagrams, they have different meanings. We hope that the difference should be clear from context. In particular, we always draw wavefunctional diagrams as ending on the final time surface.}

\begin{figure}
\begin{center}
\resizebox{9.5cm}{!}{
\begin{tikzpicture}

	\node[font=\Large] at (0,0) {$\langle \gamma^{3}\rangle\supset$};

	\begin{scope}[xshift=2cm]

	\node at (0,0) [vertex,line width=.8pt,minimum size=.5cm](vertexL) {};
	
	\coordinate (dotLeftL) at ([shift=({210:1.5})]vertexL) {};
	\coordinate (dotLeftHalfL) at ([shift=({210:.9})]vertexL) {};

	\coordinate (dotRightL) at ([shift=({-30:1.5})]vertexL) {};
	\coordinate (dotRightHalfL) at ([shift=({-30:.9})]vertexL) {};

	\coordinate (dotTopL) at ([shift=({90:1.5})]vertexL) {};
	\coordinate (dotTopHalfL) at ([shift=({90:.9})]vertexL) {};

	\node at (dotLeftHalfL) [circle,draw=black,inner sep=.1cm,line width= .2mm](mixL) {};
	\node at (dotLeftHalfL) [cross,line width=1pt,minimum size=.2cm](mixLcross) {};

	\path[]
	(dotTopL) edge[graviton,line width=1.2 pt] node  {} (vertexL)
	(dotLeftL) edge[graviton,line width=1.2 pt] node  {} (mixL)
	(mixL) edge[pmfield,line width=1.2 pt] node  {} (vertexL)
	(dotRightL) edge[graviton,line width=1.2 pt] node  {} (vertexL);

	\end{scope}

	\begin{scope}[xshift=5.5cm]
	
	\node[font=\Large] at (-1.7,0) {+};

	\node at (0,0) [vertex,line width=.8pt,minimum size=.5cm](vertexL) {};
	
	\coordinate (dotLeftL) at ([shift=({210:1.5})]vertexL) {};
	\coordinate (dotLeftHalfL) at ([shift=({210:.9})]vertexL) {};

	\coordinate (dotRightL) at ([shift=({-30:1.5})]vertexL) {};
	\coordinate (dotRightHalfL) at ([shift=({-30:.9})]vertexL) {};

	\coordinate (dotTopL) at ([shift=({90:1.5})]vertexL) {};
	\coordinate (dotTopHalfL) at ([shift=({90:.9})]vertexL) {};

	\node at (dotLeftHalfL) [circle,draw=black,inner sep=.1cm,line width= .2mm](mixL) {};
	\node at (dotLeftHalfL) [cross,line width=1pt,minimum size=.2cm](mixLcross) {};

	\node at (dotRightHalfL) [circle,draw=black,inner sep=.1cm,line width= .2mm](mixR) {};
	\node at (dotRightHalfL) [cross,line width=1pt,minimum size=.2cm](mixRcross) {};

	\path[]
	(dotTopL) edge[graviton,line width=1.2 pt] node  {} (vertexL)
	(dotLeftL) edge[graviton,line width=1.2 pt] node  {} (mixL)
	(mixL) edge[pmfield,line width=1.2 pt] node  {} (vertexL)
	(dotRightL) edge[graviton,line width=1.2 pt] node  {} (mixR)
	(mixR) edge[pmfield,line width=1.2 pt] node  {} (vertexL);

	\end{scope}

	\begin{scope}[xshift=9cm]
	
	\node[font=\Large] at (-1.7,0) {+};

	\node at (0,0) [vertex,line width=.8pt,minimum size=.5cm](vertexL) {};
	
	\coordinate (dotLeftL) at ([shift=({210:1.5})]vertexL) {};
	\coordinate (dotLeftHalfL) at ([shift=({210:.9})]vertexL) {};

	\coordinate (dotRightL) at ([shift=({-30:1.5})]vertexL) {};
	\coordinate (dotRightHalfL) at ([shift=({-30:.9})]vertexL) {};

	\coordinate (dotTopL) at ([shift=({90:1.5})]vertexL) {};
	\coordinate (dotTopHalfL) at ([shift=({90:.9})]vertexL) {};

	\node at (dotLeftHalfL) [circle,draw=black,inner sep=.1cm,line width= .2mm](mixL) {};
	\node at (dotLeftHalfL) [cross,line width=1pt,minimum size=.2cm](mixLcross) {};

	\node at (dotRightHalfL) [circle,draw=black,inner sep=.1cm,line width= .2mm](mixR) {};
	\node at (dotRightHalfL) [cross,line width=1pt,minimum size=.2cm](mixRcross) {};

	\node at (dotTopHalfL) [circle,draw=black,inner sep=.1cm,line width= .2mm](mixT) {};
	\node at (dotTopHalfL) [cross,line width=1pt,minimum size=.2cm](mixTcross) {};

	\path[]
	(dotLeftL) edge[graviton,line width=1.2 pt] node  {} (mixL)
	(mixL) edge[pmfield,line width=1.2 pt] node  {} (vertexL)
	(dotRightL) edge[graviton,line width=1.2 pt] node  {} (mixR)
	(mixR) edge[pmfield,line width=1.2 pt] node  {} (vertexL)
	(dotTopL) edge[graviton,line width=1.2 pt] node  {} (mixT)
	(mixT) edge[pmfield,line width=1.2 pt] node  {} (vertexL);

	\end{scope}

\end{tikzpicture}
}
\end{center}
\caption{Partially massless spin-2 contributions to $\langle \gamma^{3}\rangle$. Here, a mixing vertex \mixing~corresponds to a factor of $\re\langle T\Sigma\rangle$ while a PM line, \pmline, is a factor of $\re\langle \Sigma^{2}\rangle^{-1}$. }
\label{fig:GravitonnPointFunctionsFromPMFields}
\end{figure}
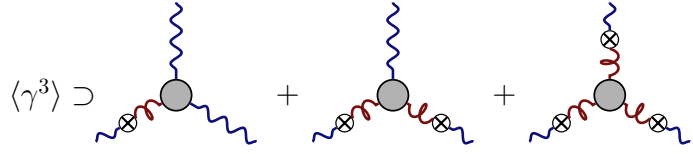

We now introduce the partially massless spin-2 fields, for which we use the notation $\varphi\to \sigma_{ij}$, $\mathcal{O}\to \Sigma_{ij}$, and denote them in graphs by \,\pmline~.  By including self-interactions and cubic couplings between the graviton and the partially massless field, we can generate non-trivial $\langle \Sigma T^{2}\rangle$, $\langle T\Sigma^{2}\rangle$, and $\langle \Sigma^{3}\rangle$ coefficients corresponding to the diagrams in Fig.~\ref{fig:GravitonPM3PointCoefficients}.  If there exists a mixing term $\langle T\Sigma \rangle$ in the wavefunction, then these cubic coefficients can imprint upon the graviton bispectrum as in Fig.~\ref{fig:GravitonnPointFunctionsFromPMFields}.\footnote{Similar mixing terms were considered in \cite{Lee:2016vti,Baumann:2017jvh} where non-Gaussianities involving scalar fluctuations were computed.}  For instance, the middle diagram in Fig.~\ref{fig:GravitonnPointFunctionsFromPMFields} gives a contribution to $\langle\gamma^{3}\rangle$ of the schematic form
\begin{align}
\langle \gamma^{3}\rangle\sim \frac{1}{\re\langle T^{2}\rangle^{3}}\frac{\re \langle T\Sigma\rangle^{2}}{\re \langle \Sigma^{2}\rangle^{2}}\re\langle T\Sigma^{2}\rangle\label{Sizeofgamma3FromMiddleDiagram}\ .
\end{align}

The contributions from partially massless fields to the graviton bispectrum can in principle be larger than the contributions from graviton self-interactions, {\it i.e.}~those dictated by General Relativity.
We can estimate the size of the various contributions to graviton correlators on scales $k\sim H$ by dimensional analysis and diagrammatics. The contributions to $\langle \gamma^{3}\rangle\sim E^{-6}$, from various interactions, can be estimated as follows:\footnote{Strictly speaking, the \textit{primed} wavefunction coefficient $\langle \gamma^{3}\rangle'$ which has had its momentum conserving delta function removed scales as $E^6$, but in this introduction we omit the primes in order to leave the presentation uncluttered.}

\vspace{.15cm}
\begin{itemize}
\item \textbf{Graviton self-interactions}: The $\langle T^{2}\rangle$ coefficient depicted on the left in Fig.~\ref{fig:Graviton2And3PointCoefficients} is generated by the Einstein--Hilbert operator and hence scales as  $\langle T^{2}\rangle\propto \MP ^{2}$. 
The cubic coefficient, $\langle T^{3}\rangle$, can be generated by the Einstein--Hilbert vertex, or by a higher derivative $\sim \MP^2W^{3}/\Lambda_{W^{3}}^{4}$ Weyl-cubed vertex \cite{Maldacena:2011nz}, and thus has contributions $\langle T^{3}\rangle\propto \MP ^{2}$ and $\langle T^{3}\rangle\propto \MP^2/\Lambda_{W^{3}}^{4}$. Therefore, from \eqref{SchematicPowerBispectrumGraviton} and dimensional analysis, it follows that on scales $k\sim H$ the bispectrum is of the form\footnote{Here, only factors of $H$ are used to fix the dimensions.  This relies on the assumption that $\langle \gamma^{3}\rangle$ is time-independent at superhorizon scales (as is suggested by the freeze-out behavior of the graviton mode function) which precludes factors of $\tau$ from appearing in the analysis. This assumption does not always hold.  For instance, a similar analysis of the parity violating $\sim \epsilon WWW$ operator would give an estimate of the form \eqref{Sizeofgamma3InGR}, while the precise calculation demonstrates that this operator only produces dS non-Gaussianity which decays in time \cite{Soda:2011am,Maldacena:2011nz}.}
\be
\langle \gamma^{3}\rangle_{\rm self-int.}\sim \frac{1}{H^{6}}\left (\frac{H}{\MP }\right )^{4}\left [1+\left (\frac{H}{\Lambda_{W^{3}}}\right )^{4}\right ]\,,\label{Sizeofgamma3InGR}
\ee
so that for $\Lambda_{W^3} \lesssim H$ the higher-derivative shape can be of the same order as the Einstein--Hilbert contribution.

\item \textbf{Interactions with PM spin-2 fields}: For concreteness, we estimate the contribution of the middle diagram in Fig.~\ref{fig:GravitonnPointFunctionsFromPMFields} to $\langle\gamma^{3}\rangle$.  We can take $\langle \Sigma^{2}\rangle\propto \MP ^{2}$ and further write the mixing coefficient as $\langle T\Sigma\rangle\propto \Lambda_{\rm mix}^{2}$.   From \eqref{Sizeofgamma3FromMiddleDiagram}, this yields
\begin{align}
\langle \gamma^{3}\rangle_{\rm PM-int.}\sim \frac{1}{H^{6}}\left (\frac{H}{\MP }\right )^{2}\left (\frac{\Lambda_{\rm mix}}{\MP }\right )^{4}\frac{\re\langle T\Sigma^{2}\rangle}{H^{3}}\ .
\end{align}
The coefficients $\langle T\Sigma\rangle$ and $\langle T\Sigma^{2}\rangle$ determine the size of PM-induced non-Gaussianity.  
Both coefficients also correct the graviton power spectrum, as shown in Fig.~\ref{fig:OneLoopgamma2CorrectionPlusMixing}. In order for the corrections to be small we require $\Lambda_{\rm mix}/\MP\ll 1 $ (see Sec.~\ref{sec:Mixing}) and\footnote{In \eqref{TS2RoughBound}, we've estimated the loop correction at scales $k\sim H$ as
\begin{align}
\langle \gamma^{2}\rangle_{\rm 1-loop}\sim \frac{1}{\re\langle T^{2}\rangle^{2}}\int\rd^{3}k\, \frac{\re\langle T\Sigma^{2}\rangle^{2}}{\re\langle \Sigma^{2}\rangle^{2}}\sim \frac{1}{H^{3}}\left (\frac{H}{\MP }\right )^{8} \left (\frac{\re\langle T\Sigma^{2}\rangle}{H^{3}}\right )^{2}\, .
\end{align}
If we worked with canonically-normalized fields such that the Lagrangian is schematically $\mathcal{L}\sim -(\partial\gamma)^{2}-(\partial\sigma)^{2}+\Lambda_{\rm UV}\left (\frac{\partial}{\Lambda_{\rm UV}}\right )^{2n}\sigma^{2}\gamma$, then the bound \eqref{TS2RoughBound} translates to the reasonable criteria $\Lambda_{\rm UV}\gg H$.} 
\begin{align}
\frac{\re\langle T\Sigma^{2}\rangle}{H^{3}}\ll \left (\frac{\MP }{H}\right )^{3}\ .\label{TS2RoughBound}
\end{align}
These power spectrum constraints are weaker than the requirement that $\langle \gamma^{3}\rangle_{\rm PM-int.}$ remain perturbative at scales $k\sim H$:
\begin{align}
\frac{1}{H^{3}}\frac{\langle \gamma^{3}\rangle_{\rm PM-int.}}{\langle \gamma^{2}\rangle^{3/2}}\ll 1  \quad \implies \quad \left (\frac{\Lambda_{\rm mix}}{\MP }\right )^{4}\frac{\re\langle T\Sigma^{2}\rangle}{H^{3}}\ll \frac{H}{\MP}\ .\label{PMParameterComboRoughBound}
\end{align}
However, the 3pt functions are fixed even if the spin-2 sector is strongly-interacting.
\end{itemize}

\noindent
We see that it is possible for additional light spin-2 fields to boost the size of the graviton bispectrum, while only weakly affecting the power spectrum.  The ratio of the PM induced bispectrum to the GR bispectrum is approximated by
\begin{align}
\frac{\langle\gamma^{3}\rangle_{\rm PM-int.}}{\langle \gamma^{3}\rangle_{\rm self-int}}\sim \left (\frac{\MP}{H}\right )^{2}\left (\frac{\Lambda_{\rm mix}}{\MP}\right )^{2}\frac{\re \langle T\Sigma^{2}\rangle}{H^{3}}\ll \frac{\MP}{H}\ ,
\end{align}
where only the Einstein--Hilbert contribution to $\langle \gamma^{3}\rangle_{\rm self-int}$ was retained in the estimate and \eqref{PMParameterComboRoughBound} was used. The above is an extremely weak limit: theoretically, we require $\frac{H}{\MP}\ll 1$ for a weakly coupled gravitational description to be valid and, further, the current experimental bounds are $\frac{H_{\star}}{\MP}<2.5\times 10^{-5}$ (95\% CL) \cite{Akrami:2018odb}, where $H_{\star}$ is the Hubble scale at the end of inflation.

These estimates illustrate the general point that the induced graviton bispectrum due to the presence of additional spin-2 particles can easily be of the same size or much larger than the intrinsic Einstein gravity 3-point function. Essentially this is because the additional spin-2 fields can have large intrinsic non-Gaussianities without disrupting any other observable quantities.

\begin{figure}
\begin{center}
\resizebox{16cm}{!}{
\begin{tikzpicture}

\begin{scope}[xshift=0cm]

\node[font=\Large] at (-1.5,0) {$\langle \gamma^{2}\rangle=$};

	\coordinate (dotL) at (0,0) {};

	\coordinate (dotR) at (4,0) {};

\path[]
	(dotL) edge[graviton,line width=1.2 pt] node  {} (dotR);

	\end{scope}

	\begin{scope}[xshift=6cm]
	
	\node[font=\Large] at (-1,0) {$+$};

	\coordinate (dotL) at (0,0) {};
	\coordinate (dotR) at (4,0) {};

	\node at ([shift=({0:1.2})]dotL) [circle,draw=black,inner sep=.1cm,line width= .2mm](mixL) {};
	\node at ([shift=({0:1.2})]dotL) [cross,line width=1pt,minimum size=.2cm](mixLcross) {};

	\node at ([shift=({180:1.2})]dotR) [circle,draw=black,inner sep=.1cm,line width= .2mm](mixR) {};
	\node at ([shift=({180:1.2})]dotR) [cross,line width=1pt,minimum size=.2cm](mixRcross) {};

\path[]
	(dotL) edge[graviton,line width=1.2 pt] node  {} (mixL)
	(mixL) edge[pmfield,line width=1.2 pt] node  {} (mixR)
	(mixR) edge[graviton,line width=1.2 pt] node  {} (dotR);

	\end{scope}

	\begin{scope}[xshift=12cm]

	\node[font=\Large] at (-1,0) {$+$};
	
	\coordinate (dotL) at (0,0) {};
	\coordinate (dotR) at (4,0) {};

	\node at ([shift=({0:1.2})]dotL) [vertex,line width=.8pt,minimum size=.5cm](vertexL) {};
	\node at ([shift=({180:1.2})]dotR) [vertex,line width=.8pt,minimum size=.5cm](vertexR) {};

	\path[]
	(dotL) edge[graviton,line width=1.2 pt] node  {} (vertexL)
	(dotR) edge[graviton,line width=1.2 pt] node  {} (vertexR)
	(vertexL) edge[pmfield,line width=1.2 pt,bend left=50] node  {} (vertexR)
	(vertexR) edge[pmfield,line width=1.2 pt,bend left=50] node  {} (vertexL);
	
	\node[font=\Large] at (5,0) {$+\ldots$};
	
	\end{scope}

\end{tikzpicture}
}
\end{center}
\caption{Corrections to the graviton power spectrum due to insertions of $\langle T\Sigma\rangle$ (middle) and $\langle T\Sigma^{2}\rangle$ (right). The first diagram is represents the familiar result $\langle \gamma^{2}\rangle  \sim \frac{H^2}{\MP^2k^{3}}$.}
\label{fig:OneLoopgamma2CorrectionPlusMixing}
\end{figure}

\subsection{Outline and conventions}

We begin in Sec.~\ref{sec:FreeSpin2sAndQuadWavefunction} by reviewing the physics of spin-2 fields on de Sitter space, where we derive the bulk-to-boundary propagator for general masses and compute the quadratic wavefunction.  The cubic wavefunction computation only requires knowledge of the action evaluated \textit{on-shell}, so in Section~\ref{sec:CubicVertices} we enumerate the possible on-shell cubic vertices involving spin-2 fields. 
In Section~\ref{sec:CubicWavefunctionCoefficients} we use these ingredients to compute the cubic wavefunction coefficients depicted in Fig.~\ref{fig:GravitonPM3PointCoefficients}, and in Section~\ref{sec:non-Gaussianity} we use these to compute the contributions to the graviton non-Gaussianities depicted in Fig.~\ref{fig:GravitonnPointFunctionsFromPMFields}.  We conclude in Section~\ref{sec:Discussion}. We collect some useful technical results in a number of appendices. In Appendix~\ref{app:TheWavefunction} we give a brief orientation to the wavefunctional approach in cosmology. In Appendices~\ref{app:Projectors} and~\ref{app:PolarizationTensors} we list and give some properties of a set of spatial projection tensors which we use in a number of places in the text and give an explicit basis of helicity-2 polarizations, which we use to evaluate correlation functions. In Sec.~\ref{app:OnShellCubicTerms} we describe a method for deriving gauge invariant on-shell cubic interactions and use this to derive all consistent cubic interactions between massless and partially massless fields in arbitrary dimensions.
The wavefunction coefficients are sensitive to both integrations-by-parts and field redefinitions and we discuss these procedures and their relationship in App.~\ref{app:IBPAndTheWavefunction}. Finally in Appendix~\ref{app:CFTTwoPointFunctions} we review the structure of CFT 2-point functions for spinning fields.

\vspace{.2cm}

\noindent
{\bf Conventions:} We use mostly plus signature, work in $(d+1)$ spacetime dimensions, and use the curvature conventions $R^{\rho}{}_{\sigma\mu\nu}=\partial_{\mu}\Gamma^{\rho}_{\nu\sigma}+\ldots$ and $R_{\mu\nu}=R^{\rho}{}_{\mu\rho\nu}$.  On de Sitter space, ${\rm dS}_{d+1}$, the Hubble scale is denoted by $H$, where $R=d(d+1)H^{2}$.  We work exclusively in the flat slicing of de Sitter
\be
\rd s^2  = \frac{1}{H^2\tau^2}\left(-\rd\tau^2 + \rd\bfx^2\right),
\label{dSMetricConformalTime}
\ee
where $\tau\in (-\infty,0)$ is the proper time.
Greek letters $\mu,\nu,\rho,\ldots$ are used for spacetime indices, while lower case Latin letters $i,j,k,\ldots$ are reserved for spatial indices.  Spacetime indices are always raised and lowered with the full metric $g_{\mu\nu}$.  Spatial indices will always be manipulated using the flat $\delta_{ij}$ metric, {\it e.g.},~$\partial^{i}h_{ij}=\delta^{ik}\partial_{i}h_{kj}$.  Tensors are symmetrized and anti-symmetrized with unit weight, {\it e.g.},~$T_{(\mu\nu)}=\frac{1}{2!}(T_{\mu\nu}+T_{\nu\mu})$ and $T_{[\mu\nu]}=\frac{1}{2!}(T_{\mu\nu}-T_{\nu\mu})$.   Spatial vectors are bolded as in $ \vec{x}\equiv\bfx$ or $\vec{k}\equiv\bfk$. We use the following Fourier conventions: $f_{\bfk}(\tau)\equiv \int\rd^{d}x\, e^{-i\bfk\cdot\bfx}f(\tau,\bfx)$, $f(\tau,\bfx)= \int\rd^{d}\tilde{k}\, e^{i\bfk\cdot\bfx}f_{\bfk}(\tau)$, where we define $\tilde{k}\equiv k/(2\pi)$ and $\tilde{\delta}^{d}(\bfk)\equiv (2\pi)^{d}\delta^{d}(\bfk)$ to minimize the number of explicit $2\pi$ factors appearing.  In three-point correlators, we denote the sum of magnitudes of momenta by $k_{T}\equiv k_{1}+k_{2}+k_{3}$.

\section{Free spin-2 fields on ${\rm dS}_{d+1}$\label{sec:FreeSpin2sAndQuadWavefunction}}
We begin by reviewing the physics of free spin-2 fields with general mass on de Sitter space and then derive their propagators and superhorizon two-point functions.

\subsection{The quadratic action}
The degrees of freedom of a 
massive spin-2 field of mass $m$ on $(d+1)$-dimensional de Sitter space are carried by a symmetric $2$-index tensor, $h_{\mu\nu}$, with the following quadratic action,
\begin{align}
\label{QuadraticSpin2ActiondS}
S_{2}&=\frac{\MP ^{d-1}}{4}\int\rd^{d+1}x\, \sqrt{-g}\left(-\frac{1}{2}\nabla_{\rho}h_{\mu\nu}\nabla^{\rho}h^{\mu\nu} -\frac{1}{2}\bigg (m^{2}+2H^{2}\right )h_{\mu\nu}h^{\mu\nu} \\
&~~~~~~~~~~~~~~~~~~~~~~~~~~\quad +\nabla^{\rho}h_{\rho\mu}\nabla_{\nu}h^{\nu\mu}-\nabla_{\mu}h\nabla_{\nu}h^{\mu\nu}+\frac{1}{2}\nabla_{\mu}h\nabla^{\mu}h+\frac{1}{2}\left (m^{2}-H^{2}(d-2)\right )h^{2}\bigg),
\nonumber
\end{align}
where $h\equiv h^{\mu}_{\mu}$ and all covariant derivatives and contractions are defined with respect to the background ${\rm dS}_{d+1}$ metric $g_{\mu\nu}$.  We work in planar inflationary coordinates where the background metric takes the form~\eqref{dSMetricConformalTime}.
The field, $h_{\mu\nu}$, in~\eqref{QuadraticSpin2ActiondS} is dimensionless, and the action is normalized such that the $m\to 0$ limit of~\eqref{QuadraticSpin2ActiondS} coincides with the expansion of the Einstein--Hilbert action, with a cosmological constant, about ${\rm dS}_{d+1}$.\footnote{Specifically, the Einstein--Hilbert action which admits~\eqref{dSMetricConformalTime} as a solution is given by
\be
  S_{\rm EH}= \frac{\MP ^{2}}{2}\int\rd^{d+1}x\, \sqrt{-g}\left (R-d(d-1)H^{2}\right )\ .\label{EinsteinHilbertActiondS}
\ee
Perturbing the metric as $\rd s^{2}=\left (g_{\mu\nu}+h_{\mu\nu}\right )\rd x^{\mu}\rd x^{\nu}$, with $g_{\mu\nu}$ the background dS metric in \eqref{dSMetricConformalTime}, gives $S_{\rm EH}=\lim_{m\to 0}S^{(2)}$ at $\mathcal{O}(h^{2})$ (after integrations by parts).}  In \eqref{QuadraticSpin2ActiondS}, the derivatives have been organized that the entire second line vanishes when $h_{\mu\nu}$ is transverse and traceless, $\nabla^{\mu}h_{\mu\nu}=h=0$, as our solutions will always obey these properties.

\subsubsection{Distinguished mass values and equations of motion}
The linear equations of motion obtained from the action~\eqref{QuadraticSpin2ActiondS} are
\be
\square h_{\mu\nu} - 2\nabla_{\mu}\nabla_{(\mu}h^\mu_{\nu)}+ g_{\mu\nu}\left(\nabla_\mu\nabla_\nu h^{\mu\nu}- \square h\right)+\nabla_\mu\nabla_\nu h +(2dH^2+m^2)h_{\mu\nu}-(dH^2+m^2)h g_{\mu\nu}= 0.
\label{GenericMasshEOM}
\ee
The number of degrees of freedom described by this equation  depends on the value of the mass parameter, $m^2$. On de Sitter space, there are two distinguished points:
\begin{itemize}
\item  When $m^{2}=0$, the spin-2 is \textit{massless} and $h_{\mu\nu}$ only propagates tensor degrees of freedom.  The massless quadratic action enjoys the usual linearized diffeomorphism gauge symmetry, with a vector gauge parameter, $\xi_{\nu}$,
\begin{align}
h_{\mu\nu}&\mapsto h_{\mu\nu}+2\nabla_{(\mu}\xi_{\nu)}\ .\label{MasslesSymmetry}
\end{align}
The two conditions $\nabla^{\mu}h_{\mu\nu}=0$ and $h=0$ may be imposed as on-shell gauge conditions~\cite{Higuchi:1991tn} and the resulting wave equation for the physical degrees of freedom is
\be
\left (\square-2H^{2}\right )h_{\mu\nu}=0 .\label{MasslesshEOM}
\ee
\item When $m^{2}=(d-1)H^{2}$, the spin-2 is \textit{partially massless} \cite{Deser:1983tm} and $h_{\mu\nu}$ only propagates tensor and vector degrees of freedom.  The partially massless quadratic action enjoys a gauge symmetry with a scalar gauge parameter, $\chi$,
\be
h_{\mu\nu}\mapsto h_{\mu\nu}+\left (\nabla_{\mu}\nabla_{\nu}+H^{2}g_{\mu\nu}\right )\chi\ .\label{PMSymmetry}
\ee
It is possible to enforce $h=0$ as a gauge choice~\cite{Deser:1983mm}, after which the equations of motion further imply that $\nabla^{\nu}h_{\mu\nu}=0$. In this gauge, the resulting equation of motion is
\be
\left (\square-(d+1)H^{2}\right )h_{\mu\nu}=0 .\label{PMhEOM}
\ee
\end{itemize}
For other mass values, the spin-2 is \textit{massive} and $h_{\mu\nu}$ propagates tensor, vector and scalar degrees of freedom.  At generic values of the mass, by taking traces and divergences of~\eqref{GenericMasshEOM} one finds the equations $\nabla^{\mu}h_{\mu\nu}=h=0$, which then simplify~\eqref{GenericMasshEOM} to
\be
 \left (\square-m^2-2H^{2}\right )h_{\mu\nu}=0\ .\label{MassivehEOM}
\ee

With the exception of the massless and partially massless points, the theory is only unitary for $m^{2}>(d-1)H^{2}$, the so-called Higuchi bound~\cite{Higuchi:1986py}.
Unitary  spin-2 fields on ${\rm dS}$ must therefore belong to one of three categories: $m^{2}>(d-1)H^{2}$, $m^2 = 0$ or $m^2 = (d-1)H^2$. In the language of de Sitter representation theory, massless and partially massless fields belong to the exceptional series (which coincides with the discrete series in $d=3$), while massive fields can either belong to the complementary or the principal series, depending on the value of their mass: spin-2 fields in the mass range $(d-1) < \frac{m^2}{H^2} \leq \frac{d^2}{4}$ belong to the complementary series, while fields with masses $\frac{m^2}{H^2} > \frac{d^2}{4}$ belong to the principal series.  All other mass values correspond to non-unitary representations \cite{Thomas,Newton,Basile:2016aen}.

\subsection{Mode functions\label{sec:ModeFunctions}}
In this section we solve the linear equations of motion for generic mass spin-2 fields on ${\rm dS}_{d+1}$ in Fourier space.\footnote{A similar calculation in ${\rm AdS}_{d+1}$ can be found in~\cite{Polishchuk:1999nh}.}
These solutions are typically called mode functions in the cosmology literature.
The massless and PM cases can then be obtained as limits of the general solutions.
In the following section, these results will be re-packaged into the \textit{bulk-to-boundary} propagator, which is a solution to the equations of motion with some specified Dirichlet boundary conditions and which plays an important role in the computation of the late-time wavefunctional.

\subsubsection{Generic mass solutions\label{subsubsec:GenericMassSolutions}}

In the generic equations of motion, we trade the mass, $m$, in favor of a parameter, $\mu$, defined below, after which the equations take on the form
\be
\left (\square-H^2\left (2+\frac{d^{2}}{4}+\mu^{2}\right )\right )h_{\mu\nu}=0\ , \qquad \nabla^{\mu}h_{\mu\nu}=0 \ , \qquad h =0\ , \qquad i\mu\equiv \sqrt{ \frac{d^2}{4}-\frac{m^2}{H^2}} \ .\label{GenericEOM}
\ee
The massless and PM cases correspond to $i\mu=\frac{d}{2}$ and $i\mu=\frac{d-2}{2}$, respectively.
As discussed previously, the $h=0$ and $\nabla^{\mu}h_{\mu\nu}=0$ conditions appear as equations of motion for generic $\mu$, but correspond to a gauge choice in the massless and PM cases.

We first decompose $h_{\mu\nu}$ in ADM-like variables,
\begin{subequations}
\label{ADMLikeDecomposition}
\begin{align}
h_{00} &= -\frac{1}{H^2\tau^2}\delta N,\\
h_{0i} &= \frac{1}{H^2\tau^2}\delta N_i,\\
h_{ij} &= \frac{1}{H^2\tau^2}\varphi_{ij},
\end{align}
\end{subequations}
and then further split $\delta N_{i}(\tau,\bfx)$ and $\varphi_{ij}(\tau,\bfx)$ into scalar, vector and tensor components which are irreducible with respect to the spatial SO$(d)$ symmetries. This decomposition is most naturally expressed in Fourier space: 
\begin{subequations}
\label{SVTDecomp}
\begin{align}
\varphi_{ij}^{\bfk} &= {\varphi}^{TT,\,\bfk}_{ij}+2ik_{(i}V^{T,\,\bfk}_{j)}+S^{\bfk} \delta_{ij}+Q^{\bfk} \left (\frac{k_{i}k_{j}}{k^{2}}-\frac{\delta_{ij}}{d}\right )\\
\delta N_{i}^{\bfk}&=\delta{N}^{T,\,\bfk}_{i}+ik_{i}\delta N_L^{\bfk}\,.
\end{align}
\end{subequations}
The various components are transverse and traceless according to
\be
k^{i}\delta{N}_{i}^{T,\,\bfk}=\delta^{ij}{\varphi}^{TT,\,\bfk}_{ij}=k^{i}{\varphi}_{ij}^{TT,\,\bfk}=k^{i}V_{i}^{T,\,\bfk}=0\,,
\ee
and the temporal dependence of all components is being suppressed.
The decomposition of $\varphi$ is defined such that the projectors introduced in Appendix \ref{app:Projectors} isolate the components shown in \eqref{SVTDecomp}
\begin{align}
 (\Pi_{TT}^{\bfk} ){}_{ij}{}^{lm}\varphi^{\bfk}_{lm}&= {\varphi}_{ij}^{TT,\,\bfk}\ , &  (\Pi_{V}^{\bfk} ){}_{ij}{}^{lm}\varphi^{\bfk}_{lm}&= 2ik_{(i}V^{T,\,\bfk}_{j)}\ , \quad \nn
(\Pi_{S}^{\bfk} ){}_{ij}{}^{lm}\varphi^{\bfk}_{lm}&= S^{\bfk} \delta_{ij} \ , &  (\Pi_{Q}^{\bfk} ){}_{ij}{}^{lm}\varphi^{\bfk}_{lm}&=Q^{\bfk}\left (\frac{k_{i}k_{j}}{k^{2}}-\frac{\delta_{ij}}{d}\right )\ .
\end{align}

We now turn to solving the equations of motion.
The trace condition, $h=0$, fixes the lapse, $\delta N$, in terms of $S$:
\be
\delta N^{\bfk}=-d \,S^{\bfk}\ .\label{LapseConstraint}
\ee  
Writing  $\nabla^{\mu}h_{\mu\nu}=\mathcal{E}_{\nu}$, the divergence constraint contains the two scalar conditions: $\mathcal{E}_{0}^{\bfk}=k^{i}\mathcal{E}_{i}^{\bfk}=0$. Written in terms of the variables in~\eqref{SVTDecomp}, these become
\begin{align}
&k^{2}\delta N_L^{\bfk}-\frac{d(d+1)}{\tau}S^{\bfk}+d\,S'^{\bfk}=0\nn
&ik^{2}S^{\bfk}+\frac{i(d-1)}{d}k^{2}Q^{\bf k}+\frac{i(d+1)}{\tau}k^{2}\delta N_L^{\bfk}-ik^{2}\delta N_L'^{\bfk}=0\, .
\end{align} 
These equations can be solved for $\delta N_L^\bfk$ and $Q^\bfk$:
\begin{align}
\delta N_L^{\bfk}&=\frac{d}{k^{2}\tau^{2}}\left(\frac{(d+1)}{\tau}S^{\bfk}-S'^{\bfk}\right )\nn 
Q^{\bfk}&=-\frac{d}{(d-1)k^{2}}\left(\frac{d(d+1)(d+2)+k^{2}\tau^{2}}{\tau^{2}}S^{\bfk}-\frac{2d(d+1)}{\tau}S'^{\bfk}+d\, S''^{\bfk}\right)\ .\label{PsiTildeSigmaConstraints}
\end{align}
The divergence condition also contains the following vector constraint,\footnote{This is isolated by  applying the transverse projector $\pi_{ij}^{\bfk}$ to $\mathcal{E}_{i}^{\bfk}$, where
\begin{align}
\pi_{ij}^{\bfk} = \delta_{ij} - \frac{k_i k_j}{k^2}\ .\label{TransverseProjector}
\end{align}  
}
\be
-k^{2}V^{T,\,\bfk}_{i}+\frac{(d+1)}{\tau}\delta {N}^{T,\,\bfk}_{i}-\partial_\tau\delta{N}^{T,\,\bfk}_{i}=0\ ,\label{ScalarDivergenceConstraints}
\ee
The constraint \eqref{ScalarDivergenceConstraints} shows that $V^{T,\,\bfk}_{i}$ is determined by $\delta {N}^{T,\,\bfk}_{i}$.  It proves fruitful at this point to express the remaining $\delta {N}_{i}^{T,\,\bfk}, {\varphi}^{TT,\,\bfk}_{ij }, V^{T,\,\bfk}_{i}$, and $S^{\bfk}$ components as time-dependent functions multiplying time-independent tensor structures.  Since the vector constraint implies that $V^{T,\,\bfk}_{i}\propto\delta {N}^{T,\,\bfk}_{i}$, these two components also share the same tensor structure, and hence we can write
\be
{\varphi}^{TT,\,\bfk}_{ij}\equiv f_{TT}(\tau,k)\bar\varphi_{ij}^{TT,\,\bfk}, \quad V^{T,\,\bfk}_{i}\equiv f_{V}(\tau,k) \bar V_{i}^{T,\,\bfk}, \quad S^{\bfk}\equiv f_{S}(\tau,k) \bar S^{\bfk} , \quad \delta{N}^{T,\,\bfk}_{i}\equiv f_{N^T}(\tau,k)\bar V_i^{T,\,\bfk}, 
\ee
where $\bar\varphi,\bar V$ and $\bar S$ are $\tau$-independent.  After these replacements, \eqref{ScalarDivergenceConstraints} reads
\be
f_{V}(\tau,k)=\frac{1}{k^{2}}\left (\frac{(d+1)}{\tau}f_{N^T}(\tau,k)-f'_{N^T}(\tau,k)\right )\ .\label{fVConstraint}
\ee
Using the above constraints and definitions, the remaining wave equation 
\be
{\cal E}_{\mu\nu}\equiv \left (\square-H^2\left (2+\frac{d^{2}}{4}+\mu^{2}\right )\right )h_{\mu\nu}=0
\ee
can be solved straightforwardly. Only some of its components give non-trivial relations:
\begin{itemize}
\item The $\mathcal{E}^{\bfk}_{00}$ component gives an equation for $f_{S}$ alone:
\be
f_{S}''-\frac{(d+3)}{\tau}f_{S}'+\left (k^{2}+\frac{(d+4)^{2}+4\mu^{2}}{4\tau^{2}}\right )f_{S}=0\, .\label{fSEOM}
\ee
\item The $\pi^{\bfk}_{ij}\mathcal{E}^{\bfk}_{0j}$ component gives an equation for $f_{N^T}$ alone:
\be
f_{N^T}''-\frac{(d+1)}{\tau}f_{N^T}'+\left (k^{2}+\frac{(d+2)^{2}+4\mu^{2}}{4\tau^{2}}\right )f_{N^T}=0\, .\label{fdeltaNEOM}
\ee
\item Finally, the transverse, traceless components  $(\Pi_{TT}^\bfk)_{ij}{}^{lm}\mathcal{E}^{\bfk}_{lm}$ give an equation for $f_{TT}$ alone:
\begin{align}
f_{TT}''-\frac{(d-1)}{\tau}f_{TT}'+\left (k^{2}+\frac{d^{2}+4\mu^{2}}{4\tau^{2}}\right )f_{TT}=0\, .\label{fTEOM}
\end{align}
\end{itemize} 
The general solutions to \eqref{fSEOM}, \eqref{fdeltaNEOM}, and \eqref{fTEOM} are expressions involving Hankel functions of first and second kinds.  However, the Bunch--Davies vacuum condition requires solutions which behave as $\sim e^{+ik\tau}$ at early times, which corresponds to only using the Hankel function of the second kind.  The desired solutions to the linear equations are then given by
\begin{align}
f_{S}(\tau,k)&=(-k\tau)^{\frac{d+4}{2}}H^{(2)}_{ i\mu}(-k\tau),\nn
f_{N^T}(\tau,k)&=(-k\tau)^{\frac{d+2}{2}}H^{(2)}_{ i\mu}(-k\tau),\nn
f_{V}(\tau,k)&=\frac{\tau}{2}\left (-k\tau\right )^{\frac{d-2}{2}}\left [2k\tau H^{(2)}_{ i\mu-1}(-k\tau)+(d+2 i\mu)H^{(2)}_{ i\mu}(-k\tau)\right], \nn
f_{TT}(\tau,k)&=(-k\tau)^{\frac{d}{2}}H^{(2)}_{ i\mu}(-k\tau),\label{fSolns}
 \end{align} 
 \noindent
 where $f_{V}$ was determined through \eqref{fVConstraint}. The constrained fields are then determined using~\eqref{PsiTildeSigmaConstraints} and~\eqref{LapseConstraint}. Making the definitions
\begin{align}
\delta N^\bfk &\equiv f_N(\tau, k) \bar S^\bfk,\\
Q^\bfk &\equiv f_Q(\tau, k)\bar S^\bfk,\\
\delta N_L^\bfk &\equiv f_{N_L}(\tau, k)\bar S^\bfk,
\end{align}
we find the following expressions for the mode functions:
\begin{align}
f_{N}(\tau,k)&=-d\, (-k\tau)^{\frac{d+4}{2}}H^{(2)}_{ i\mu}(-k\tau)\nn
f_{Q}(\tau,k)&=-d\frac{(-k\tau)^{\frac{d}{2}}}{4(d-1)}\Bigg[\Big(-d  \left(4 d   i\mu +(d -2) d -4
   \left(k ^2 \tau ^2+ i\mu \right)-4  \mu ^2\right)-4 k ^2 \tau ^2\Big )H^{(2)}_{ i\mu-2}(-k\tau)\nn
&\quad+\Big (2 k ^2
   \tau ^2-2 (d +2  i\mu -2) \left(d  \left(d  ( i\mu -1)-2
   \left(k ^2 \tau ^2+ i\mu \right)-2  \mu ^2\right)\right)\Big)\frac{H^{(2)}_{ i\mu-1}(-k\tau)}{k\tau}\Bigg]\nn
f_{N_L}&=\frac{d\, \tau }{2}\left (-k\tau\right )^{\frac{d}{2}}\left [2k\tau H^{(2)}_{ i\mu-1}(-k\tau)+(d+2 i\mu-2)H^{(2)}_{ i\mu}(-k\tau)\right ]\,.\label{ConstrainedSolutionsGeneric}
\end{align}
Putting all this together, the generic momentum-space solution for $h_{\mu\nu}$ can be written as
\begin{subequations}
 \label{ADMLikeSolutionGeneric}
\begin{align}
h_{00}^\bfk &= -\frac{1}{H^2\tau^2} f_{N} \bar S^\bfk,\\
h_{0i}^\bfk &= \frac{1}{H^2\tau^2} \left(f_{N^T} \bar V_i^\bfk+ik_i f_{N_L}\bar S^\bfk\right),\\
h_{ij}^\bfk &= \frac{1}{H^2\tau^2}\left (f_{TT}\bar\varphi_{ij}^{TT,\,\bfk}+2if_{V}k_{(i}\bar V_{j)}^{T\,,\bfk}+f_{S}\bar S^{\bfk}\delta_{ij}+f_{Q}\bar S^{\bfk}\left (\frac{k_{i}k_{j}}{k^{2}}-\frac{\delta_{ij}}{d}\right )\right ),
\end{align}
\end{subequations}
where the $(\tau,k)$-dependence of the $f_{i}$'s has been suppressed.
The generic, massive solution is determined by the boundary data contained in the $(d+1)(d-2)/2$ components of the transverse, traceless tensor, $\bar\varphi_{ij}^{TT,\,\bfk}$, the $(d-1)$ components of the transverse vector, $\bar V_i^{\bfk}$, and the one component of $\bar S^{\bfk}$, totaling $(d(d+1)-2)/2$ degrees of freedom, which is the expected counting for a massive spin-2 \cite{Hinterbichler:2011tt}.

\subsubsection{Partially massless solutions}
We now discuss simplifications which can be made in the partially massless limit, where $i\mu=\frac{d-2}{2}$.  After choosing the gauge such that $h=\nabla^{\mu}h_{\mu\nu}=0$, there remain residual gauge-transformations of the form
\be
\delta h_{\mu\nu}=\left (\nabla_{\mu}\nabla_{\nu}+H^{2}g_{\mu\nu}\right )\chi \, ,\qquad\quad \left (\square+(d+1)H^{2}\right )\chi=0\, .\label{ResidualPMSymmetry}
\ee
We can solve this equation in Fourier space in the same way as in the previous Section. We find that
the solution for a residual $\chi$ is
\begin{align}
\chi^{\bfk}(\tau)&=(-k\tau)^{\frac{d}{2}}H^{(2)}_{\frac{d+2}{2}}(-k\tau)\bar \chi^{\bfk}\, ,\label{ChiResdiualExplicit}
\end{align}
for some $\tau$-independent $\bar\chi^{\bfk}$. By choosing $\bar\chi^{\bfk}=\frac{d}{H^{2}}\bar{S}^{\bfk}$, we can use $\chi$ to remove $\bar S^{\bfk}$ everywhere in the solution \eqref{ADMLikeSolutionGeneric}, yielding the PM solution
\begin{subequations}
\label{ADMLikeSolutionPM}
\begin{align}
h_{00}^\bfk &= 0,\\
h_{0i}^\bfk &= \frac{1}{H^2\tau^2} f_{N^T} \bar V_i^\bfk,\\
h_{ij}^\bfk &= \frac{1}{H^2\tau^2}\left (f_{TT}\bar\varphi_{ij}^{TT,\,\bfk}+2if_{V}k_{(i}\bar V_{j)}^{T\,,\bfk}\right ),
\end{align}
\end{subequations}
where all the mode functions should be understood to be evaluated at $i\mu = \frac{d-2}{2}$.
The PM solution is determined by the $(d+1)(d-2)/2$ components of the transverse, traceless $\bar\varphi_{ij}^{TT,\,\bfk}$ and the $(d-1)$ components of the transverse $\bar V_i^{T,\,\bfk}$, totaling $(d(d+1)-4)/2$ degrees of freedom.

\subsubsection{Massless solutions}
Even more drastic simplifications can be made in the massless limit, where $i\mu=\frac{d}{2}$.  After choosing the gauge such that $h=\nabla^{\mu}h_{\mu\nu}=0$, there remain residual gauge-transformations of the form
\be
\delta h_{\mu\nu}=2\nabla_{(\mu}\xi_{\nu)} \ ,\qquad\quad \left (\square+d\, H^{2}\right )\xi_{\mu}=0\ , \qquad\quad \nabla^{\mu}\xi_{\mu}=0\ .\label{ResidualPMSymmetry}
\ee
Using the same techniques, we can find the solution for a residual gauge transformation as a function of boundary data:
\begin{subequations}
\begin{align}
\xi_0^\bfk &= (-k\tau)^{\frac{d}{2}}H^{(2)}_{\frac{d+2}{2}}(-k\tau)\bar\xi^\bfk\\
\xi_i^\bfk &=(-k\tau)^{\frac{d-2}{2}}H^{(2)}_{\frac{d+2}{2}}(-k\tau)\bar\xi_i^{T,\,\bfk}+\frac{ik_i}{k^{2}\tau}\left (d(-k\tau)^{\frac{d}{2}}H^{(2)}_{\frac{d+2}{2}}(-k\tau)-(-k\tau)^{\frac{d+2}{2}}H^{(2)}_{\frac{d}{2}}(-k\tau)\right )\bar\xi^\bfk,
\end{align}
\end{subequations}
for some $\tau$-independent $\bar{\xi}^{\bfk}$ and $\bar\xi_i^{T,\,\bfk}$, where the latter is transverse . By choosing $\bar{\xi}^{\bfk}=\frac{d\, k}{2H^{2}}\bar S^{\bfk}$ and $\bar{\xi}_i^{T,\,\bfk}=\frac{k}{H^{2}}\bar V_i^{T,\,\bfk}$, we can use $\xi$ to remove $\bar S^{\bfk}$ and $\bar V_i^{T,\,\bfk}$ everywhere in the solution~\eqref{ADMLikeSolutionGeneric},  yielding the massless solution
\begin{subequations}
\label{ADMLikeSolutionMassless}
\begin{align}
h_{0\mu} &= 0,\\
h_{ij} &= \frac{1}{H^2\tau^2} f_{TT} \bar\varphi_{ij}^{TT,\,\bfk}.
\end{align}
\end{subequations}
The massless solution is determined by the $(d+1)(d-2)/2$ components of the transverse, traceless $\bar\varphi_{ij}^{TT,\,\bfk}$.

\subsection{Bulk-to-boundary propagators\label{sec:BulkBoundaryPropagators}}
The perturbative approach to computing the wavefunction we are taking amounts to computing the on-shell action as a function of boundary data on some $\tau=\tau_\star$ surface. The solutions to the linear equations of motion which satisfy some specified set of Dirichlet boundary conditions are known as {\it bulk-to-boundary propagators}.
It is straightforward to re-interpret the solutions from the previous Section as these objects.
In particular, we will write this relation in the form
\begin{align}
h_{\mu\nu}^{\bfk}(\tau)&=\frac{1}{H^{2}\tau^{2}}\mathcal{K}_{\mu\nu}^{\bfk}{}^{lm}(\tau)\bar{\varphi}^{\bfk}_{lm} \ ,
\label{BulkToBoundaryDefinition}
\end{align}
where $\bar{\varphi}_{ij}^{\bfk}$ is the boundary value of the $\varphi_{ij}$ field appearing in \eqref{ADMLikeDecomposition} and $\mathcal{K}_{\mu\nu}^{\bfk}{}^{ij}(\tau)$ is the bulk-to-boundary propagator. Since we require $\bar\varphi_{ij}$ to be the boundary value of the field $h_{\mu\nu}$, we therefore require the spatial parts of the bulk-to-boundary propagator to satisfy a relation of the schematic form
\begin{align}
\mathcal{K}^{\bfk}(\tau)\xrightarrow{\tau\to\ts}\mathds{1}\ ,
\end{align}
where $\mathds{1}$ is a type of identity matrix whose precise form will depend on the case at hand.

We now build the massless and PM propagators, which are relatively simple, and end with the massive case, which is slightly more cumbersome. In doing this, it is extremely useful to use the spatial projection tensors defined in Appendix~\ref{app:Projectors} to organize the bulk-to-boundary propagator into its irreducible components.

\subsubsection{The massless bulk-to-boundary propagator}
We first construct the massless bulk-to-boundary propagator.
 From \eqref{ADMLikeSolutionMassless}, the massless propagator is purely spatial, $\mathcal{K}^{\bfk}_{0\nu}{}^{ij}=0$. Its components are given by
\begin{tcolorbox}[colframe=white,arc=0pt,colback=greyish2]
\vspace{-.35cm}
\begin{align}
\mathcal{K}^{\bfk}_{ij}{}^{lm}(\tau)&=\left (\frac{\tau}{\ts}\right )^{\frac{d}{2}}\frac{H_{\frac{d}{2}}^{(2)}(-k\tau)}{H_{\frac{d}{2}}^{(2)}(-k\ts)} (\Pi_{TT}^{\bfk}){}_{ij}{}^{lm}\ .\label{MasslessBulkBoundary}
 \end{align} 
 \end{tcolorbox}
At the boundary, $\mathcal{K}^{\bfk}_{ij}{}^{lm}(\tau)$ reduces to the identity matrix for transverse, traceless tensors by construction, {\it i.e.},~$\mathcal{K}(\tau)\xrightarrow{\tau\to\ts}\Pi_{TT}$.  In the massless case, we define $\sigma_{ij}\to \gamma_{ij}$ in order to conform to standard notation for massless spin-2's in the literature and to distinguish the graviton from the PM and massive spin-2 fields.  Hence, the bulk solution will be written as
 \begin{align}
h_{\mu\nu}^{\bfk}(\tau)=\frac{1}{H^{2}\tau^{2}} \mathcal{K}^{\bfk}_{\mu\nu}{}^{lm}(\tau)\bar{\gamma}_{lm}^{\bfk}\ ,
 \end{align}
 rather than as \eqref{BulkToBoundaryDefinition}, and $\bar{\gamma}_{ij}^{\bfk}$ is required to be transverse and traceless.\footnote{A sidenote: since ${\cal K}$ is transverse and traceless, so too will the wavefunctional coefficients be. However, we know that these coefficients should have non-transverse-traceless pieces dictated by the stress tensor Ward identities. In fact, these additional pieces can be reconstructed using the Ward identities~\cite{Bzowski:2013sza}. Alternatively, these pieces can be computed directly in a different gauge where the bulk-to-boundary propagator is not transverse-traceless.}

\subsubsection{The partially massless bulk-to-boundary propagator}

Next we consider the bulk-to-boundary propagator of a partially massless field.
 From \eqref{ADMLikeSolutionPM}, the PM propagator has both spatial and temporal components.  These are given by
\begin{tcolorbox}[colframe=white,arc=0pt,colback=greyish2]
\vspace{-.35cm}
\begin{align}
 \mathcal{K}^{\bfk}_{ij}{}^{lm}(\tau)&=\left (\frac{\tau}{\ts}\right )^{\frac{d}{2}}\frac{H_{\frac{d-2}{2}}^{(2)}(-k\tau)}{H_{\frac{d-2}{2}}^{(2)}(-k\ts)}(\Pi_{TT}^{\bfk} ){}_{ij}{}^{lm}+\left (\frac{\tau}{\ts}\right )^{\frac{d}{2}}\left [\frac{H^{(2)}_{\frac{d-2}{2}}(-k\tau)-k\tau H^{(2)}_{\frac{d}{2}}(-k\tau)}{H^{(2)}_{\frac{d-2}{2}}(-k\ts)-k\ts H^{(2)}_{\frac{d}{2}}(-k\ts)}\right ](\Pi_{V}^{\bfk} ){}_{ij}{}^{lm}\nn
 \mathcal{K}^{\bfk}_{0j}{}^{lm}(\tau)&=\left (\frac{\tau}{\ts}\right )^{\frac{d+2}{2}}\frac{-k\ts H^{(2)}_{\frac{d-2}{2}}(-k\tau)}{H^{(2)}_{\frac{d-2}{2}}(-k\ts)-k\ts H^{(2)}_{\frac{d}{2}}(-k\ts)}\frac{ik^{i}}{k}(\Pi_{V}^{\bfk} ){}_{ij}{}^{lm}\label{PMBulkBoundary} \ .
 \end{align} 
 \end{tcolorbox}

 At the boundary, $\mathcal{K}^{\bfk}_{ij}{}^{lm}(\tau)$ reduces to the identity matrix for symmetric, two-index tensors which only carry tensor and vector components by construction, {\it i.e.},~$\mathcal{K}(\tau)\xrightarrow{\tau\to\ts}\Pi_{TT}+\Pi_{V}$. In the PM case, we will redefine $\varphi_{ij}\to \sigma_{ij}$ in order to distinguish the PM from the massless and massive spin-2 fields.   As previously stated, for PM fields, the boundary data $\bar{\sigma}_{ij}^{\bfk}$ is required to be free of scalar components, because they can always be gauged to zero: $\Pi_{S}^{\bfk}\cdot\bar{\sigma}^{\bfk}=\Pi_{Q}^{\bfk}\cdot\bar{\sigma}^{\bfk}=0$, in condensed notation.   Hence, the bulk solution will be written as
 \begin{align}
h_{\mu\nu}^{\bfk}(\tau)=\frac{1}{H^{2}\tau^{2}} \mathcal{K}^{\bfk}_{\mu\nu}{}^{lm}(\tau)\bar{\sigma}_{lm}^{\bfk}\ .
 \end{align}

\subsubsection{The massive bulk-to-boundary propagator}

We now turn to the fully massive bulk-to-boundary propagator.
 From \eqref{ADMLikeSolutionGeneric}, we see that the generic mass propagator has both spatial and temporal components.  In this case, all of the components of the boundary data $\bar{\varphi}_{ij}$ are generically non-zero, but they cannot all be independently chosen---they have to be consistent with the constraint structure of the theory---as was seen in Sec.~\ref{subsubsec:GenericMassSolutions}.  Specifically, the $\Pi_{Q}^{\bfk}\cdot \bar{\varphi}^{\bfk}$ component of the boundary data is constrained in terms of the $\Pi_{S}^{\bfk}\cdot \bar{\varphi}^{\bfk}$ component.  This makes the construction of the bulk-to-boundary propagator slightly more complicated than the massless and PM cases. Due to this fact, we will define $\bar\phi_{ij}^{\bfk}$, which contains only the unconstrained components of $\bar{\varphi}_{ij}^{\bfk}$, {\it i.e.},~$\bar{\phi}_{ij}^{\bfk}$ obeys $\Pi_{Q}\cdot \bar\phi=0$ but coincides with $\bar{\varphi}_{ij}$ otherwise,\footnote{Explicitly, using the projectors in App.~\ref{app:Projectors} we can write
 \be
 \bar{\phi}^\bfk_{ij} \equiv (\Pi_{TT}^{\bfk} ){}_{ij}{}^{lm}\bar\varphi^\bfk_{lm}+(\Pi_{V}^{\bfk} ){}_{ij}{}^{lm}\bar\varphi^\bfk_{lm}+(\Pi_{S}^{\bfk} ){}_{ij}{}^{lm}\bar\varphi^\bfk_{lm}.
 \ee
 } 
 and the bulk solution will be expressed as 
\be
 h_{\mu\nu}^{\bfk}(\tau)=\mathcal{K}_{\mu\nu}^{\bfk}{}^{lm}\bar\phi_{lm}^{\bfk}\ .
\ee
 The components of the bulk-to-boundary propagator are then given by
\begin{tcolorbox}[colframe=white,arc=0pt,colback=greyish2]
\vspace{-.35cm}
\begin{align}
 \mathcal{K}^{\bfk}_{ij}{}^{lm}(\tau)&=\left (\frac{\tau}{\ts}\right )^{\frac{d}{2}}\frac{H_{i\mu}^{(2)}(-k\tau)}{H_{i\mu}^{(2)}(-k\ts)}(\Pi_{TT}^{\bfk} ){}_{ij}{}^{lm}\nn
 &\quad +\left (\frac{\tau}{\ts}\right )^{\frac{d}{2}}\left [\frac{2k\tau H^{(2)}_{ i\mu-1}(-k\tau)+(d+2i\mu)H^{(2)}_{ i\mu}(-k\tau)}{2k\ts H^{(2)}_{ i\mu-1}(-k\ts)+(d+2i\mu)H^{(2)}_{ i\mu}(-k\ts)}\right ](\Pi_{V}^{\bfk}){}_{ij}{}^{lm}\nn
 &\quad +\left (\frac{\tau}{\ts}\right )^{\frac{d+4}{2}}\frac{H^{(2)}_{i\mu}(-k\tau)}{H^{(2)}_{i\mu}(-k\ts)}(\Pi_{S}^{\bfk}){}_{ij}{}^{lm}\nn
 &\quad - \frac{f_{Q}(\tau,k)}{(-k\ts)^{\frac{d+4}{2}}H^{(2)}_{i\mu}(-k\ts)}\left (\frac{k_{i}k_{j}}{k^{2}}-\frac{\delta_{ij}}{d}\right ) \delta^{no} (\Pi_{S}^{\bfk}){}_{no}{}^{lm}\nn
 \mathcal{K}^{\bfk}_{0j}{}^{lm}(\tau)&=\left (\frac{\tau}{\ts}\right )^{\frac{d+2}{2}}\frac{-2k\ts H^{(2)}_{i\mu }(-k\tau)}{2k\ts H^{(2)}_{i\mu -1}(-k\ts)+(d+2i\mu)H^{(2)}_{i\mu}(-k\ts)}\frac{ik^{i}}{k}(\Pi_{V}^{\bfk}){}_{ij}{}^{lm}\nn
 &\quad+\left (\frac{\tau}{\ts}\right )^{\frac{d+2}{2}}\frac{d\left [(d+2i\mu -2)H^{(2)}_{i\mu}(-k\tau)+2k\tau H^{(2)}_{i\mu-1}(-k\tau)\right ]}{2k\ts H^{(2)}_{i\mu}(-k\ts)}ik^{i}(\Pi_{S}^{\bfk}){}_{ij}{}^{lm}\nn
 \mathcal{K}^{\bfk}_{00}{}^{lm}(\tau)&= \left (\frac{\tau}{\ts}\right )^{\frac{d+4}{2}}\frac{H^{(2)}_{i\mu}(-k\tau)}{H^{(2)}_{i\mu}(-k\ts)}\delta^{ij} (\Pi_{S}^{\bfk}){}_{ij}{}^{lm},\label{MassiveBulkBoundary}
 \end{align} 
 \end{tcolorbox}
 \noindent
 where $f_{Q}(\tau,k)$ is the lengthy expression in \eqref{ConstrainedSolutionsGeneric}.
 We have chosen $\mathcal{K}^{\bfk}_{ij}{}^{lm}(\tau)$ such that its scalar, vector, and tensor components reduce to their respective identity matrices when $\tau=\ts$,  {\it i.e.},
 \begin{align}
 (\Pi_{S}+\Pi_{V}+\Pi_{TT})\cdot \mathcal{K}(\tau)\xrightarrow{\tau\to\ts} (\Pi_{S}+\Pi_{V}+\Pi_{TT})\ .
 \end{align}

\subsection{Quadratic on-shell action and two-point functions\label{sec:TwoPointFunctions}}
In this section we evaluate the quadratic on-shell action and derive the superhorizon power spectrum for the transverse, traceless components of spin-2 fields in general dimensions.

 Conforming to the standard variables used in cosmology, we will compute the correlators and wavefunctions corresponding to the field $\varphi_{ij}^{\bfk}(\tau)$, rather than $h_{ij}^{\bfk}(\tau)$, where
\begin{align}
h_{ij}^{\bfk}(\tau)\equiv \frac{1}{H^{2}\tau^{2}}\varphi^{\bfk}_{ij}(\tau)\ ,\label{hTovarphiRelation}
\end{align}
and the notation $\varphi\to \gamma$ and $\varphi\to \sigma$ will be used for the massless and PM cases, respectively.  Because we are only focusing on the transverse, traceless components of correlators, our on-shell solutions for $\varphi^{\bfk}_{ij}(\tau)$ are related to the boundary data $\bar{\varphi}_{ij}^{\bfk}(\tau)$ via
\begin{align}
\varphi_{ij}^{\bfk}(\tau)&={\mathcal{K}}_{ij}^{TT,\,\bfk\,}{}^{lm}(\tau)\bar{\varphi}_{lm}^{\bfk}\nn
{\mathcal{K}}_{ij}^{TT,\,\bfk\,}{}^{lm}(\tau)&\equiv \mathcal{K}_{ij}^{\bfk}{}^{no}(\tau) (\Pi_{TT}^{\bfk}){}_{no}{}^{lm}=\left (\frac{\tau}{\ts}\right )^{\frac{d}{2}}\frac{H_{i\mu}^{(2)}(-k\tau)}{H_{i\mu}^{(2)}(-k\ts)}(\Pi_{TT}^{\bfk}){}_{ij}{}^{lm}\ .\label{OnShellTTSolutions}
\end{align}
For later convenience, we explicitly write the form of the transverse, traceless $d=3$ solutions for the massless and partially massless fields:
\begin{tcolorbox}[colframe=white,arc=0pt,colback=greyish2]
\vspace{-.425cm}
\begin{align}
({\rm massless } \ d=3)\quad\quad\gamma_{ij}^{\bfk}(\tau)&=\frac{(1-ik\tau)}{(1-ik\ts)}e^{ik(\tau-\ts)}(\Pi_{TT}^{\bfk}){}_{ij}{}^{lm}\bar{\gamma}^{\bfk}_{lm}\nn
({\rm PM } \ d=3)\quad\quad\sigma_{ij}^{\bfk}(\tau)&=\frac{\tau}{\ts}e^{ik(\tau-\ts)}(\Pi_{TT}^{\bfk}){}_{ij}{}^{lm}\bar{\sigma}^{\bfk}_{lm}\label{Explicitd3Solutions}\ .
\end{align}
\end{tcolorbox}

\subsubsection{The on-shell action and generic two-point functions}
We can obtain the transverse, traceless two-point function for generic spin-2 modes by evaluating the quadratic action on the solution \eqref{OnShellTTSolutions}.

Starting from the quadratic action \eqref{QuadraticSpin2ActiondS}, we integrate by parts to obtain 
\be
S_{2}=\int_{\tau=\ts}\rd^{d}x\, \sqrt{\bar{g}}\, \left (-\frac{1}{2}n_{\mu}h_{\nu\sigma}\frac{\partial\mathcal{L}}{\partial \nabla_{\mu}h_{\nu\sigma}}\right )+\int\rd^{d+1}x\,\sqrt{-g}\,\frac{1}{2}h_{\mu\nu}\mathcal{E}^{\mu\nu}_{\rho\sigma}h^{\rho\sigma} \ .
\ee
Here $\bar g_{ij}$ is the induced metric on the $\tau=\ts$ boundary, $n^\mu$ is the outward-pointing normal vector to the boundary, and $\mathcal{E}^{\mu\nu}_{\rho\sigma}$ is the differential operator appearing in the free equation of motion: $\mathcal{E}_{\mu\nu}^{\rho\sigma}h^{\rho\sigma}=0$. When evaluated on-shell, the second term vanishes and we are left with
\begin{align*}
S_{2}&=\frac{\MP ^{d-1}}{8}\int_{\tau=\ts}\rd^{d}x\, \sqrt{\bar{g}}\, \Big(h^{\nu \rho}n^{\mu}\nabla_{\mu}h_{\nu \rho}-hn^{\mu}\nabla_{\mu}h+h_{\mu}{}^{\nu}n^{\mu}\nabla_{\nu}h+h n^{\mu}\nabla_{\rho}h_{\mu}{}^{\rho}-2h_{\mu}{}^{\nu}n^{\mu}\nabla_{\rho}h_{\nu}{}^{\rho}\Big)\ .
\end{align*}
When we further use the $\nabla^{\mu}h_{\mu\nu}=h=0$ restriction, all terms but the first vanish and the on-shell action reduces to
\begin{align}
S_{2}\Big|_{\nabla^{\mu}h_{\mu\nu}=h=0}=\frac{\MP ^{d-1}}{8}\int_{\tau=\ts}\rd^{d}x\, \sqrt{\bar{g}}\, h^{\nu \rho}n^{\mu}\nabla_{\mu}h_{\nu \rho}\ .
\end{align}
Finally, if we convert from $h_{\mu\nu}$ to $\varphi_{ij}$ via \eqref{hTovarphiRelation}, the on-shell action for the transverse, traceless modes becomes
\begin{align}
S_{2}=\frac{\MP ^{d-1}}{8H^{d-1}}\int_{\tau=\ts}\rd^{d}x\, \tau^{d-1} \varphi^{ij}\partial_{\tau}\varphi_{ij}.
\end{align}

The action is obtained as a function of boundary data by using \eqref{OnShellTTSolutions} for $\varphi_{ij}$:
\begin{align}
S_{2}\left[\bar\varphi\right]&=-\frac{\MP ^{d-1}}{16H^{d-1}\left (-\ts\right )^{d}}\int\rd^{d}\tilde{k}\, \left(\frac{d\,  H_{i \mu }^{(2)}(-k  \ts )+k  \ts  (H_{i \mu +1}^{(2)}(-k  \ts )-H_{i \mu -1}^{(2)}(-k  \ts ))}{H_{i \mu }^{(2)}(-k 
   \ts )}\right)\bar{\varphi}^{\bfk,ij}\bar{\varphi}^{-\bfk}_{ij}\ .
\end{align}
The wavefunctional is written in terms of the 
on-shell action as $\Psi\sim e^{iS}$ and we parameterize its quadratic component as
\be
\Psi[\bar\varphi] \simeq \exp\left(-\frac{1}{2}\int{\rm d}^d\tilde k\,\langle\Phi^{ij}_{\bfk}\Phi^{lm}_{-\bf k}\rangle'\bar{\varphi}^{\bfk}_{ij}\bar{\varphi}^{-\bfk}_{lm}\right),
\ee
where the Gaussian wavefunctional coefficient is given by
\be
\langle\Phi^{ij}_{\bfk}\Phi^{lm}_{-\bf k}\rangle'=\frac{\MP ^{d-1}}{8H^{d-1}\left (-\ts\right )^{d}} \left(\frac{d\,  H_{i \mu }^{(2)}(-k  \ts )+k  \ts  (H_{i \mu +1}^{(2)}(-k  \ts )-H_{i \mu -1}^{(2)}(-k  \ts ))}{H_{i \mu }^{(2)}(-k 
   \ts )}\right)(\Pi^\bfk_{TT})^{ijlm},
\ee
and where $\Pi^{\bfk}_{TT}$ is the projector onto symmetric, transverse, traceless tensors \eqref{Projectors}.
Only the real part of the wavefunctional coefficient (respectively the imaginary component of the action) affects correlation functions of $\varphi_{ij}$, as they are governed by the probability distribution $\lvert\Psi\rvert^2$.  The real part of the coefficient is\footnote{Here, we have used $\left (H^{(2)}_{i\mu}(-k\tau)\right )^{*}=e^{-\pi(\mu+\mu ^{*})/2}H^{(1)}_{i\mu}(-k\tau)$ which holds for both real and imaginary $\mu$ \cite{Arkani-Hamed:2015bza}.}
\be
{\rm Re}\langle\Phi^{ij}_{\bfk}\Phi^{lm}_{-\bf k}\rangle'=\frac{\MP ^{d-1}}{2H^{d-1}\left (-\ts\right )^{d}} \left (\frac{1}{\pi H^{(1)}_{i\mu }(-k\ts)H^{(2)}_{i\mu }(-k\ts)}\right )(\Pi^\bfk_{TT})^{ijlm}\ .\label{ImaginaryQuadAction}
\ee
When $i\mu\in\mathbb{R}^{+}$, as in the cases of a massless or partially massless spin-2 field, the superhorizon $k\ts\to 0$ limit of \eqref{ImaginaryQuadAction} is given by
\be
\lim _{k\ts\to 0}{\rm Re}\langle\Phi^{ij}_{\bfk}\Phi^{lm}_{-\bf k}\rangle'=\frac{\MP ^{d-1}\Gamma\left (1-|\mu|\right )^{2}\sin(\pi|\mu|)^{2}}{2^{2+2|\mu|}\pi H^{d-1}\left (-\ts\right )^{d} } \left (-k\ts\right )^{2|\mu|}(\Pi^\bfk_{TT})^{ijlm}.
\label{QuadraticOnShellActionImMu}
\ee
We now specialize to the cases of a massless spin-2 field and a partially massless spin-2 field.

\subsubsection{The massless spin-2 two-point function}
In the massless limit, we write the quadratic wavefunction as\footnote{We write the ``dual operator" to the massless $\gamma_{ij}^{\bfk}$ field as $T^{ij}_{\bfk}$ and that of the partially massless $\sigma_{ij}^{\bfk}$ field as $\Sigma^{ij}_{\bfk}$, using the language of holography.}
\begin{align}
\Psi[\bar\gamma]&\simeq\exp\left (-\frac{1}{2}\int\rd^{d}\tilde{k}_{1}\rd^{d}\tilde{k}_{2}\, \re \langle T^{ij}_{\bfk_{1}}T^{lm}_{\bfk_{2}}\rangle\bar{\gamma}^{\bfk_{1}}_{ij}\bar{\gamma}^{\bfk_{2}}_{lm}\right )\nn
\re \langle T^{ij}_{\bfk}T^{lm}_{-\bfk}\rangle'&=\left (\frac{\MP }{H}\right )^{d-1}\left (\frac{\sin^{2}\left (d\pi/2\right )\Gamma(1-\frac{d}{2})^{2}}{2^{d+1}\pi }\right )k^{d}(\Pi^\bfk_{TT})^{ijlm} \ ,
\label{ReTTMassless}
\end{align}
where the second expression follows from evaluating \eqref{QuadraticOnShellActionImMu} with $i\mu=\frac{d}{2}$.
 By squaring and integrating the wavefunctional over $\bar\gamma$, and using~\eqref{ReTTMassless} and \eqref{CorrelatorExpressions}, we find the superhorizon two-point function is given by
\begin{align}
\langle\gamma^{\bfk}_{ij}(\ts)\gamma^{-\bfk}_{lm}(\ts)\rangle'&= \left (\frac{H}{\MP }\right )^{d-1}\left (\frac{2^{d}\pi }{\sin^{2}\left( (d\pi/2\right )\Gamma(1-\frac{d}{2})^{2}}\right )\frac{1}{k^{d}}(\Pi^\bfk_{TT})_{ijlm} \ .
\end{align}
We are primarily interested in the $d\to 3$ limit, where we recover the well-known result
\be
\langle \gamma^{\bfk}_{ij}(\ts)\gamma^{-\bfk}_{lm}(\ts)\rangle'=\frac{2H^{2}}{\MP ^{2}k^{3}}(\Pi^\bfk_{TT})_{ijlm}.\label{MasslessSpin2TwoPointd3}
\ee

\subsubsection{The partially massless spin-2 two-point function}
In the partially massless limit, we write the quadratic wavefunction as
\begin{align}
\Psi[\bar\sigma]&\simeq\exp\left (-\frac{1}{2}\int\rd^{d}\tilde{k}_{1}\rd^{d}\tilde{k}_{2}\, \re \langle \Sigma^{ij}_{\bfk_{1}}\Sigma^{lm}_{\bfk_{2}}\rangle\bar{\sigma}^{\bfk_{1}}_{ij}\bar{\sigma}^{\bfk_{2}}_{lm}\right )\nn
\re \langle \Sigma^{ij}_{\bfk}\Sigma^{lm}_{-\bfk}\rangle'&=\left (\frac{\MP }{H}\right )^{d-1}\left (\frac{(d-2)^{2}\sin^{2}\left (d\pi/2\right )\Gamma(1-\frac{d}{2})^{2}}{2^{d+1}\pi \ts^{2}}\right )k^{d-2}(\Pi^\bfk_{TT})^{ijlm} \ ,
\label{ReQuadSPartiallyMassless}
\end{align}
where the second expression follows from evaluating \eqref{QuadraticOnShellActionImMu} with $i\mu=\frac{d-2}{2}$.
We have only written the part of the wavefunctional corresponding to tensor polarizations.
This result is sufficient for the interests of this paper, but in general a partially massless spin-2 solution will carry both tensor and vector modes which propagate.\footnote{That is, this calculation has only included the $\Pi_{TT}$ parts of \eqref{PMBulkBoundary}.  Had we also included the $ \Pi_{V}$ vector mode components, then $\langle \Sigma^{2}\rangle$ would also contain a piece $\propto \Pi_V$.  However, because we are ultimately interested in the induced non-Gaussianity for $\gamma_{ij}$, all such vector components are projected out since the final answer can only be proportional to $ \Pi_{TT}$ tensors.  For this reason, vector modes are ignored throughout this paper. If desired, the $ \Pi_V$ components of the PM two-point function can be restored using the results of Appendix~\ref{app:CFTTwoPointFunctions}, as they are tied to the $\Pi_{TT}$ terms by conformal symmetry.}

Using~\eqref{ReTTMassless} and \eqref{CorrelatorExpressions}, we can compute the transverse, traceless parts of the superhorizon two-point function for a PM field
\begin{align}
\langle\sigma^{\bfk}_{ij}(\ts)\sigma^{-\bfk}_{lm}(\ts)\rangle'&=\frac{1}{k^{d-2}}\left (\frac{H}{\MP }\right )^{d-1}\left (\frac{2^{d}\pi \ts^{2}}{(d-2)^{2}\sin^{2}\left (d\pi/2\right )\Gamma(1-\frac{d}{2})^{2}}\right )(\Pi^\bfk_{TT})_{ijlm} \ .
\end{align}
We are primarily interested in the $d\to 3$ limit where we obtain:
\begin{align}
\langle \sigma^{\bfk}_{ij}(\ts)\sigma^{-\bfk}_{lm}(\ts)\rangle'&=\frac{2H^{2}\ts^{2}}{\MP ^{2}k}(\Pi^\bfk_{TT})_{ijlm}.\label{PartiallyMasslessSpin2TwoPointd3}
\end{align}

\section{On-shell cubic vertices\label{sec:CubicVertices}}

We now turn to interactions. Our focus is on three-point correlation functions involving mixtures of spin-2 fields. There is a certain universality in these objects; in the de Sitter limit, their form is completely fixed by conformal invariance up to a finite number of constants \cite{Costa:2011mg}, much as the structure of three-point on-shell scattering amplitudes is fixed by Lorentz invariance~\cite{Benincasa:2007xk}. 
In order to compute three-point correlators, only the cubic component of the on-shell action is required.
For instance, the GR contribution to the massless tensor bispectrum $\langle \gamma^{3}\rangle$ comes from the $\langle T^{3}\rangle$ wavefunction coefficient, which is related to the cubic on-shell action via
\begin{align}
\log\Psi[\bar\gamma] \supset-\frac{1}{3!}\int\rd^{d}\tilde{k}_{1}\rd^{d}\tilde{k}_{2}\rd^{d}\tilde{k}_{3}\,\langle T^{ij}_{\bfk_{1}}T^{lm}_{\bfk_{2}}T^{no}_{\bfk_{3}}\rangle\,\bar{\gamma}^{\bfk_1}_{ij}\bar{\gamma}^{\bfk_2}_{lm}\bar{\gamma}^{\bfk_3}_{no} &= i\int\rd^{d+1}x\, \sqrt{-g}\, \mathcal{L}^{(3)}_{\rm int}[\gamma]\Big|_{\gamma=\tilde{\cal K}\bar{\gamma}} \ .
\label{CubicCoefficientCorrespondence}
\end{align}

The universality of three-point correlation functions manifests itself in the fact that there are only a finite number of independent on-shell cubic interactions for a given set of fields of various spins. 
Therefore, we can compute the most general, model-independent, three-point correlation function by finding a basis for these cubic structures. In this Section, we describe such a basis---the detailed construction of these interactions is discussed in Appendix~\ref{app:OnShellCubicTerms}.

\subsection{Generic spin-2 on-shell interactions}

We begin by  describing a basis for generic mass cubic spin-2 operators in ${\rm dS}_{d+1}$. Throughout, the vertices are on-shell, meaning we are imposing the conditions $\nabla^{\mu}h_{\mu\nu}=h=0$ and $\square h_{\mu\nu}\propto h_{\mu\nu}$ when identifying independent operators.

\begin{itemize}
\item {\bf Basis of three different spin-2 fields:}
We begin with the most general parity preserving cubic interactions amongst three distinguishable spin-2 fields $h_{(i)\mu\nu}$, $i\in\{1,2,3\}$, which are each taken to have unique and generic masses.  There are 11 parity preserving CFT structures involving three spin-2 primaries with generic weights~\cite{Costa:2011mg,Kravchuk:2016qvl}.  Hence, we expect there to be 11 independent on-shell cubic interactions.  A basis of eleven operators which are independent is given by:
\begin{align}
\mathcal{L}_{3}[h_{(1)},h_{(2)},h_{(3)}]&=c_{1}h_{(1)}^{\mu }{}_{\nu}h_{(2)}^{\nu}{}_{\sigma}h_{(3)}^{\sigma}{}_{\mu }\nn
&\quad+c_{2a}h_{(1)}^{\rho \sigma }\nabla_{\rho }h_{(2)}^{\mu \nu }\nabla_{\sigma }h_{(3) \mu \nu }+c_{2b}h_{(2)}^{\rho \sigma }\nabla_{\rho }h_{(3)}^{\mu \nu }\nabla_{\sigma }h_{(1) \mu \nu }+c_{2c}h_{(3)}^{\rho \sigma }\nabla_{\rho }h_{(1)}^{\mu \nu }\nabla_{\sigma }h_{(2) \mu \nu }\nn
&\quad +c_{3a}h_{(1)}^{\mu \rho }\nabla_{\nu }h_{(2)}^{\sigma }{}_{\rho}\nabla_{\sigma }h_{(3)}^{\nu }{}_{\mu}+c_{3b}h_{(2)}^{\mu \rho }\nabla_{\nu }h_{(3)}^{\sigma }{}_{\rho}\nabla_{\sigma }h_{(1)}^{\nu }{}_{\mu}+c_{3c}h_{(3)}^{\mu \rho }\nabla_{\nu }h_{(1)}^{\sigma }{}_{\rho}\nabla_{\sigma }h_{(2)}^{\nu }{}_{\mu}\nn
&\quad +c_{4a}\nabla_{(\rho  }\nabla_{\sigma )}h_{(1)}^{\mu  \nu }\nabla_{\mu }h_{(2)}^{\kappa \rho  }\nabla_{\nu }h_{(3)}^{\sigma}{}_{\kappa}+c_{4b}\nabla_{(\rho  }\nabla_{\sigma )}h_{(2)}^{\mu  \nu }\nabla_{\mu }h_{(3)}^{\kappa \rho  }\nabla_{\nu }h_{(1)}^{\sigma}{}_{\kappa}\nn
&\quad+c_{4c}\nabla_{(\rho  }\nabla_{\sigma )}h_{(3)}^{\mu  \nu }\nabla_{\mu }h_{(1)}^{\kappa \rho  }\nabla_{\nu }h_{(2)}^{\sigma}{}_{\kappa}+c_{5}\nabla_{\mu }\nabla_{\nu }h_{(1)}^{\lambda \kappa }\nabla_{\rho }\nabla_{\sigma }h_{(2)}^{\mu \nu }	\nabla_{\lambda }\nabla_{\kappa }h_{(3)}^{\rho \sigma }\ .\label{ThreehOperatorBasis}
\end{align}
Any cubic term in the action can be written as a linear combination of these vertices after integrations by parts and using the conditions on $h_{\mu\nu}$ stated at the start of this section.

\item {\bf Basis when two fields are identical:}
When two out of the three spin-2 fields are identical, the action is symmetric under their interchange and some of the operators in \eqref{ThreehOperatorBasis} degenerate.  Only eight independent interactions remain in this case.  Again, this is the expected counting from CFT or $S$-matrix considerations~\cite{Costa:2011mg,Kravchuk:2016qvl}. Our basis for one $h_{(1)}$ interacting with two $h_{(2)}$ fields is
\begin{align}
\mathcal{L}_{3}[h_{(1)},h_{(2)}]&=b_{1}h_{(1)}^{\mu }{}_{\nu}h_{(2)}^{\nu}{}_{\sigma}h_{(2)}^{\sigma}{}_{\mu }\nn
&\quad+b_{2a}h_{(1)}^{\rho \sigma }\nabla_{\rho }h_{(2)}^{\mu \nu }\nabla_{\sigma }h_{(2) \mu \nu }+b_{2b}h_{(2)}^{\rho \sigma }\nabla_{\rho }h_{(2)}^{\mu \nu }\nabla_{\sigma }h_{(1) \mu \nu }\nn
&\quad +b_{3a}h_{(1)}^{\mu \rho }\nabla_{\nu }h_{(2)}^{\sigma }{}_{\rho}\nabla_{\sigma }h_{(2)}^{\nu }{}_{\mu}+b_{3b}h_{(2)}^{\mu \rho }\nabla_{\nu }h_{(2)}^{\sigma }{}_{\rho}\nabla_{\sigma }h_{(1)}^{\nu }{}_{\mu}\nn
&\quad +b_{4a}\nabla_{\mu }h_{(2)}^{\kappa \rho  }\nabla_{\nu }h_{(2)}^{\sigma}{}_{\kappa}\nabla_{(\rho  }\nabla_{\sigma )}h_{(1)}^{\mu  \nu }+b_{4b}\nabla_{\mu }h_{(1)}^{\kappa \rho  }\nabla_{\nu }h_{(2)}^{\sigma}{}_{\kappa}\nabla_{(\rho  }\nabla_{\sigma )}h_{(2)}^{\mu  \nu }\nn
&\quad+b_{5}\nabla_{\mu }\nabla_{\nu }h_{(1)}^{\lambda \kappa }\nabla_{\rho }\nabla_{\sigma }h_{(2)}^{\mu \nu }	\nabla_{\lambda }\nabla_{\kappa }h_{(2)}^{\rho \sigma }\ .\label{TwohOperatorBasis}
\end{align}

\item {\bf Basis for a single spin-2 field:}

Finally, we consider the self-interactions of a single spin-2, $h^{(i)}_{\mu\nu}=h_{\mu\nu}$.  In this case, the interactions have to be totally symmetric, which further reduces the number of independent operators to five. Again, this is the expected counting from CFT or $S$-matrix considerations~\cite{Costa:2011mg,Kravchuk:2016qvl,Hinterbichler:2017qyt}, and the basis we use for self-interactions is 
\begin{align}
\mathcal{L}_{3}[h]&=a_{1}h^{\mu }{}_{\nu}h^{\nu}{}_{\sigma}h^{\sigma}{}_{\mu }\nn
&\quad+a_{2}h^{\rho \sigma }\nabla_{\rho }h^{\mu \nu }\nabla_{\sigma }h_{\mu \nu }  +a_{3}h^{\mu \rho }\nabla_{\nu }h^{\sigma }{}_{\rho}\nabla_{\sigma }h^{\nu }{}_{\mu} \nn
&\quad +a_{4}\nabla_{\mu }h^{\kappa \rho  }\nabla_{\nu }h^{\sigma}{}_{\kappa}\nabla_{(\rho  }\nabla_{\sigma )}h^{\mu  \nu } \nn
&\quad+a_{5}\nabla_{\mu }\nabla_{\nu }h^{\lambda \kappa }\nabla_{\rho }\nabla_{\sigma }h^{\mu \nu }	\nabla_{\lambda }\nabla_{\kappa }h^{\rho \sigma }\ .\label{OnehOperatorBasis}
\end{align}

\end{itemize}

The operator bases~\eqref{ThreehOperatorBasis},~\eqref{TwohOperatorBasis}, and~\eqref{OnehOperatorBasis} are written for generic mass spin-2 fields.  
When one or more of the $h_{(i)\mu\nu}$ fields are massless or partially massless, gauge invariance places further constraints on the operator combinations.
This reduces the number of independent cubic interaction terms.  In Appendix~\ref{app:OnShellCubicTerms} we describe how to implement on-shell gauge invariance at the level of the action. In what follows, we merely catalog the results of this procedure.

\subsection{Massless self-interactions}
We first consider the self-interactions of a single massless field. Before imposing gauge invariance, the possible interactions are of the form \eqref{OnehOperatorBasis}. 
In general dimension there exists a three-parameter family of on-shell gauge-invariant combinations of these interactions, where $a_1$ and $a_2$ are fixed to be
\begin{subequations}
\label{eq:Massless3Soln}
\begin{align}
a_1 &= a_{5} (3-d  (4 d +9)) H^6-\frac{1}{3} a_{4} (d  (d +3)+3) H^4+\frac{1}{3} a_{3} (2 d +3) H^2,\\
a_2 &= 3 a_{5}(1-d) H^4-\frac{1}{2} a_{4} (d+2) H^2+\frac{a_{3}}{2},
\end{align}
\end{subequations}
and $a_3,a_4,a_5$ are free. It can be verified that the set of interactions described by \eqref{eq:Massless3Soln} is on-shell equivalent to a sum of the Einstein--Hilbert term \eqref{EinsteinHilbertActiondS}, the Weyl tensor squared, and the Weyl tensor cubed, up to integrations-by-parts.  Furthermore, this counting matches the well-known counting of massless spin-2 scattering amplitudes or of stress tensor 3-point functions in general dimensions.

In $d=3$, it is instead fruitful to map \eqref{eq:Massless3Soln} to a sum of the Einstein--Hilbert term \eqref{EinsteinHilbertActiondS}, the Gauss--Bonnet combination, and the Weyl tensor cubed.  In particular, the $d=3$ Gauss--Bonnet term $\mathcal{L}_{\rm GB}=R^{2}-4R_{\mu\nu}^{2}+R_{\mu\nu\rho\sigma}^{2}$ corresponds to
\be
a^{d=3,{\rm GB}}_1 = 8H^4~~~~~~~~~~~~~a^{d=3,{\rm GB}}_3 = 5H^2,
\label{eq:GBparamvalues}
\ee
with $a_2=a_4=a_5 = 0$,
which can be shown by direct computation.  The Gauss--Bonnet combination is a total derivative in $d=3$.

\subsection{Partially massless self-interactions}

In this section, we present the cubic self-interactions of partially massless fields in arbitrary dimensions. Again the fields are identical, so the initial basis of operators is of the form \eqref{OnehOperatorBasis} and we find two branches of solutions for the $a_{i}$: one branch which is gauge invariant for arbitrary $d$ and one which only exists in $d=3$.

\paragraph{Arbitrary $d$ Solution:}  On one branch of solutions, $d$ can be kept arbitrary.  In this case, we find a two-parameter family of solutions, three of the coefficients are fixed to be
\begin{subequations}
\label{eq:ddimPM}
\begin{align}
a_1 &=\frac{a_5}{2}  (d +1) (d  (d +8)+13) H^6+\frac{1}{4} a_{4} (d^{2}-1) H^4,\\
a_2 &=12 a_{5} H^4-a_{4} H^2,\\
a_3 &= \frac{3a_5}{4} (d  (d +12)+3) H^4+a_{4} (d +1) H^2,
\end{align}
\end{subequations}
where $a_4,a_5$ are arbitrary.

\paragraph{$d=3$ Solution:}  On the other branch of solutions we are forced to set $d=3$, in which case we find a three-parameter family of solutions where we fix
\begin{subequations}
\label{eq:3dPM}
\begin{align}
a_1 &= -16 a_{5} H^6-10 a_{4} H^4+3 a_{3} H^2,\\
a_2 &=-6a_{5}H^{4}-3a_{3}H^{2}+\frac{1}{2}a_{3},
\end{align}
\end{subequations}
with free coefficients, $a_3, a_4, a_5$.

Our $d$-dimensional result~\eqref{eq:ddimPM} matches the counting found in~\cite{Joung:2012rv}: there is one operator whose highest derivative components are $\mathcal{O}(\nabla^{6})$ and one whose are $\mathcal{O}(\nabla^{4})$.  In the $d\to 3$ limit of~\eqref{eq:ddimPM}, the four-derivative $a_{4}$ interaction degenerates to the Gauss--Bonnet total derivative which is invariant under the partially massless gauge symmetry only in $d=3$:
\begin{align}
\left (\nabla_{\mu}\nabla_{\nu}+H^{2}g_{\mu\nu}\right )\frac{\delta }{\delta g_{\mu\nu}}\int\rd^{d+1}x\,\sqrt{-g}\, \mathcal{L}_{\rm GB}&=\frac{H^{2}}{2}\left (3-d\right )\sqrt{-g}\mathcal{L}_{\rm GB}\ .
\end{align}
On the $d=3$ branch of solutions \eqref{eq:3dPM}, we have one additional operator whose highest derivative components are only $\mathcal{O}(\nabla^{2})$. One combination of the $a_{i}$'s reproduces the six-derivative interaction (one six-derivative combination is on-shell equivalent to the cubic interactions coming from a $W_{\mu\nu\rho\sigma}^{3}$ term) of \eqref{eq:ddimPM}, one combination of the $a_{4}$ and $a_{3}$ terms corresponds to $\mathcal{L}_{\rm GB}$, and another combination of $a_{4}$ and $a_{3}$ reproduces the cubic self-interaction found in \cite{deRham:2013wv}.

\subsection{PM-massless-massless interactions\label{sec:PMM0M0CubicVerts}}

Next we consider cubic interactions between two massless fields and one partially massless spin-2.
The basis of operators is of the form \eqref{TwohOperatorBasis}, where $h_{(1)\mu\nu}$ is partially massless and the two $h_{(2)\mu\nu}$'s are massless. We find two branches of solutions for the $b_{i}$'s: one which holds for arbitrary $d$ and another which requires $d=3$.

\paragraph{Arbitrary $d$ solution:} On the first branch of solutions, $d$ can be kept arbitrary.  In this case, we find a one-parameter family of solutions for the $b_{i}$, where
\begin{subequations}
\label{eq:pmmmd}
\begin{align}
b_1 &=\frac{b_5}{8} \left(d  \left(9 d ^2+51 d +55\right)+61\right) H^6,\\
b_{2a} &=\frac{3b_5}{2} (1-d) H^4,\\
b_{2b} &=0\\
b_{3a} &= \frac{3b_5}{4} (d +1) (d +7) H^4,\\
b_{3b} &= \frac{1b_5}{2} (d  (5 d +26)+17) H^4,\\
b_{4a} & =\frac{b_5}{2} (3 d +5) H^2,\\
b_{4b} &= b_{5} (d +7) H^2,
\end{align}
\end{subequations}
while $b_5$ is free.

\paragraph{$d=3$ solution:} On the other branch of solutions we are forced to set $d=3$, in which case we find a two-parameter family of solutions for the $b_{i}$:
\begin{subequations}
\label{eq:pmmm3}
\begin{align}
b_1 &=36 b_{5} H^6+8 b_{4b} H^4,\\
b_{2a} &=2 b_{5} H^4-\frac{b_{4b} H^2}{2},\\
b_{2b} &=0\\
b_{3a} &= 3 b_{4b} H^2,\\
b_{3b} &=  30 b_{5} H^4+4 b_{4b}H^2,\\
b_{4a} & =2b_{5}H^{2}+\frac{1}{2}b_{4b},
\end{align}
\end{subequations}
where the free coefficients are $b_{4b}$ and $b_5$.

Our $d$-dimensional result~\eqref{eq:pmmmd} matches the counting found in~\cite{Joung:2012rv}: there is one operator whose highest derivative components are $\mathcal{O}(\nabla^{6})$.  
On the $d=3$ branch of solutions~\eqref{eq:pmmm3}, we have one additional operator whose highest derivative components are only $\mathcal{O}(\nabla^{4})$. One combination of $b_{i}$ in~\eqref{eq:pmmm3} reproduces the six-derivative interaction of \eqref{eq:pmmmd}.  

\subsection{Massless-PM-PM interactions}
In this section we consider the cubic interactions between two partially massless fields and one massless spin-2.
The basis of operators is of the form \eqref{TwohOperatorBasis}, where $h_{(1)\mu\nu}$ is massless and the two $h_{(2)\mu\nu}$'s are partially massless.   
Imposing gauge invariance, we find a single branch of solutions which holds in arbitrary dimensions.  
The family of solutions depend on five free parameters, the coefficients
\begin{subequations}
\label{eq:pmpmmd}
\begin{align}
\nonumber
b_1 &=b_{5} \left(3-4 d ^2-d \right) H^6-\frac{H^4}{4} (2 b_{4a} (5 d +2)+b_{4b} (d  (2 d +5)+4))+\frac{H^2}{4} (4 b_{3a}
   (d +1)+2 b_{3b} (d +2)),\\
b_{2a} &=b_{5} (d +3) H^4-\frac{H^2}{4}  (4 b_{4a}+b_{4b} (d +2))+\frac{1}{4} (2 b_{3b}-2 b_{3a}),\\
b_{2b} &=6 b_{5} (1-d) H^4+\frac{H^2}{2} (-2 b_{4a}-b_{4b}-b_{4b} d )+    b_{3a},
\end{align}
\end{subequations}
are fixed, while the 5 parameters $b_{3a}, b_{3b}, b_{4a}, b_{4b}$ and $b_5$ are free.
 The above results match the counting of \cite{Joung:2012rv}: there is one operator whose highest derivative components are $\mathcal{O}(\nabla^{6})$, two whose are $\mathcal{O}(\nabla^{4})$, and two whose are $\mathcal{O}(\nabla^{2})$.

\subsection{Massive-massless-massless interactions}
In this section we consider the cubic interactions between two massless fields and one massive spin-2.
The basis of operators is of the form \eqref{TwohOperatorBasis}, where $h_{(1)\mu\nu}$ is massive and the two $h_{(2)\mu\nu}$'s are massless.   
Imposing gauge invariance, we find a single branch of solutions which holds in arbitrary dimensions.  
The family of solutions depend on two free parameters, the coefficients
\begin{subequations}
\label{eq:Mm0m0}
\begin{align}
\nonumber
b_{1}&=  b_{5} (-7 d -3) H^6+\frac{H^4}{8}  \left(4 b_{4b} (d +1) (d +3)+8 b_{5} (d  (d +4)+7)
   m ^2\right)\nn
   &\quad+\frac{H^2}{8}  \left(8 b_{5} m ^4-2 b_{4b} (d +4) m ^2\right)-\frac{b_{4b} m ^4}{8}   \\
b_{2a}&= \frac{1}{2} b_{5} d  H^2 m ^2-\frac{1}{4} m ^2 \left(b_{4b}+b_{5} m ^2\right) \\
b_{2b}&=  0  \\
b_{3a}&= 2 b_{5} (d -1) H^4+\frac{H^2}{2} \left(b_{4b} (d +2)-4 b_{5} m ^2\right)+\frac{b_{4b} m ^2}{4} \\
b_{3b}&=  4 b_{5} (d -1) H^4+H^2 \left(b_{4b} (d +2)+b_{5} (2 d +5) m ^2\right)-\frac{b_{4b} m ^2}{2}  \\
b_{4a}&=  \frac{b_{4b}}{2}+b_{5} m ^2,
\end{align}
\end{subequations}
are fixed, while the 2 parameters $b_{4b}$ and $b_{5}$ are free. This matches the counting of 3-point scattering amplitudes found, for example, in~\cite{Bonifacio:2017nnt}, and the counting of CFT correlators \cite{Costa:2011mg,Kravchuk:2016qvl}.

\subsection{Massless-massive-massive interactions}
In this section we consider the cubic interactions between two massive fields and one massless spin-2.
The basis of operators is of the form \eqref{TwohOperatorBasis}, where $h_{(1)\mu\nu}$ is massless and the two $h_{(2)\mu\nu}$'s are massive.   
Imposing gauge invariance, we find a single branch of solutions which holds in arbitrary dimensions.  
The family of solutions depend on six free parameters, the coefficients
\begin{subequations}
\label{eq:m0MM}
\begin{align}
\nonumber
b_{1}&=  b_{5} (1-d  (4 d +7)) H^6+\frac{H^4}{4} \left(-4 b_{4a}-2 b_{4b} d ^2-2 b_{4a} d -3 b_{4b} d +16 b_{5}
   m ^2\right)\nn
   &\quad+\frac{H^2}{4}  \left(8 b_{2a}+4 b_{3a} (d +2)+2 b_{3b}
   d -8 b_{4a} m ^2\right) \\
b_{2b}&= b_{3a}-2 b_{5} (d -1) H^4+H^2 \left(-b_{4a}+\frac{1}{2} b_{4b} (-d -1)-4 b_{5} m ^2\right)
\end{align}
\end{subequations}
are fixed, while the 6 parameters $b_{2a}$, $b_{3a}, b_{3b}, b_{4a}, b_{4b}$ and $b_5$ are free. This again matches the counting of scattering amplitudes~\cite{Bonifacio:2017nnt} and the counting of CFT correlators \cite{Costa:2011mg,Kravchuk:2016qvl}.

\subsection{Massive-PM-PM interactions}
In this section we consider the cubic interactions between two partially massless fields and one massive spin-2.
The basis of operators is of the form \eqref{TwohOperatorBasis}, where $h_{(1)\mu\nu}$ is massive and the two $h_{(2)\mu\nu}$'s are partially massless.   
Imposing gauge invariance, we find a single branch of solutions which holds in arbitrary dimensions.  
The family of solutions depend on five free parameters, the coefficients
\begin{subequations}
\label{eq:Mpmpm}
\begin{align}
\nonumber
b_{1}&=  \frac{H^6}{8}  \left(40 b_{5}-32 b_{5} d ^2+40 b_{5} d \right)\nn
&\quad+\frac{H^4}{8}  \left(-12 b_{4b}-4 b_{4b}
   d ^2+2 d  \left(-10 b_{4a}-3 b_{4b}+16 b_{5} m ^2\right)-16 b_{5} m ^2\right)\nn
&\quad+\frac{H^2}{8} \left(8
   b_{3b}+8 b_{3a} d +4 b_{3b} d -4 b_{4a} d  m ^2-16 b_{5} m ^4+6 b_{4a}
   m ^2-3 b_{4b} m ^2\right)\nn
&\quad+\frac{1}{8} \left(b_{4b} m ^4-4 b_{3a} m ^2+2 b_{3b} m ^2\right)  \\
b_{2a}&=  -2 b_{5} (d -1) H^4+\frac{H^2}{4} \left(-6 b_{4a}-b_{4b}-2 b_{4b} d -2 b_{5} (d +2)
   m ^2\right)\nn
&\quad+\frac{1}{4} \left(2 b_{3b}+m ^2 \left(-2 b_{4a}+b_{4b}+b_{5} m ^2\right)\right)  \\
b_{2b}&= 8 b_{5} H^4-b_{4b} H^2
\end{align}
\end{subequations}
are fixed, while the 5 parameters $b_{3a}, b_{3b}, b_{4a}, b_{4b}$ and $b_5$ are free.
(Here \cite{Joung:2012rv} find instead a 6 parameter family.)

\subsection{PM-massive-massive interactions}
In this section we consider the cubic interactions between two massive fields and one partially massless spin-2.
The basis of operators is of the form \eqref{TwohOperatorBasis}, where $h_{(1)\mu\nu}$ is partially massless and the two $h_{(2)\mu\nu}$'s are massive.   
Imposing gauge invariance, we find a single branch of solutions which holds in arbitrary dimensions.  
The family of solutions depend on five free parameters, the coefficients
\begin{subequations}
\label{eq:pmMM}
\begin{align}
\nonumber
b_{1}&=  b_{5} \left(-\left(2 d ^2+d -9\right)\right) H^6+\frac{1}{4} H^4 \left(b_{4a} (-d  (d +2)-9)+b_{4b} (-d -1)
   (2 d +3)+16 b_{5} m ^2\right)\nn
&\quad+\frac{1}{4} H^2 \left(4 b_{3a} (d +1)+2 b_{3b} (d +1)-4 b_{4a}
   m ^2\right)  \\
b_{2a}&= 4 b_{5} H^4-b_{4a} H^2   \\
b_{2b}&= b_{3a}+\frac{1}{4} b_{5} \left(17-d ^2\right) H^4+H^2 \left(-b_{4a}+\frac{1}{2} b_{4b} (-d -2)-3 b_{5}
   m ^2\right)
\end{align}
\end{subequations}
are fixed, while the 5 parameters $b_{3a}, b_{3b}, b_{4a}, b_{4b}$ and $b_5$ are free.
(The case of two different masses is studied in \cite{Joung:2012rv} and they find instead an 8 parameter family.)

\subsection{Massless-PM-massive interactions}
In this section we consider the cubic interactions between two massless fields and one massive spin-2.
The basis of operators is of the form \eqref{ThreehOperatorBasis}, where $h_{(1)\mu\nu}$ is massless, $h_{(2)\mu\nu}$ is partially massless, and $h_{(1)\mu\nu}$ is massive.
Imposing gauge invariance, we find a single branch of solutions which holds in arbitrary dimensions.  
The family of solutions depend on two free parameters, the coefficients
\begin{subequations}
\label{eq:mPMM}
\begin{align}
\nonumber
c_{1}&=   \frac{H^4}{4} \left(2 c_{4c} \left(3 d ^2+d -6\right)-c_{4a} (d -7) (d +1)\right)+\frac{H^2m^2}{2} (c_{4a}
   (d -3)-c_{4c} (d +2))-\frac{c_{4a} m ^4}{4} \\
c_{2a}&=  0    \\
c_{2b}&=-c_{4c} H^2 \\
c_{2c}&=  \frac{c_{4c}}{2}  (d -3) H^2-\frac{c_{4c} m ^2}{2}  \\
c_{3a}&=  \frac{H^2}{2}  (2 c_{4a}+3 c_{4c} (d -1))-\frac{c_{4c} m ^2}{2} \\
c_{3b}&= \frac{H^2}{2}  (c_{4a} (d +2)+2 c_{4c} d +c_{4c})-\frac{c_{4a} m ^2}{2} \\
c_{3c}&= \frac{H^2}{2}  (-c_{4a} (d -4)+2 c_{4c} d +c_{4c})+\frac{c_{4a} m ^2}{2} \\
c_{4b}&=  c_{4c}  \\
c_{5}&=  0 ,
\end{align}
\end{subequations}
are fixed, while the 2 parameters $c_{4a}$ and $c_{4c}$ are free.
(\cite{Joung:2012rv} find a 3 parameter family.)

\section{Cubic wavefunction coefficients\label{sec:CubicWavefunctionCoefficients}}
We now have all of the elements required to compute three-point correlation functions involving arbitrary admixtures of massless and massive fields. Through the use of the bulk-to-boundary propagators listed in Section~\ref{sec:BulkBoundaryPropagators} and the on-shell cubic vertices enumerated in Section~\ref{sec:CubicVertices}, we can, in principle, compute three-point correlation functions in full generality.  However, we will we make a number of simplifications in our concrete calculations:
\begin{itemize}
\item We only consider cases involving massless and partially massless fields.\footnote{It is possible to move perturbatively away from the PM point, and consider $m^2 = 2H^2+\alpha$. In this case, closed form expressions for correlation functions can be computed perturbatively in the parameter $\alpha$, but there is not any qualitative difference with the PM case, so we focus on that here.}
\item We work in $d=3$.
\item We only consider helicity-2 polarizations for the fields (equivalently, we restrict to transverse, traceless boundary data for all fields).
\end{itemize}
The first two restrictions are chosen for technical convenience in order to obtain closed form solutions.  In general dimensions, or for general graviton masses, the $\tau$ integrals involved in computing the wavefunctional do not evaluate to closed-form expressions, but there is no in-principle obstruction to evaluating them numerically for any given case. Additionally, partially massless spin-2 fields are the lightest unitary massive spin-2 representations; we therefore may expect them to be of the most phenomenological interest.  The final restriction arises as our primary interest is to understand how the presence of additional fields can modify the correlation functions of the massless graviton, this can be relaxed straightforwardly.  Since the massless graviton only carries helicity-2 modes, only the helicity-2 modes of other particles can affect the graviton bispectrum.

\subsection{Presentation of results\label{sec:PresentationOfCubicCoefficients}}
Even in the simplified cases that we consider, the resulting correlation functions are rather complex, so it is worthwhile to briefly describe how the cubic wavefunction coefficients will be presented.  

On de Sitter space, spin-2 degrees of freedom can always be diagonalized so that the late-time wavefunctional takes the following form:
\begin{align}
 \Psi[\bar{\gamma},\bar{\sigma},\ts]=\exp\bigg(&-\frac{1}{2}\int\rd^{d}\tilde{k}_{1}\rd^{d}\tilde{k}_{2}\,\langle T^{ij}_{\bfk_{1}}T^{lm}_{\bfk_{2}}\rangle \bar{\gamma}_{ij}^{\bfk_{1}}\bar{\gamma}_{lm}^{\bfk_{2}}-	\frac{1}{2}\int\rd^{d}\tilde{k}_{1}\rd^{d}\tilde{k}_{2}\,\langle \Sigma^{ij}_{\bfk_{1}}\Sigma^{lm}_{\bfk_{2}}\rangle \bar{\sigma}_{ij}^{\bfk_{1}}\bar{\sigma}_{lm}^{\bfk_{2}}\nn
 &\quad-\frac{1}{3!}\int\rd^{d}\tilde{k}_{1}\rd^{d}\tilde{k}_{2}\rd^{d}\tilde{k}_{3}\, \langle T^{ij}_{\bfk_{1}}T^{lm}_{\bfk_{2}}T^{no}_{\bfk_{3}}\rangle \bar{\gamma}_{ij}^{\bfk_{1}}\bar{\gamma}_{lm}^{\bfk_{2}}\bar{\gamma}_{no}^{\bfk_{3}}
 \nn
 &\quad-\frac{1}{3!}\int\rd^{d}\tilde{k}_{1}\rd^{d}\tilde{k}_{2}\rd^{d}\tilde{k}_{3}\, \langle \Sigma^{ij}_{\bfk_{1}}\Sigma^{lm}_{\bfk_{2}}\Sigma^{no}_{\bfk_{3}}\rangle \bar{\sigma}_{ij}^{\bfk_{1}}\bar{\sigma}_{lm}^{\bfk_{2}}\bar{\sigma}_{no}^{\bfk_{3}}
 \nn
 &\quad-\frac{1}{2}\int\rd^{d}\tilde{k}_{1}\rd^{d}\tilde{k}_{2}\rd^{d}\tilde{k}_{3}\, \langle \Sigma^{ij}_{\bfk_{1}}T^{lm}_{\bfk_{2}}T^{no}_{\bfk_{3}}\rangle \bar{\sigma}_{ij}^{\bfk_{1}}\bar{\gamma}_{lm}^{\bfk_{2}}\bar{\gamma}_{no}^{\bfk_{3}}
 \nn
 &\quad-\frac{1}{2}\int\rd^{d}\tilde{k}_{1}\rd^{d}\tilde{k}_{2}\rd^{d}\tilde{k}_{3}\, \langle T^{ij}_{\bfk_{1}}\Sigma^{lm}_{\bfk_{2}}\Sigma^{no}_{\bfk_{3}}\rangle \bar{\gamma}_{ij}^{\bfk_{1}}\bar{\sigma}_{lm}^{\bfk_{2}}\bar{\sigma}_{no}^{\bfk_{3}}+\cdots\bigg)\ .
 \label{eq:wavefunctionconventions}
 \end{align}
In the above, we have specialized to the case with one massless and one PM field, with $\bar\gamma_{ij}$ and $\bar\sigma_{ij}$ the transverse, traceless boundary values of these respective fields on the $\tau=\tau_\star$ time slice, but similar expressions would hold in the presence of additional spin-2 particles. The variables $T$ and $\Sigma$ transform under the de Sitter isometries as conformal primaries with weights $\Delta_T = d$, $\Delta_\Sigma = d-1$.

In~\eqref{eq:wavefunctionconventions}, the $\bar\gamma_{ij}$ and $\bar\sigma_{ij}$ tensors play a role analogous to that of polarization tensors in scattering amplitudes.
In $d=3$, it can sometimes be inconvenient to keep $\bar\gamma_{ij}$ and $\bar\sigma_{ij}$ arbitrary because of the presence of Gram/Schouten identities---examples of which are discussed in Appendix~\ref{app:PolarizationTensors}.
It is therefore convenient to choose an explicit basis of polarizations and compute the wavefunctional coefficients in this basis. There is no loss of information in doing this. In principle the wavefunctional coefficient for arbitrary boundary data can be reconstructed as a linear combination of the values in an explicit basis.

We use the same basis of polarizations considered in~\cite{Maldacena:2011nz} and choose the $\bar{\gamma}_{ij}^{\bfk}$'s to be one of the explicit polarization tensors $\epsilon^{\bfk P}_{ij}$ and $\epsilon^{\bfk X}_{ij}$, which are defined in Appendix~\ref{app:PolarizationTensors}.  The polarization $\epsilon^{\bfk P}_{ij}$ is parity even and $\epsilon^{\bfk X}_{ij}$ is parity odd. Since the interactions we consider preserve parity, only contractions with three $\epsilon ^{P}$'s or one $\epsilon ^{P}$ and two $\epsilon^{X}$ are non-vanishing. Hence we define, for example
 \begin{align}
 \langle T^{P}_{\bfk_{1}}T^{P}_{\bfk_{2}}T^{P}_{\bfk_{3}}\rangle'&\equiv \epsilon^{\bfk_{1}P}_{ij}\epsilon^{\bfk_{2}P}_{lm}\epsilon^{\bfk_{3}P}_{no}\langle T^{ij}_{\bfk_{1}}T^{lm}_{\bfk_{2}}T^{no}_{\bfk_{3}}\rangle' ,\\
  \langle T^{P}_{\bfk_{1}}T^{X}_{\bfk_{2}}T^{X}_{\bfk_{3}}\rangle'&\equiv \epsilon^{\bfk_{1}P}_{ij}\epsilon^{\bfk_{2}X}_{lm}\epsilon^{\bfk_{3}X}_{no}\langle T^{ij}_{\bfk_{1}}T^{lm}_{\bfk_{2}}T^{no}_{\bfk_{3}}\rangle' , \label{PPPandPXXCorrelatorDefs}
 \end{align}
 and similar for the $\langle \Sigma^{3}\rangle$, $\langle \Sigma T^{2}\rangle$, and $\langle T\Sigma^{2}\rangle$ terms.  It is these combinations which we calculate. Also, we will report only the {\it real} part of these wavefunctional coefficients, as these are the only parts which contribute to $\lvert\Psi\rvert^2$ and hence to correlation functions for the graviton. This is merely to make the final expressions manageable.
 
Another subtlety to address is that our bases for interactions~\eqref{ThreehOperatorBasis}, \eqref{TwohOperatorBasis}, and~\eqref{OnehOperatorBasis} are ambiguous up to integrations-by-parts.
Other, equally valid, operator bases hence will differ by boundary terms and we must therefore keep track of possible contributions to correlation functions coming from such operators.
Remarkably, all possible boundary terms seem to produce the same shape in correlation functions when evaluated in an explicit basis of polarizations---at least in the case where all fields are the same, or two fields are the same and one is different.\footnote{We have computed all possible boundary terms up to $\mathcal{O}(\nabla^{3})$ and find that in all cases the only non-trivial wavefunction contributions produce the same shape as~\eqref{h1Twoh2BoundaryTerm} or~\eqref{ThreehBoundaryTerm}. We do not have a completely intuitive explanation for this fact, but see Appendix \ref{app:IBPAndTheWavefunction} for additional discussion.}
We may therefore take as boundary term representatives the interaction
 \begin{align}
 S_{\rm bdy}[h_{(1)},h_{(2)}]&=\lambda\int_{\tau=\ts}\rd^{3}x \sqrt{\bar{g}}\, n^{\mu}\nabla_{\mu}\left (h^{\alpha}_{(1)\beta}h^{\beta}_{(2)\rho}h^{\rho}_{(2)\alpha}\right ),\label{h1Twoh2BoundaryTerm}
 \end{align} 
 in the case where two different fields interact or
 \begin{align}
 S_{\rm bdy}[h]&=\lambda\int_{\tau=\ts}\rd^{3}x \sqrt{\bar{g}}\, n^{\mu}\nabla_{\mu}\left (h^{\alpha}_{\beta}h^{\beta}_{\rho}h^{\rho}_{\alpha}\right ),\label{ThreehBoundaryTerm}
 \end{align}
 in the case of self-interactions, where $\lambda$ is an arbitrary coefficient.  As discussed in Appendix~\ref{app:IBPAndTheWavefunction}, the resulting shapes are also those associated to local redefinitions of the fields.
 
When presenting results for wavefunction coefficients, we therefore give both the result which follows from our choice of basis as well as the shape produced by~\eqref{h1Twoh2BoundaryTerm} or~\eqref{ThreehBoundaryTerm}, which represent the ambiguous parts of the coefficients.  It turns out that in our special case of interest where $d=3$, the boundary term shapes are redundant with those arising from our bulk interactions due to the existence of dimension-dependent Gauss--Bonnet total derivatives. However, we still give the form of the boundary shapes in order to demonstrate this fact.\footnote{In fact, this coincidence is absolutely crucial for the Gauss--Bonnet contribution to the three-point function to vanish, as it relies on the cancellation between the bulk GB vertex and its associated boundary term.}    When non-Gaussianities are computed in Sec.~\ref{sec:non-Gaussianity}, we will similarly compute the shapes that follow from our choice of basis as well as the shapes produced by boundary terms.

One way to fix the boundary term ambiguity in general is to demand that the wavefunction transform correctly under gauge transformations on the $\tau=\ts$ surface~\cite{Hui:2018cag}.
Indeed, one can think of Maldacena's consistency relation~\cite{Maldacena:2002vr} as fixing the contact term ambiguity in the cosmological wavefunction, as $f_{\rm NL}^{\rm local}$ corresponds precisely to a contact term. For pure GR, it was demonstrated in~\cite{Pimentel:2013gza} that demanding the gauge-invariance of the wavefunctional fully fixes the ambiguous terms, which in turn fixes the squeezed limit of $\langle \gamma^{3}\rangle$.
However, our cubic on-shell analysis only requires the lowest-order gauge transformations of $\gamma_{ij}$ and $\sigma_{ij}$, and performing a similar analysis in our cases would require knowing how the gauge transformations are deformed at next order---such an off-shell analysis is beyond the scope of the present paper.\footnote{For instance, due to the interactions with the PM spin-2 particle, the graviton's linear transformation law is not expected to simply be of the  usual schematic $\gamma\mapsto \gamma+\partial\epsilon+\epsilon\partial\gamma$ diffeomorphism form, but will also include $\mathcal{O}(\sigma)$ terms.  The analysis of the present paper is not sensitive to these corrections.}

\subsection{The $\langle T^{3}\rangle$ coefficient\label{sec:WavefunctionCoefficientT3}}

 We begin by computing the cubic wavefunctional coefficient for three massless gravitons, $\langle T^{ij}_{\bfk_{1}}T^{lm}_{\bfk_{2}}T^{no}_{\bfk_{3}}\rangle'$, which arises from the interactions~\eqref{OnehOperatorBasis} with coefficients~\eqref{eq:Massless3Soln}.

 \subsubsection{Bulk interactions}
 Here we list the contribution to the wavefunctional coefficients proportional to each of the free parameters---in this case there are 3 free parameters, each of which multiply a particular linear combination of the interactions~\eqref{OnehOperatorBasis}.
 \begin{itemize}
 \item The shape proportional to $a_{3}$ is:
 \begin{align}
\re  \langle T^{P}_{\bfk_{1}}T^{P}_{\bfk_{2}}T^{P}_{\bfk_{3}}\rangle'_{a_{3}}&=\frac{a_{3}\prod_{i}\left (k_{T}-2k_{i}\right )}{16H^{2}k_{T}\prod_{i}k_{i}^{2}} \left (\sum_{i}k_{i}^{4}+\sum_{i\neq j}3k_{i}^{2}k_{j}^{2}\right )\left(\sum_{i}k_{i}^{3}+\sum_{i\neq j}2k_{i}^{2}k_{j}+2\prod_{i}k_{i}\right),\nn
\re  \langle T^{P}_{\bfk_{1}}T^{X}_{\bfk_{2}}T^{X}_{\bfk_{3}}\rangle'_{a_{3}}&=-\frac{a_{3}\left (3k_{1}^{2}+k_{2}^{2}+k_{3}^{2}\right )\prod_{i}\left (k_{T}-2k_{i}\right )}{4H^{2}k_{1}k_{T}\prod_{i}k_{i}} \left(\sum_{i}k_{i}^{3}+\sum_{i\neq j}2k_{i}^{2}k_{j}+2\prod_{i}k_{i}\right ).
 \end{align}
  \item The shape proportional to $a_{4}$ is
   \begin{align}
\re  \langle T^{P}_{\bfk_{1}}T^{P}_{\bfk_{2}}T^{P}_{\bfk_{3}}\rangle'_{a_{4}}&= -\frac{a_{4}\prod_{i}\left (k_{T}-2k_{i}\right )}{16k_{T}\prod_{i}k_{i}^{2}} \Big[\sum_{i}3k_{i}^{7}+\sum_{i\neq j}\big(6k_{i}^{6}k_{j}+36k_{i}^{5}k_{j}^{2}+59k_{i}^{4}k_{j}^{3}\big)\nn
&\quad+\sum_{i\neq j\neq l}\big(66k_{i}^{4}k_{j}^{2}k_{l}+26k_{i}^{2}k_{j}^{2}k_{l}+3k_{i}^{5}k_{j}k_{l}+73k_{i}^{3}k_{j}^{2}k_{l}^{2}\big)\Big],\nn
\re  \langle T^{P}_{\bfk_{1}}T^{X}_{\bfk_{2}}T^{X}_{\bfk_{3}}\rangle'_{a_{4}}&=\frac{a_{4}\prod_{i}\left (k_{T}-2k_{i}\right )}{2k_{1}k_{T}\prod_{i}k_{i}} \Big[7 k_{1}^5+k_{1}^4 (14 k_{2}+14 k_{3})\nn
&\quad+k_{1}^3 \left(17
   k_{2}^2+14 k_{2} k_{3}+17 k_{3}^2\right)+k_{1}^2 \left(12
   k_{2}^3+20 k_{2}^2 k_{3}+20 k_{2} k_{3}^2+12
   k_{3}^3\right)\nn
&\quad+k_{1} \left(4 k_{2}^4+4 k_{2}^3 k_{3}+10
   k_{2}^2 k_{3}^2+4 k_{2} k_{3}^3+4 k_{3}^4\right)\nn
&\quad+2
   k_{2}^5+4 k_{2}^4 k_{3}+7 k_{2}^3 k_{3}^2+7 k_{2}^2
   k_{3}^3+4 k_{2} k_{3}^4+2 k_{3}^5\Big]\ .
 \end{align}
\item The shape proportional to $a_{5}$ is:
 \begin{align}
\re  \langle T^{P}_{\bfk_{1}}T^{P}_{\bfk_{2}}T^{P}_{\bfk_{3}}\rangle'_{a_{5}}&= \frac{3a_{5}H^{2}\prod_{i}\left (k_{T}-2k_{i}\right )}{2k_{T}^{3}\prod_{i}k_{i}^{2}} \bigg[\sum_{i}3k_{i}^{9}+\sum_{i\neq j}\big (12k_{i}^{8}k_{j}+18k_{i}^{7}k_{j}^{2}+16k_{i}^{6}k_{j}^{3}\nn
&\quad+15k_{i}^{5}k_{j}^{4}\big )+\sum_{i\neq j\neq l}\Big (18k_{i}^{7}k_{j}k_{l}+ 36 k_{i}^{6}k_{j}^{2}k_{l}+28  k_{i}^{5}k_{j}^{3}k_{l}+5k_{i}^{5}k_{j}^{2}k_{l}^{2}\nn
&\quad+16k_{i}^{4}k_{j}^{4}k_{l}-4k_{i}^{4}k_{j}^{3}k_{l}^{2}+16 k_{i}^{3}k_{j}^{3}k_{l}^{3}\Big )\bigg],\nn
\re  \langle T^{P}_{\bfk_{1}}T^{X}_{\bfk_{2}}T^{X}_{\bfk_{3}}\rangle'_{a_{5}}&=\frac{3a_{5}H^{2}\prod_{i}\left (k_{T}-2k_{i}\right )}{2k_{T}^{3}k_{1}\prod_{i}k_{i}}\Big[-k_{1}^7+k_{1}^6 (-4 k_{2}-4 k_{3})\nn
&\quad+k_{1}^5 \left(2
   k_{2}^2-12 k_{2} k_{3}+2 k_{3}^2\right)+k_{1}^4 \left(15
   k_{2}^3+20 k_{2}^2 k_{3}+20 k_{2} k_{3}^2+15
   k_{3}^3\right)\nn
&\quad+k_{1}^3 \left(3 k_{2}^4+40 k_{2}^3 k_{3}-38
   k_{2}^2 k_{3}^2+40 k_{2} k_{3}^3+3
   k_{3}^4\right)\nn
&\quad+k_{1}^2 \left(-22 k_{2}^5-28 k_{2}^4 k_{3}+46
   k_{2}^3 k_{3}^2+46 k_{2}^2 k_{3}^3-28 k_{2} k_{3}^4-22
   k_{3}^5\right)\nn
&\quad+k_{1} \left(-20 k_{2}^6-60 k_{2}^5 k_{3}-44
   k_{2}^4 k_{3}^2-8 k_{2}^3 k_{3}^3-44 k_{2}^2 k_{3}^4-60
   k_{2} k_{3}^5-20 k_{3}^6\right)\nn
&\quad-5 k_{2}^7-20 k_{2}^6
   k_{3}-26 k_{2}^5 k_{3}^2-13 k_{2}^4 k_{3}^3-13 k_{2}^3
   k_{3}^4-26 k_{2}^2 k_{3}^5-20 k_{2} k_{3}^6-5 k_{3}^7\Big].
 \end{align}
 \end{itemize}
 All other non-trivial correlators follow from permutations of the above results.  In Sec.~\ref{sec:BispectrumFromT3} we will verify explicitly that the above results reproduce the results of~\cite{Maldacena:2011nz}. Note that each of these coefficients is singular in the $k_T\to 0$ limit, where the total energy of the bulk interaction is conserved. The residue of this singularity is precisely the flat space scattering amplitude. We can think of this as a signature of the fact that these correlation functions came from local interactions in the bulk~\cite{Maldacena:2011nz,Raju:2012zr,Arkani-Hamed:2017fdk,Arkani-Hamed:2018kmz}.

As expected, each of these coefficients is time-independent, reflecting the fact that graviton perturbations freeze out at long wavelengths.
Something worth noting is that if we compute the wavefunctional for the {\it wrong} linear combinations of operators, {\it i.e.},~if combinations other than those in \eqref{eq:Massless3Soln} are used, then the wavefunction will have
 $\sim \log k\ts$ factors.  However, these all cancel once the appropriate gauge-invariant combinations are used.  Similar results hold for the wavefunction coefficients calculated in later sections: logarithms appear when non-gauge invariant values for the coefficients are used, but cancel once gauge-invariance is imposed on the interactions.
 
\subsubsection{Boundary terms}

Using the boundary term \eqref{ThreehBoundaryTerm}, converting from $h_{\mu\nu}\mapsto \gamma_{ij}$, and evaluating the above result on the solution \eqref{Explicitd3Solutions}, the result is
 \begin{align}
  S_{\rm bdy}^{\gamma^{3}}[\bar\gamma]\supset i\lambda\int\rd^{3}\tilde{k}_{1}\rd^{3}\tilde{k}_{2}\rd^{3}\tilde{k}_{3}\, \tilde{\delta}^{3}(\sum_{i}\bfk_{i} )\,\frac{\sum_{i}k_{i}^{3}}{H^{2}}\, \bar{\gamma}^{\bfk_{1} i }_{j}\bar{\gamma}^{\bfk_{2} l }_{i}\bar{\gamma}^{\bfk_{3} j }_{l} \ ,
  \label{Massless3BoundaryTerm}
 \end{align}
 where we have only kept the leading imaginary terms in the superhorizon limit.   This result is in agreement with the expectations of Appendix~\ref{app:IBPAndTheWavefunction}, as~\eqref{Massless3BoundaryTerm} is the shape that arises from a local field-redefinition of the form $\gamma_{ij}(\bfx)\mapsto \gamma_{ij}(\bfx)+c \,\gamma_{il}(\bfx)\gamma^{l}{}_{j}(\bfx)$ for constant $c$.   Evaluating \eqref{Massless3BoundaryTerm} on the explicit polarization tensors gives:
\begin{align}
\re  \langle T^{P}_{\bfk_{1}}T^{P}_{\bfk_{2}}T^{P}_{\bfk_{3}}\rangle'_{\rm bdy}&= \lambda\frac{3k_{T}\prod_{i}\left (k_{T}-2k_{i}\right )}{4H^{2}\prod_{i}k_{i}^{2}}\left (\sum_{i}k_{i}^{2}\right )\left (\sum_{i}k_{i}^{3}\right ) \ , \nn
\re  \langle T^{P}_{\bfk_{1}}T^{X}_{\bfk_{2}}T^{X}_{\bfk_{3}}\rangle'_{\rm bdy}&=-\lambda\frac{3k_{T}\prod_{i}\left (k_{T}-2k_{i}\right )}{2H^{2}k_{1}\prod_{i}k_{i}}\left (\sum_{i}k_{i}^{3}\right ) \ . \label{Massless3BdyTermCorrelators}
 \end{align}
Note that these terms are completely regular in the $k_T\to0$ limit, which is consistent with them having an intrinsically boundary origin.

\subsubsection{Comments on the Gauss--Bonnet term}
In $d=3$, the Gauss--Bonnet combination is a total derivative. This manifests itself in terms of the wavefunctional coefficients as the fact that the following sum vanishes
\begin{align}
0&=\re\langle T^{ij}_{\bfk_{1}}T^{lm}_{\bfk_{2}}T^{no}_{\bfk_{3}}\rangle'_{a_{3}}+ \re\langle T^{ij}_{\bfk_{1}}T^{lm}_{\bfk_{2}}T^{no}_{\bfk_{3}}\rangle'_{a_{4}}+ \re\langle T^{ij}_{\bfk_{1}}T^{lm}_{\bfk_{2}}T^{no}_{\bfk_{3}}\rangle'_{a_{5	}}+ \re\langle T^{ij}_{\bfk_{1}}T^{lm}_{\bfk_{2}}T^{no}_{\bfk_{3}}\rangle'_{\rm bdy},\label{GBCorrelatorRelation}
 \end{align} 
when evaluated on the following parameter values
 \be
a_{5}=0,~~~~~~~~~ a_{4}=1,~~~~~~~~~ a_{3}=5H^{2},~~~~~~~~~ \lambda=-H^{2}/6.
 \ee
 This can be explicitly seen by substituting the above values of the $a_{i}$s into \eqref{eq:Massless3Soln}, which gives a result proportional to \eqref{eq:GBparamvalues}.    Similar Gauss--Bonnet-like combinations exist for the other cubic interactions of interest;\footnote{For instance, the Gauss--Bonnet--like cubic interaction for an interaction involving one $h_{(1)\mu\nu}$ field with two $h_{(2)\mu\nu}$ fields can be derived from taking the $\mathcal{O}(h^{3}_{\mu\nu})$ terms in $\mathcal{L}_{\rm GB}$, replacing $h_{\mu\nu}\mapsto h_{(1)\mu\nu}+h_{(2)\mu\nu}$ everywhere and extracting the $\mathcal{O}(h_{(1)\mu\nu}h^{2}_{(2)\mu\nu})$ terms.  } hence in the following we will find additional combinations of correlators that are degenerate.  These identities make our results somewhat insensitive to the ambiguities regarding choosing a basis of interactions and integrations by parts, since the effect of boundary terms can be traded for shifts in the values of the bulk interaction coefficients.  
 
 Note that the vanishing of~\eqref{GBCorrelatorRelation} requires a contribution from a boundary term.
 A direct calculation of the cubic wavefunction coefficient induced by the Gauss--Bonnet term, $\mathcal{L}_{\rm GB}$, proceeds similarly.  If we compute the cubic coefficient for the variable $\gamma_{ij}$ introduced as $h_{ij}=\frac{1}{(H\tau)^{2}}\gamma_{ij}$ without performing any integrations-by-parts in $\mathcal{L}_{\rm GB}$, the result is non-zero and of the form \eqref{Massless3BdyTermCorrelators}.  This non-zero result can be understood from the fact when a manifold has a boundary, such as the $\tau=\ts$ surface, integrating $\mathcal{L}_{\rm GB}$ over the manifold only produces a topological invariant if an appropriate boundary term is added to the action \cite{Myers:1987yn}.  If the boundary term is added to the action, then the bulk result is completely cancelled and the wavefunction coefficient induced by the Gauss--Bonnet term vanishes, as it should.\footnote{Alternatively, if we instead introduce $\gamma_{ij}$ via $h_{ij}=\frac{1}{(H\tau)^{2}}\exp[\gamma]_{ij}$, then the bulk and boundary contributions to the wavefunction coefficient both separately vanish because the field redefinition introduces the correct boundary term to cancel the bulk contribution.}

\subsection{The $\langle \Sigma^{3}\rangle$ coefficient\label{sec:WavefunctionCoefficientSigma3}}
In this Section, we compute the cubic wavefunctional coefficients $\langle \Sigma^{ij}_{\bfk_{1}}\Sigma^{lm}_{\bfk_{2}}\Sigma^{no}_{\bfk_{3}}\rangle'$ for a self-interacting partially massless field in $d=3$, corresponding to the interactions~\eqref{OnehOperatorBasis} with parameter values~\eqref{eq:3dPM}.
 
\subsubsection{Bulk interactions}
 
 We find the following contributions to $\langle \Sigma^{3}\rangle$, there is a three-parameter family of shapes
 \begin{itemize}
 \item The shape proportional to $a_{3}$ is:
 \begin{align}
\re  \langle \Sigma^{P}_{\bfk_{1}}\Sigma^{P}_{\bfk_{2}}\Sigma^{P}_{\bfk_{3}}\rangle'_{a_{3}}&= \frac{a_{3}\prod_{i}\left (k_{T}-2k_{i}\right )}{16H^{2}\ts^{2}\prod_{i}k_{i}^{2}} \Big[\sum_{i}k_{i}^{4}+\sum_{i\neq j}3k_{i}^{2}k_{j}^{2}\Big],\nn
\re  \langle \Sigma^{P}_{\bfk_{1}}\Sigma^{X}_{\bfk_{2}}\Sigma^{X}_{\bfk_{3}}\rangle'_{a_{3}}&=-\frac{a_{3}\prod_{i}\left (k_{T}-2k_{i}\right )}{4H^{2}\ts^{2}k_{1}\prod_{i}k_{i}} \left (3k_{1}^{2}+k_{2}^{2}+k_{3}^{2}\right )\ .\label{Sigma3Froma3}
 \end{align}
  \item The shape proportional to $a_{4}$ is:
   \begin{align}
\re  \langle \Sigma_{\bfk_{1}}^{P}\Sigma_{\bfk_{2}}^{P}\Sigma_{\bfk_{3}}^{P}\rangle'_{a_{4}}&=-\frac{a_{4}\prod_{i}\left (k_{T}-2k_{i}\right )}{8\ts^{2}\prod_{i}k_{i}^{2}} \bigg[-\sum_{i}k_{i}^{4}+\sum_{i\neq j}\big(-3k_{i}^{3}k_{j}+3k_{i}^{2}k_{j}^{2}\big)\bigg], \nn
\re  \langle \Sigma_{\bfk_{1}}^{P}\Sigma_{\bfk_{2}}^{X}\Sigma_{\bfk_{3}}^{X}\rangle'_{a_{4}}&=-\frac{a_{4}\prod_{i}\left (k_{T}-2k_{i}\right )}{4\ts^{2}k_{1}\prod_{i}k_{i}}\Big[-9 k_{1}^2+k_{1} (6 k_{2}+6 k_{3})-k_{2}^2+6 k_{2}
   k_{3}-k_{3}^2\Big] \ . \label{Sigma3Froma4}
\end{align}
\item The shape proportional to $a_{5}$ is:
 \begin{align}
\re  \langle \Sigma^{P}_{\bfk_{1}}\Sigma^{P}_{\bfk_{2}}\Sigma^{P}_{\bfk_{3}}\rangle'_{a_{5}}&=- \frac{3a_{5}H^{2}\prod_{i}\left (k_{T}-2k_{i}\right )}{2\ts^{2}k_{T}^{2}\prod_{i}k_{i}^{2}}\bigg[\sum_{i}2k_{i}^{6}+\sum_{i\neq j}\left (5k_{i}^{5}k_{j}+14k_{i}^{4}k_{j}^{2}+11k_{i}^{3}k_{j}^{3}\right )\nn
&\quad+\sum_{i\neq j\neq l}\left (\frac{13}{2}k_{i}^{4}k_{j}k_{l}+14k_{i}^{3}k_{j}^{2}k_{l}+18k_{i}^{2}k_{j}^{2}k_{l}^{2}\right )\bigg],\nn
\re  \langle \Sigma^{P}_{\bfk_{1}}\Sigma^{X}_{\bfk_{2}}\Sigma^{X}_{\bfk_{3}}\rangle'_{a_{5}}&=\frac{3a_{5}H^{2}\prod_{i}\left (k_{T}-2k_{i}\right )}{2\ts^{2}k_{T}^{2}k_{1}\prod_{i}k_{i}} \bigg[23 k_{1}^4+k_{1}^3 (28 k_{2}+28 k_{3})\nn
&\quad+k_{1}^2 \left(18
   k_{2}^2+120 k_{2} k_{3}+18 k_{3}^2\right)\nn
&\quad+k_{1} \left(20
   k_{2}^3+12 k_{2}^2 k_{3}+12 k_{2} k_{3}^2+20
   k_{3}^3\right)\nn
&\quad+7 k_{2}^4+16 k_{2}^3 k_{3}+18 k_{2}^2
   k_{3}^2+16 k_{2} k_{3}^3+7 k_{3}^4\bigg].\label{Sigma3Froma5}
 \end{align}
 \end{itemize}  
 All other non-trivial correlators can be obtained from permutations of these.  
 
Note that each of these coefficients scales as $\sim \tau_\star^{-2}$. This is not the scaling that one would naively expect
 from the fact that the partially massless mode functions decay as $\sim \ts$ on superhorizon scales. 
 Instead we would expect the partially massless three-point correlator to scale as $\langle \sigma^{3}\rangle \sim \ts^{3}$.  We can translate this into a scaling for the wavefunctional coefficient using
\begin{align}
 \langle \sigma^{3}\rangle \sim \frac{1}{\re\langle\Sigma^{2}\rangle^{3}}\re\langle \Sigma^{3}\rangle
 \end{align}
 and we can see from~\eqref{PartiallyMasslessSpin2TwoPointd3} that $\re\langle \Sigma^{2}\rangle \propto \ts^{-2}$. Therefore, the natural expectation is that $\re\langle \Sigma^{3}\rangle\propto \ts^{-3}$, while the explicit answer scales as $\re\langle \Sigma^{3}\rangle\propto \ts^{-2}$, resulting in a faster-than-expected decay of correlation functions.\footnote{The imaginary part of $\langle \Sigma^{3}\rangle$ does however contain $\mathcal{O}(\ts^{-3})$ terms that are the components that would be of interest in the analogous AdS/CFT computation.}  This also indicates that we should not expect self-interactions of partially massless fields to imprint themselves on the cosmologically interesting part of the graviton bispectrum, as will be seen in Sec.~\ref{sec:BispectrumFromS3}.

\subsubsection{Boundary terms} 

 Using the boundary term \eqref{ThreehBoundaryTerm}, converting from $h_{\mu\nu}\mapsto \sigma_{ij}$, and evaluating the above result on the solution \eqref{Explicitd3Solutions}, the result is
 \begin{align}
 S_{\rm bdy}^{\sigma^{3}}[\bar\sigma]\supset i\lambda\int\rd^{3}\tilde{k}_{1}\rd^{3}\tilde{k}_{2}\rd^{3}\tilde{k}_{3}\, \tilde{\delta}^{3}(\sum_{i}\bfk_{i} )\,\frac{k_{T}}{H^{2}\ts^{2}}\, \bar{\sigma}^{\bfk_{1} i }_{j}\bar{\sigma}^{\bfk_{2} l }_{i}\bar{\sigma}^{\bfk_{3} j }_{l} \ ,
 \label{PM3BoundaryTerm}
 \end{align}
 where we have only kept the leading imaginary terms in the superhorizon limit.   The result \eqref{PM3BoundaryTerm} again corresponds to the shape arising from a local field-redefinition of the form $\sigma_{ij}(\bfx)\mapsto \sigma_{ij}(\bfx)+c\, \sigma_{il}(\bfx)\sigma^{l}{}_{j}(\bfx)$.  Evaluating \eqref{PM3BoundaryTerm} for explicit polarization tensors gives: 
 \begin{align}
\re  \langle \Sigma^{P}_{\bfk_{1}}\Sigma^{P}_{\bfk_{2}}\Sigma^{P}_{\bfk_{3}}\rangle'_{\rm bdy}&= \lambda\frac{3k_{T}^{2}\prod_{i}\left (k_{T}-2k_{i}\right )}{4H^{2}\ts ^{2}\prod_{i}k_{i}^{2}}\left (\sum_{i}k_{i}^{2}\right ) \ , \nn
\re  \langle \Sigma^{P}_{\bfk_{1}}\Sigma^{X}_{\bfk_{2}}\Sigma^{X}_{\bfk_{3}}\rangle'_{\rm bdy}&=-\lambda\frac{3k_{T}^{2}\prod_{i}\left (k_{T}-2k_{i}\right )}{2H^{2}\ts ^{2}k_{1}\prod_{i}k_{i}}\ .\label{PM3BdyTermCorrelators}
 \end{align}
 It can be checked that there is another Gauss--Bonnet relation of the type \eqref{GBCorrelatorRelation} when the results \eqref{PM3BdyTermCorrelators} are added to the above coefficients \eqref{Sigma3Froma3}, \eqref{Sigma3Froma4}, and \eqref{Sigma3Froma5}.

\subsection{The $\langle \Sigma T^{2}\rangle$ coefficient\label{sec:WavefunctionCoefficientSigmaT2}}
Next we consider the interaction between two massless spin-2 fields and a partially massless spin-2 field, $\langle \Sigma^{ij}_{\bfk_{1}}T^{lm}_{\bfk_{2}}T^{no}_{\bfk_{3}}\rangle'$. This corresponds to substituting the parameter values~\eqref{eq:pmmm3} into the interactions~\eqref{TwohOperatorBasis}.
 
\subsubsection{Bulk interactions}
 In this case, the entire leading late-time answer is
 \begin{align}
\re\langle \Sigma^{ij}_{\bfk_{1}}T^{lm}_{\bfk_{2}}T^{no}_{\bfk_{3}}\rangle'\bar{\sigma}^{\bfk_{1}}_{ij}\bar{\gamma}^{\bfk_{2}}_{lm}\bar{\gamma}^{\bfk_{3}}_{no}&=\left (b_{4b}+12H^{2}b_{5}\right )\frac{k_{1}}{\ts^{2}}\bar{\sigma}^{\bfk_{1} i }_{j}\bar{\gamma}^{\bfk_{2} l }_{i}\bar{\gamma}^{\bfk_{3} j }_{l}\ ,\label{SigmaT2WavefunctionCoefficient}
   \end{align}
which scales as $\mathcal{O}(\tau_\star^{-2})$.
   The above turns out to be exactly the shape produced by boundary terms. An analysis similar to that performed at the end of the previous section shows that the expected scaling of this coefficient is $\re \langle \Sigma T^{2}\rangle\propto \ts^{-1}$, rather than $\ts^{-2}$ as above.   However, keeping the subleading terms in the calculation, it is found that all $\mathcal{O}(\ts^{-1})$ pieces vanish entirely.\footnote{ The $\mathcal{O}(\ts^{-1})$ terms in $\im \langle \Sigma T^{2}\rangle$ are not vanishing, even when evaluated on~\eqref{eq:pmmm3}, which are again terms which would be interesting in the AdS/CFT analysis of the same bulk fields.} This implies that the $\langle \Sigma T^{2}\rangle$ coefficient does not source the time-independent component of the superhorizon graviton bispectrum; see Sec.~\ref{sec:BispectrumFromST2}.

\subsubsection{Boundary terms}  

Using the boundary term \eqref{h1Twoh2BoundaryTerm}, converting from $h_{(1)\mu\nu}\mapsto \sigma_{ij}$ and $h_{(2)\mu\nu}\mapsto\gamma_{ij}$, and evaluating the above result on the solutions \eqref{Explicitd3Solutions} and \eqref{Explicitd3Solutions}, the result is
 \begin{align}
  S_{\rm bdy}^{\sigma\gamma^{2}}[\bar\sigma,\bar\gamma]\supset i\lambda\int\rd^{3}\tilde{k}_{1}\rd^{3}\tilde{k}_{2}\rd^{3}\tilde{k}_{3}\, \tilde{\delta}^{3}(\sum_{i}\bfk_{i} )\,\frac{k_{1}}{H^{2}\ts^{2}}\, \bar{\sigma}^{\bfk_{1} i }_{j}\bar{\gamma}^{\bfk_{2} l }_{i}\bar{\gamma}^{\bfk_{3} j }_{l} \ ,
  \label{PMMassless2BoundaryTerm}
 \end{align}
 where we have only kept the leading imaginary terms in the superhorizon limit.  This is the same form as \eqref{SigmaT2WavefunctionCoefficient}, as discussed. The result \eqref{PMMassless2BoundaryTerm} corresponds to the shape arising from a local field-redefinition of the form\footnote{In \eqref{PMMassless2BoundaryTerm} there are also subleading $\sim k^{3}\sigma\gamma^{2}$ contributions which correspond to redefining  $\gamma\mapsto\gamma+c\sigma\gamma$.} $\sigma_{ij}(\bfx)\mapsto \sigma_{ij}(\bfx)+c\, \gamma_{il}(\bfx)\gamma^{l}{}_{j}(\bfx)$.  Evaluating \eqref{PMMassless2BoundaryTerm} on the explicit polarization tensors gives: 
\begin{align}
\re  \langle \Sigma^{P}_{\bfk_{1}}T^{P}_{\bfk_{2}}T^{P}_{\bfk_{3}}\rangle'_{\rm bdy}&= \lambda\frac{k_{1}k_{T}\prod_{i}\left (k_{T}-2k_{i}\right )}{4H^{2}\ts ^{2}\prod_{i}k_{i}^{2}}\left (\sum_{i}k_{i}^{2}\right ), \nn
\re  \langle \Sigma^{P}_{\bfk_{1}}T^{X}_{\bfk_{2}}T^{X}_{\bfk_{3}}\rangle'_{\rm bdy}&=-\lambda\frac{k_{T}\prod_{i}\left (k_{T}-2k_{i}\right )}{2H^{2}\ts ^{2}\prod_{i}k_{i}},\nn
\re  \langle \Sigma^{X}_{\bfk_{1}}T^{P}_{\bfk_{2}}T^{X}_{\bfk_{3}}\rangle'_{\rm bdy}&=-\lambda\frac{k_{T}\prod_{i}\left (k_{T}-2k_{i}\right )}{2H^{2}\ts ^{2}k_{2}^{2}k_{3}}\ .\label{PMMassless2BdyTermCorrelators}
 \end{align}

\subsection{The $\langle T\Sigma^{2}\rangle$ coefficient\label{sec:WavefunctionCoefficientT1Sigma2}}
Finally, we consider the situation with two partially massless fields interacting with a single massless field, leading to the cubic wavefunctional coefficient $\langle T^{ij}_{\bfk_{1}}\Sigma^{lm}_{\bfk_{2}}\Sigma^{no}_{\bfk_{3}}\rangle'$. This corresponds to the interactions~\eqref{TwohOperatorBasis} with the parameter values~\eqref{eq:pmpmmd}.

\subsubsection{Bulk interactions}
The parameter choices~\eqref{eq:pmpmmd} leave 5 free parameters, corresponding to the following shapes
 \begin{itemize}
 \item The shape proportional to $b_{3a}$ is:
 \begin{align}
\re  \langle T^{P}_{\bfk_{1}}\Sigma^{P}_{\bfk_{2}}\Sigma^{P}_{\bfk_{3}}\rangle'_{b_{3a}}&=-\frac{b_{3a}}{16 H^{2}\ts^{2}}\frac{\prod_{i}\left (k_{T}-2k_{i}\right )}{k_{T}\prod_{i}k_{i}^{2}}\Big[-2 k_{1}^5+k_{1}^4 (k_{2}+k_{3})+k_{1}^3 \left(-8
   k_{2}^2-8 k_{3}^2\right)\nn
   &\quad+k_{1}^2 \left(-2 k_{2}^3-2 k_{2}^2
   k_{3}-2 k_{2} k_{3}^2-2 k_{3}^3\right)+k_{1} \left(2
   k_{2}^4+12 k_{2}^2 k_{3}^2+2
   k_{3}^4\right)\nn
   &\quad+k_{2}^5+k_{2}^4 k_{3}+6 k_{2}^3
   k_{3}^2+6 k_{2}^2 k_{3}^3+k_{2} k_{3}^4+k_{3}^5\Big],\nn
\re  \langle T^{P}_{\bfk_{1}}\Sigma^{X}_{\bfk_{2}}\Sigma^{X}_{\bfk_{3}}\rangle'_{b_{3a}}&=-\frac{b_{3a}}{4 H^{2}\ts^{2}}\frac{\prod_{i}\left (k_{T}-2k_{i}\right )}{k_{1}k_{T}\prod_{i}k_{i}}\Big[4 k_{1}^3+k_{1}^2 (k_{2}+k_{3})+k_{1} \left(-2 k_{2}^2-2
   k_{3}^2\right)\nn
   &\quad-k_{2}^3-k_{2}^2 k_{3}-k_{2}
   k_{3}^2-k_{3}^3\Big],\nn
\re  \langle T^{X}_{\bfk_{1}}\Sigma^{P}_{\bfk_{2}}\Sigma^{X}_{\bfk_{3}}\rangle'_{b_{3a}}&=-\frac{b_{3a}}{2 H^{2}\ts^{2}}\frac{ k_{1}^{3}\prod_{i}\left (k_{T}-2k_{i}\right )}{k_{T}k_{2}\prod_{i}k_{i}}\ .
 \end{align}
  \item The shape proportional to $b_{3b}$ is:
 \begin{align}
\re  \langle T^{P}_{\bfk_{1}}\Sigma^{P}_{\bfk_{2}}\Sigma^{P}_{\bfk_{3}}\rangle'_{b_{3b}}&=-\frac{b_{3b}}{16 H^{2}\ts^{2}}\frac{\prod_{i}\left (k_{T}-2k_{i}\right )}{k_{T}\prod_{i}k_{i}^{2}}\Big[k_{1}^4 (-k_{2}-k_{3})+k_{1}^3 \left(-2 k_{2}^2-2
   k_{3}^2\right)\nn
   &\quad+k_{1}^2 \left(-2 k_{2}^3-2 k_{2}^2 k_{3}-2
   k_{2} k_{3}^2-2 k_{3}^3\right)+k_{1} \left(-2 k_{2}^4-12
   k_{2}^2 k_{3}^2-2 k_{3}^4\right)\nn
   &\quad-k_{2}^5-k_{2}^4
   k_{3}-6 k_{2}^3 k_{3}^2-6 k_{2}^2 k_{3}^3-k_{2}
   k_{3}^4-k_{3}^5\Big],\nn
\re  \langle T^{P}_{\bfk_{1}}\Sigma^{X}_{\bfk_{2}}\Sigma^{X}_{\bfk_{3}}\rangle'_{b_{3b}}&=-\frac{b_{3b}}{4H^{2}\ts^{2}}\frac{\prod_{i}\left (k_{T}-2k_{i}\right )}{k_{1}k_{T}\prod_{i}k_{i}}\Big[k_{1}^3+k_{1}^2 (k_{2}+k_{3})+k_{1} \left(2 k_{2}^2+2
   k_{3}^2\right)\nn
   &\quad+k_{2}^3+k_{2}^2 k_{3}+k_{2}
   k_{3}^2+k_{3}^3\Big],\nn
\re  \langle T^{X}_{\bfk_{1}}\Sigma^{P}_{\bfk_{2}}\Sigma^{X}_{\bfk_{3}}\rangle'_{b_{3b}}&=-\frac{b_{3b}}{8 H^{2}\ts^{2}}\frac{\prod_{i}\left (k_{T}-2k_{i}\right )}{k_{2}k_{T}\prod_{i}k_{i}}\Big[k_{1}^2 (k_{2}+k_{3})+k_{1} \left(6 k_{2}^2+2
   k_{3}^2\right)\nn
   &\quad+3 k_{2}^3+3 k_{2}^2 k_{3}+k_{2}
   k_{3}^2+k_{3}^3\Big]\ .
 \end{align}
  \item The shape proportional to $b_{4a}$ is:
 \begin{align}
\re  \langle T^{P}_{\bfk_{1}}\Sigma^{P}_{\bfk_{2}}\Sigma^{P}_{\bfk_{3}}\rangle'_{b_{4a}}&=-\frac{b_{4a}}{8 \ts^{2}}\frac{\prod_{i}\left (k_{T}-2k_{i}\right )}{k_{T}\prod_{i}k_{i}^{2}}\Big[k_{1}^5+k_{1}^3 \left(5 k_{2}^2-16 k_{2} k_{3}+5
   k_{3}^2\right)+k_{1}^2 \left(2 k_{2}^3+2
   k_{3}^3\right)\nn
   &\quad+k_{1} \left(-4 k_{2}^3 k_{3}+8 k_{2}^2
   k_{3}^2-4 k_{2} k_{3}^3\right)-2 k_{2}^4 k_{3}+2
   k_{2}^3 k_{3}^2+2 k_{2}^2 k_{3}^3-2 k_{2} k_{3}^4\Big],\nn
\re  \langle T^{P}_{\bfk_{1}}\Sigma^{X}_{\bfk_{2}}\Sigma^{X}_{\bfk_{3}}\rangle'_{b_{4a}}&=-\frac{b_{4a}}{4 \ts^{2}}\frac{\prod_{i}\left (k_{T}-2k_{i}\right )}{k_{1}k_{T}\prod_{i}k_{i}}\Big[-k_{1}^3+k_{1}^2 (-3 k_{2}-3 k_{3})\nn
   &\quad+k_{1} \left(-2
   k_{2}^2+4 k_{2} k_{3}-2 k_{3}^2\right)-k_{2}^3+k_{2}^2
   k_{3}+k_{2} k_{3}^2-k_{3}^3\Big],\nn
\re  \langle T^{X}_{\bfk_{1}}\Sigma^{P}_{\bfk_{2}}\Sigma^{X}_{\bfk_{3}}\rangle'_{b_{4a}}&=-\frac{b_{4a}}{8 \ts^{2}}\frac{\prod_{i}\left (k_{T}-2k_{i}\right )}{k_{2}k_{T}\prod_{i}k_{i}}\Big[-4 k_{1}^3+k_{1}^2 (11 k_{2}-k_{3})\nn
   &\quad+k_{1} \left(-14
   k_{2}^2+8 k_{2} k_{3}-2 k_{3}^2\right)-7 k_{2}^3-3
   k_{2}^2 k_{3}+3 k_{2} k_{3}^2-k_{3}^3\Big]\ .
 \end{align}
  \item The shape proportional to $b_{4b}$ is:
 \begin{align}
\re  \langle T^{P}_{\bfk_{1}}\Sigma^{P}_{\bfk_{2}}\Sigma^{P}_{\bfk_{3}}\rangle'_{b_{4b}}&=-\frac{b_{4b}}{32 \ts^{2}}\frac{\prod_{i}\left (k_{T}-2k_{i}\right )}{k_{T}\prod_{i}k_{i}^{2}}\Big[4 k_{1}^5+k_{1}^4 (k_{2}+k_{3})+k_{1}^3 \left(22 k_{2}^2+8
   k_{2} k_{3}+22 k_{3}^2\right)\nn
   &\quad+k_{1}^2 \left(10 k_{2}^3+2
   k_{2}^2 k_{3}+2 k_{2} k_{3}^2+10 k_{3}^3\right)\nn
   &\quad+k_{1}
   \left(2 k_{2}^4-16 k_{2}^3 k_{3}+44 k_{2}^2 k_{3}^2-16
   k_{2} k_{3}^3+2 k_{3}^4\right)\nn
   &\quad+k_{2}^5-7 k_{2}^4
   k_{3}+14 k_{2}^3 k_{3}^2+14 k_{2}^2 k_{3}^3-7 k_{2}
   k_{3}^4+k_{3}^5\Big],\nn
\re  \langle T^{P}_{\bfk_{1}}\Sigma^{X}_{\bfk_{2}}\Sigma^{X}_{\bfk_{3}}\rangle'_{b_{4b}}&=-\frac{b_{4b}}{8 \ts^{2}}\frac{\prod_{i}\left (k_{T}-2k_{i}\right )}{k_{1}k_{T}\prod_{i}k_{i}}\Big[-21 k_{1}^3+k_{1}^2 (-7 k_{2}-7 k_{3})\nn
   &\quad+k_{1} \left(-6
   k_{2}^2+8 k_{2} k_{3}-6 k_{3}^2\right)-3
   k_{2}^3+k_{2}^2 k_{3}+k_{2} k_{3}^2-3 k_{3}^3\Big],\nn
\re  \langle T^{X}_{\bfk_{1}}\Sigma^{P}_{\bfk_{2}}\Sigma^{X}_{\bfk_{3}}\rangle'_{b_{4b}}&=-\frac{b_{4b}}{16 \ts^{2}}\frac{\prod_{i}\left (k_{T}-2k_{i}\right )}{k_{2}k_{T}\prod_{i}k_{i}}\Big[-8 k_{1}^3+k_{1}^2 (-15 k_{2}-3 k_{3})\nn
   &\quad+k_{1} \left(-34
   k_{2}^2+16 k_{2} k_{3}-6 k_{3}^2\right)-17 k_{2}^3-9
   k_{2}^2 k_{3}+5 k_{2} k_{3}^2-3 k_{3}^3\Big]\ .
 \end{align}
  \item The shape proportional to $b_{5}$ is:
 \begin{align}
\re  \langle T^{P}_{\bfk_{1}}\Sigma^{P}_{\bfk_{2}}\Sigma^{P}_{\bfk_{3}}\rangle'_{b_{5}}&=-\frac{b_{5}H^{2}}{2\ts^{2} }\frac{\prod_{i}\left (k_{T}-2k_{i}\right )}{k_{T}^{3}\prod_{i}k_{i}^{2}}\Big[5 k_{1}^7+k_{1}^6 (8 k_{2}+8 k_{3})+k_{1}^5 \left(22
   k_{2}^2+22 k_{3}^2\right)\nn
   &\quad+k_{1}^4 \left(46 k_{2}^3+13
   k_{2}^2 k_{3}+13 k_{2} k_{3}^2+46
   k_{3}^3\right)\nn
   &\quad+k_{1}^3 \left(29 k_{2}^4+36 k_{2}^3 k_{3}+110
   k_{2}^2 k_{3}^2+36 k_{2} k_{3}^3+29
   k_{3}^4\right)\nn
   &\quad+k_{1}^2 \left(-4 k_{2}^5+2 k_{2}^4 k_{3}-34
   k_{2}^3 k_{3}^2-34 k_{2}^2 k_{3}^3+2 k_{2} k_{3}^4-4
   k_{3}^5\right)\nn
   &\quad+k_{1} \left(-8 k_{2}^6-20 k_{2}^5 k_{3}-56
   k_{2}^4 k_{3}^2-88 k_{2}^3 k_{3}^3-56 k_{2}^2
   k_{3}^4-20 k_{2} k_{3}^5-8 k_{3}^6\right)\nn
   &\quad-2 k_{2}^7-7
   k_{2}^6 k_{3}-19 k_{2}^5 k_{3}^2-36 k_{2}^4 k_{3}^3-36
   k_{2}^3 k_{3}^4-19 k_{2}^2 k_{3}^5-7 k_{2} k_{3}^6-2
   k_{3}^7\Big],\nn
\re  \langle T^{P}_{\bfk_{1}}\Sigma^{X}_{\bfk_{2}}\Sigma^{X}_{\bfk_{3}}\rangle'_{b_{5}}&=-\frac{b_{5}H^{2}}{2\ts^{2} }\frac{\prod_{i}\left (k_{T}-2k_{i}\right )}{k_{1}k_{T}^{3}\prod_{i}k_{i}}\Big[-44 k_{1}^5+k_{1}^4 (-77 k_{2}-77 k_{3})\nn
   &\quad+k_{1}^3 \left(-32
   k_{2}^2-208 k_{2} k_{3}-32 k_{3}^2\right)+k_{1}^2 \left(22
   k_{2}^3+6 k_{2}^2 k_{3}+6 k_{2} k_{3}^2+22
   k_{3}^3\right)\nn
   &\quad+k_{1} \left(28 k_{2}^4+64 k_{2}^3 k_{3}+72
   k_{2}^2 k_{3}^2+64 k_{2} k_{3}^3+28 k_{3}^4\right)\nn
   &\quad+7
   k_{2}^5+23 k_{2}^4 k_{3}+34 k_{2}^3 k_{3}^2+34 k_{2}^2
   k_{3}^3+23 k_{2} k_{3}^4+7 k_{3}^5\Big],\nn
\re  \langle T^{X}_{\bfk_{1}}\Sigma^{P}_{\bfk_{2}}\Sigma^{X}_{\bfk_{3}}\rangle'_{b_{5}}&=-\frac{b_{5}H^{2}}{2\ts^{2} }\frac{\prod_{i}\left (k_{T}-2k_{i}\right )}{k_{2}k_{T}^{3}\prod_{i}k_{i}}\Big[-20 k_{1}^5+k_{1}^4 (-41 k_{2}-41 k_{3})\nn
   &\quad+k_{1}^3 \left(-8
   k_{2}^2-16 k_{2} k_{3}-24 k_{3}^2\right)+k_{1}^2 \left(-2
   k_{2}^3-90 k_{2}^2 k_{3}+26 k_{2} k_{3}^2-6
   k_{3}^3\right)\nn
   &\quad+k_{1} \left(-20 k_{2}^4-32 k_{2}^3 k_{3}-8
   k_{2}^2 k_{3}^2-4 k_{3}^4\right)\nn
   &\quad-5 k_{2}^5-13 k_{2}^4
   k_{3}-10 k_{2}^3 k_{3}^2-2 k_{2}^2 k_{3}^3-k_{2}
   k_{3}^4-k_{3}^5\Big]\ .
 \end{align}
 \end{itemize}  All other non-trivial correlators follow from permutations of the above results.

\subsubsection{Boundary terms} 

 Using the boundary term \eqref{h1Twoh2BoundaryTerm}, converting from $h_{(1)\mu\nu}\mapsto \gamma_{ij}$ and $h_{(2)\mu\nu}\mapsto\sigma_{ij}$, and evaluating the above result on the solutions~\eqref{Explicitd3Solutions} and \eqref{Explicitd3Solutions}, the result is
 \begin{align}
  S_{\rm bdy}^{\gamma\sigma^{2}}[\bar \sigma,\bar\gamma]\supset i\lambda\int\rd^{3}\tilde{k}_{1}\rd^{3}\tilde{k}_{2}\rd^{3}\tilde{k}_{3}\, \tilde{\delta}^{3}(\sum_{i}\bfk_{i} )\,\frac{ \left (k_{2}+k_{3}\right )}{H^{2}\ts^{2}}\, \bar{\gamma}^{\bfk_{1} i }_{j}\bar{\sigma}^{\bfk_{2} l }_{i}\bar{\sigma}^{\bfk_{3} j }_{l}\label{PM2MasslessBoundaryTerm},
 \end{align}
 where we have only kept the leading imaginary terms in the superhorizon limit. The result \eqref{PM2MasslessBoundaryTerm} corresponds to the shape arising from a local field-redefinition of the form\footnote{In \eqref{PM2MasslessBoundaryTerm} there are also subleading $\sim k^{3}\gamma\sigma^{2}$ contributions which would arise from redefining $\gamma\mapsto\gamma+c\sigma^{2}$.} $\sigma_{ij}(\bfx)\mapsto \sigma_{ij}(\bfx)+c\, \sigma_{il}(\bfx)\gamma^{l}{}_{j}(\bfx)$. Evaluating \eqref{PM2MasslessBoundaryTerm} on the explicit polarization tensors gives:
  \begin{align}
\re  \langle T^{P}_{\bfk_{1}}\Sigma^{P}_{\bfk_{2}}\Sigma^{P}_{\bfk_{3}}\rangle'_{\rm bdy}&= \lambda\frac{\left (k_{2}+k_{3}\right )k_{T}\prod_{i}\left (k_{T}-2k_{i}\right )}{4H^{2}\ts ^{2}\prod_{i}k_{i}^{2}}\left (\sum_{i}k_{i}^{2}\right ), \nn
\re  \langle T^{P}_{\bfk_{1}}\Sigma^{X}_{\bfk_{2}}\Sigma^{X}_{\bfk_{3}}\rangle'_{\rm bdy}&=-\lambda\frac{\left (k_{2}+k_{3}\right )k_{T}\prod_{i}\left (k_{T}-2k_{i}\right )}{2H^{2}\ts ^{2}k_{1}\prod_{i}k_{i}},\nn
\re  \langle T^{X}_{\bfk_{1}}\Sigma^{P}_{\bfk_{2}}\Sigma^{X}_{\bfk_{3}}\rangle'_{\rm bdy}&=-\lambda\frac{\left (k_{2}+k_{3}\right )k_{T}\prod_{i}\left (k_{T}-2k_{i}\right )}{2H^{2}\ts ^{2}k_{2}\prod_{i}k_{i}}\ .\label{PM2MasslessBdyTermCorrelators}
 \end{align}
  Once again, it can be verified that there is a Gauss--Bonnet relation of the type \eqref{GBCorrelatorRelation} when the results \eqref{PMMassless2BdyTermCorrelators} are added to the sum of $\langle T\Sigma^{2}\rangle$ coefficients above.

 \section{Graviton non-Gaussianities\label{sec:non-Gaussianity}}
 The wavefunctional coefficients computed in Section~\ref{sec:CubicWavefunctionCoefficients} involve partially massless spin-2 fields on the external lines, and therefore will lead to mixed massless-PM 3-point correlation functions. Though these may be of independent phenomenological interest if the partially massless spin-2 itself couples to baryonic matter, in this section we wish to consider the effects of these wavefunctional coefficients on pure {\it graviton} correlation functions. This requires a linear mixing that can convert the PM spin-2 external lines into massless graviton legs. Such a mixing is not possible in pure de Sitter space, so we expect that such mixings will be slow-roll suppressed. Nevertheless, this suppression can be compensated by the size of the 3-point couplings, as was demonstrated in Sec.~\ref{sec:SketchOfCalculation}.
 
In this section we discuss two different effects of mixing. In the first case, we imagine that the conversion to massless gravitons happens at the {\it end} of inflation, so that the late-time wavefunctional computed in exact de Sitter space gives the leading contribution, and that the mixing occurs due to the off-diagonal 2-point function between the PM field and the graviton in the wavefunction. Here we focus on the 3-point interactions involving 2 PM fields and one massless graviton, since it happens that these are the only vertices that lead to a constant late-time graviton 3-point function. We also consider the effect of linear mixings directly in quasi-de Sitter space, which allows the conversion to take place {\it during} inflation, which can also affect graviton correlation functions at late times, particularly in soft limits. In technical terms, the former processes depend on the cubic coefficients calculated in the previous section, while the latter stem from contributions of partially massless fields to $\langle T^{3}\rangle$ via bulk-mixing diagrams which have not yet been discussed.

We give the contributions of the computed coefficients to the bispectrum $\langle \gamma^{3}\rangle$, and comment on their contributions to the trispectrum $\langle \gamma^{4}\rangle$.

\subsection{Pure graviton bispectrum\label{sec:BispectrumFromT3}}

  We begin by computing the graviton bispectrum induced by the $\langle T^{3}\rangle$ wavefunction coefficient.  These are the contributions to $\langle\gamma^{3}\rangle$ from General Relativity and higher order curvature terms, which were originally computed in \cite{Maldacena:2011nz}.  We verify here that our calculations reproduce those results.

  The relation between $\langle \gamma^{3}\rangle$ and the wavefunction coefficients follows from a straightforward generalization of \eqref{CorrelatorExpressions}.  Completing the computation, we find that the shapes due to the various components are:
  \begin{itemize}
 \item The shape proportional to $a_{3}$ is:
 \begin{align}
  \langle \gamma_{\bfk_{1}}^{P}(\ts)\gamma_{\bfk_{2}}^{P}(\ts)\gamma_{\bfk_{3}}^{P}(\ts)\rangle'_{a_{3}}&=-\frac{a_{3}H^{4}\prod_{i}\left (k_{T}-2k_{i}\right )}{\MP ^{6}k_{T}\prod_{i}k_{i}^{5}} \left (\sum_{i}k_{i}^{4}+\sum_{i\neq j}3k_{i}^{2}k_{j}^{2}\right )\nn
&\qquad\quad\times\left(\sum_{i}k_{i}^{3}+\sum_{i\neq j}2k_{i}^{2}k_{j}+2\prod_{i}k_{i}\right )\nn
  \langle \gamma_{\bfk_{1}}^{P}(\ts)\gamma_{\bfk_{2}}^{X}(\ts)\gamma_{\bfk_{3}}^{X}(\ts)\rangle'_{a_{3}}&=\frac{4a_{3}H^{4}\prod_{i}\left (k_{T}-2k_{i}\right )}{\MP ^{6}k_{1}k_{T}\prod_{i}k_{i}^{4}} \left (3k_{1}^{2}+k_{2}^{2}+k_{3}^{2}\right )\nn
&\qquad\quad\times\left(\sum_{i}k_{i}^{3}+\sum_{i\neq j}2k_{i}^{2}k_{j}+2\prod_{i}k_{i}\right )\ .\label{NGFromTTTa3}
 \end{align}
  \item The shape proportional to $a_{4}$ is
   \begin{align}
  \langle \gamma_{\bfk_{1}}^{P}(\ts)\gamma_{\bfk_{2}}^{P}(\ts)\gamma_{\bfk_{3}}^{P}(\ts)\rangle'_{a_{4}}&= \frac{a_{4}H^{6}\prod_{i}\left (k_{T}-2k_{i}\right )}{\MP ^{6}k_{T}\prod_{i}k_{i}^{5}}\Big[\sum_{i}3k_{i}^{7}+\sum_{i\neq j}\big(6k_{i}^{6}k_{j}+36k_{i}^{5}k_{j}^{2}+59k_{i}^{4}k_{j}^{3}\big)\nn
&\quad+\sum_{i\neq j\neq l}\big(66k_{i}^{4}k_{j}^{2}k_{l}+26k_{i}^{2}k_{j}^{2}k_{l}+3k_{i}^{5}k_{j}k_{l}+73k_{i}^{3}k_{j}^{2}k_{l}^{2}\big)\Big]\nn
  \langle \gamma_{\bfk_{1}}^{P}(\ts)\gamma_{\bfk_{2}}^{X}(\ts)\gamma_{\bfk_{3}}^{X}(\ts)\rangle'_{a_{4}}&=- \frac{8a_{4}H^{6}\prod_{i}\left (k_{T}-2k_{i}\right )}{\MP ^{6}k_{1}k_{T}\prod_{i}k_{i}^{4}} \Big[7 k_{1}^5+k_{1}^4 (14 k_{2}+14 k_{3})\nn
&\quad+k_{1}^3 \left(17
   k_{2}^2+14 k_{2} k_{3}+17 k_{3}^2\right)\nn
   &\quad+k_{1}^2 \left(12
   k_{2}^3+20 k_{2}^2 k_{3}+20 k_{2} k_{3}^2+12
   k_{3}^3\right)\nn
&\quad+k_{1} \left(4 k_{2}^4+4 k_{2}^3 k_{3}+10
   k_{2}^2 k_{3}^2+4 k_{2} k_{3}^3+4 k_{3}^4\right)\nn
&\quad+2
   k_{2}^5+4 k_{2}^4 k_{3}+7 k_{2}^3 k_{3}^2+7 k_{2}^2
   k_{3}^3+4 k_{2} k_{3}^4+2 k_{3}^5\Big]\, .\label{NGFromTTTa4}
 \end{align}
\item The shape proportional to $a_{5}$ is:
 \begin{align}
  \langle \gamma_{\bfk_{1}}^{P}(\ts)\gamma_{\bfk_{2}}^{P}(\ts)\gamma_{\bfk_{3}}^{P}(\ts)\rangle'_{a_{5}}&= -\frac{24a_{5}H^{8}\prod_{i}\left (k_{T}-2k_{i}\right )}{\MP ^{6}k_{T}^{3}\prod_{i}k_{i}^{5}} \Big[\sum_{i}3k_{i}^{9}+\sum_{i\neq j}\big (12k_{i}^{8}k_{j}+18k_{i}^{7}k_{j}^{2}\nn
&\quad+16k_{i}^{6}k_{j}^{3}+15k_{i}^{5}k_{j}^{4}\big )+\sum_{i\neq j\neq l}\big (18k_{i}^{7}k_{j}k_{l}+ 36 k_{i}^{6}k_{j}^{2}k_{l}+28  k_{i}^{5}k_{j}^{3}k_{l}\nn
&\quad+5k_{i}^{5}k_{j}^{2}k_{l}^{2}+16k_{i}^{4}k_{j}^{4}k_{l}-4k_{i}^{4}k_{j}^{3}k_{l}^{2}+16 k_{i}^{3}k_{j}^{3}k_{l}^{3}\big )\Big]\nn
  \langle \gamma_{\bfk_{1}}^{P}(\ts)\gamma_{\bfk_{2}}^{X}(\ts)\gamma_{\bfk_{3}}^{X}(\ts)\rangle'_{a_{5}}&=-\frac{24a_{5}H^{8}\prod_{i}\left (k_{T}-2k_{i}\right )}{\MP ^{6}k_{T}^{3}k_{1}\prod_{i}k_{i}^{4}} \Big[-k_{1}^7+k_{1}^6 (-4 k_{2}-4 k_{3})\nn
&\quad+k_{1}^5 \left(2
   k_{2}^2-12 k_{2} k_{3}+2 k_{3}^2\right)+k_{1}^4 \left(15
   k_{2}^3+20 k_{2}^2 k_{3}+20 k_{2} k_{3}^2+15
   k_{3}^3\right)\nn
&\quad+k_{1}^3 \left(3 k_{2}^4+40 k_{2}^3 k_{3}-38
   k_{2}^2 k_{3}^2+40 k_{2} k_{3}^3+3
   k_{3}^4\right)\nn
&\quad+k_{1}^2 \left(-22 k_{2}^5-28 k_{2}^4 k_{3}+46
   k_{2}^3 k_{3}^2+46 k_{2}^2 k_{3}^3-28 k_{2} k_{3}^4-22
   k_{3}^5\right)\nn
&\quad+k_{1} \big(-20 k_{2}^6-60 k_{2}^5 k_{3}-44
   k_{2}^4 k_{3}^2-8 k_{2}^3 k_{3}^3-44 k_{2}^2 k_{3}^4\nn
&\quad-60
   k_{2} k_{3}^5-20 k_{3}^6\big)-5 k_{2}^7-20 k_{2}^6
   k_{3}-26 k_{2}^5 k_{3}^2-13 k_{2}^4 k_{3}^3-13 k_{2}^3
   k_{3}^4\nn
&\quad-26 k_{2}^2 k_{3}^5-20 k_{2} k_{3}^6-5 k_{3}^7\Big]\, .\label{NGFromTTTa5}
 \end{align}
 \end{itemize}

  We also calculate the contribution to $\langle \gamma^{3}\rangle$ from the purely boundary terms \eqref{Massless3BoundaryTerm}:
  \begin{align}
  \langle \gamma_{\bfk_{1}}^{P}(\ts)\gamma_{\bfk_{2}}^{P}(\ts)\gamma_{\bfk_{3}}^{P}(\ts)\rangle'_{\rm bdy.}&= -\frac{24\lambda H^{4}}{2\MP ^{6}}\frac{k_{T}\prod_{i}(k_{T}-2k_{i})}{\prod_{i}k_{i}^{5}}\left (\sum_{i}k_{i}^{2}\right )\left (\sum_{i}k_{i}^{3}\right )\nn
    \langle \gamma_{\bfk_{1}}^{P}(\ts)\gamma_{\bfk_{2}}^{X}(\ts)\gamma_{\bfk_{3}}^{X}(\ts)\rangle'_{\rm bdy.}&=\frac{24\lambda H^{4}}{\MP ^{6}}\frac{k_{T}\prod_{i}(k_{T}-2k_{i})}{k_{1}\prod_{i}k_{i}^{4}}\left (\sum_{i}k_{i}^{3}\right )\, . \label{NGFromTTTbdy}
  \end{align}

\paragraph{Comparison to \cite{Maldacena:2011nz}:} The results in \eqref{NGFromTTTa3} reproduce the Einstein--Hilbert shape found in \cite{Maldacena:2011nz} if we set $a_{3}=\MP ^{2}/4$ and $a_{4}=a_{5}=0$.  The $W_{\mu\nu\rho\sigma}^{3}/\Lambda_{W}^{2}$ shape of \cite{Maldacena:2011nz} is reproduced by setting $a_{3}=\frac{297H^{4}}{2\Lambda_{W}^{2}}$, $a_{4}=\frac{63H^{2}}{2\Lambda_{W}^{2}}$, and $a_{5}=-\frac{3}{4\Lambda_{W}^{2}}$.  Due to the relation \eqref{GBCorrelatorRelation}, the values of the $a_{i}$'s can be shifted around in these matchings at the cost of also adding boundary term shapes to the correlator.
  
\subsection{Graviton bispectrum from interactions involving partially massless fields}

All remaining cubic wavefunction coefficients involve at least one factor of $\Sigma$, which encodes the effects of partially massless fields. In order for the these components to affect the observable graviton bispectrum $\langle \gamma^{3}\rangle$, we require a way to convert the PM spin-2 into the massless graviton: this requires the presence of a mixing term $\sim \langle T\Sigma\rangle\,\bar\gamma\bar\sigma$ in the wavefunction.  In the following sections, we discuss the generation of such a mixing term and use the result to compute the imprint of $\Sigma$-dependent coefficients on $\langle \gamma^{3}\rangle$.

\subsubsection{A Spin-2 mixing interaction\label{sec:Mixing}}

In pure de Sitter space, linear interactions between two spin-2 fields of different mass can always be diagonalized, so a mixing of the kind desired is impossible.\footnote{This is most straightforward to see from the dual perspective: the two-point function of spin-2 fields is constrained by conformal invariance. In particular, invariance under special conformal transformations requires that the spin-2 fields have the same conformal weight (equivalently the same superhorizon time dependence). This requires the fields to have the same mass.}
However, inflationary spacetimes are only quasi-de Sitter, with the departure characterized by the slow-roll parameters, the first two of which are given by
\begin{align}
\varepsilon&\equiv -\frac{1}{H^2}\frac{\rd H}{\rd t}\,,~~~~~~~~~~~~~~~ \eta\equiv \frac{1}{H\varepsilon}\frac{\rd\varepsilon}{\rd t}\, ,
\end{align}
with $t$ the proper time.
By relaxing the constraint of special conformal invariance---but retaining all other de Sitter symmetry requirements---the leading order form of the mixing  coefficient is restricted to take the form
\begin{align}
   \langle T^{ij}_{\bfk}\Sigma^{lm}_{-\bfk}\rangle'&= f(\varepsilon,\eta)(\Pi^{\bfk}_{TT})^{ ijlm}\frac{k^{2}}{\ts},\label{ExpectedMixingForm}
   \end{align}
   for some $f(\varepsilon,\eta)$ that vanishes as the slow-roll parameters approach zero.\footnote{The $k$ dependence follows from dilation invariance with $\Delta_T = 3, \Delta_\Sigma =2$. The $\ts$-dependence in \eqref{ExpectedMixingForm} follows from the late time behavior of the solutions \eqref{Explicitd3Solutions} and the form of the other quadratic wavefunction coefficients.  At late times $\gamma\sim \tau^{0}$ and $\sigma \sim \tau$ and so we expect $\langle\gamma\sigma\rangle\propto \ts$.  This correlator is built from wavefunction coefficients via $\langle\gamma\sigma\rangle\propto \re\langle T^{2}\rangle^{-1} \re\langle T\Sigma\rangle \re\langle \Sigma^{2}\rangle^{-1} $ and from \eqref{ReTTMassless} and \eqref{ReQuadSPartiallyMassless} it follows that $\langle\gamma\sigma\rangle\propto \ts^{2} \re\langle T\Sigma\rangle $, thus we expect $\re\langle T\Sigma\rangle\propto \ts^{-1}$.}  This form of mixing is realized by using, for instance, the following explicit interaction between a massless spin-2 $(h^{(\gamma)}_{\mu\nu})$, a partially massless spin-2 $(h^{(\sigma)}_{\mu\nu})$, and the inflaton $(\phi)$:\footnote{We are assuming here that this interaction can be made gauge invariant under both the massless and partially massless spin-2 gauge transformations (perhaps by including other interaction terms, adjusting the overall coefficient, and/or having $\phi$ transform non-trivially under these gauge symmetries) and that the resulting theory admits a slow-roll inflationary solution.   Verifying this explicitly is beyond the scope of the present paper, however.  See \cite{Baumann:2017jvh} for a different discussion of couplings between partially massless field and the inflaton.}
   \begin{align}
   S_{\rm mixing}=\frac{1}{{\Lambda}^{2}}\int\rd^{4}x\, \sqrt{-g}\, \nabla_{\alpha}h^{(\sigma)}_{\mu\nu}\nabla_{\beta}h^{(\gamma)\mu\nu}\nabla^{\alpha}\phi\nabla^{\beta}\phi\ .\label{MixingInteraction}
   \end{align}
   Putting $\phi$ on the inflationary background by replacing $\phi(x^{\mu})\to \bar{\phi}(\tau)$, converting from $h^{(\gamma)}_{\mu\nu}\to \gamma_{ij}$ and $h^{(\sigma)}_{\mu\nu}\to \sigma_{ij}$, and finally going on-shell, one finds
   \begin{align}
   S_{\rm mixing, cl}[\gamma_{\rm cl},\sigma_{\rm cl}]&\supset -i\int\rd^{3}\tilde{k}\, \frac{\pi\varepsilon \MP ^{2}}{{\Lambda}^{2}}\frac{k^{2}}{\ts}\bar{\gamma}_{ij}^{\bfk}\bar{\sigma}^{\bfk ij} \ ,\label{ExampleMixingAction}
   \end{align}
   where $\partial_{\tau}\bar{\phi}(\tau)\approx a(\tau)\sqrt{2\varepsilon}H \MP $ and the standard slow-roll approximations $\varepsilon\approx$ constant and $a(\tau)\approx \frac{1}{H(-\tau)}$ were used. The result \eqref{ExampleMixingAction} corresponds to \eqref{ExpectedMixingForm} with $f(\varepsilon,\eta) =\pi\varepsilon \MP ^{2}/{\Lambda}^{2}$.

     In the calculations to follow, we will take the form of $\langle T\Sigma\rangle$ to be
   \begin{align}
   \langle T_{\bfk}^{ij}\Sigma_{-\bfk}^{lm}\rangle'&\equiv \left (\frac{\Lambda_{\rm mix}}{H}\right )^{2}\frac{k^{2}}{\ts}(\Pi^{\bfk}_{TT} )^{ ijlm} \ ,
   \label{FormOfTSigmaMixing}
   \end{align}
   where all slow-roll parameters and energy scales apart from Hubble have been absorbed into $\Lambda_{\rm mix}$.\footnote{The $\sim H^{-2}$ form of the prefactor was chosen to mimic the $H^{-2}$ factors in $\langle T^{2}\rangle$ and $\langle \Sigma^{2}\rangle$ and is a convenient choice for later expressions.} In the context of the example~\eqref{MixingInteraction}, we have $\Lambda_{\rm mix}^2 = \pi\varepsilon\MP^2 H^2/\Lambda^2$. The mixing term \eqref{FormOfTSigmaMixing} corrects the graviton power spectrum through diagrams of the form shown in Fig.~\ref{fig:Graviton2PointFunctionWithMixing}. At leading order, the corrections come in the form
\begin{align}
\langle \gamma^{\bfk}_{ij}(\ts)\gamma^{-\bfk}_{lm}(\ts)\rangle'&=\frac{H^{2}}{\MP ^{2}k^{3}}(\Pi^{\bfk}_{TT} )_{ ijlm}\left (1+\mathcal{O}\left (\frac{\Lambda_{\rm mix}^{4}}{\MP ^{4}}\right )\right )\, .
\end{align}
In order for our computations to be reliable, we want these corrections to the graviton propagator to be small, so we
demand that $\Lambda_{\rm mix} \ll \MP $.

\begin{figure}
\begin{center}
\begin{tikzpicture}
\node at (-.25,0) {$\langle \gamma^{2}\rangle$=};
\draw[graviton, line width=1.2 pt] (.6,0)--(3.6,0);

	\begin{scope}[xshift=4cm]

	\node at (0,0) {$+$};

	\coordinate (dotLeftL) at (.5,0) {};

	\coordinate (dotRightL) at (4.25,0) {};

	\node at (1.6,0) [circle,draw=black,inner sep=.1cm,line width= .2mm](mixL) {};
	\node at (1.6,0) [cross,line width=1pt,minimum size=.2cm](mixLcross) {};
	
	\node at (3.25,0) [circle,draw=black,inner sep=.1cm,line width= .2mm](mixR) {};
	\node at (3.25,0) [cross,line width=1pt,minimum size=.2cm](mixRcross) {};

\path[]
	(dotLeftL) edge[graviton,line width=1.2 pt] node  {} (mixL)
	(mixL) edge[pmfield,line width=1.2 pt] node  {} (mixR)
	(mixR) edge[graviton,line width=1.2 pt] node  {} (dotRightL);

		\node at (5,0) {$+\ldots$};
	
	\end{scope}

\end{tikzpicture}
\end{center}
\caption{The $\langle T\Sigma\rangle$ term also corrects the graviton power spectrum through diagrams of the above type. These corrections are small in the limit $\Lambda_{\rm mix}/\MP \ll 1$.}
\label{fig:Graviton2PointFunctionWithMixing}
\end{figure}

\subsubsection{Graviton bispectrum from $\langle \Sigma^{3}\rangle$\label{sec:BispectrumFromS3}}
We first consider the contribution to the graviton bispectrum coming from pure PM spin-2 interactions, 
  generated by the $\langle \Sigma^{3}\rangle$ wavefunction coefficient. It turns out that the induced bispectrum decays as $\ts$, since the contribution is of the schematic form (see Fig.~\ref{fig:GravitonnPointFunctionsFromPMFields})
  \begin{align}
  \langle \gamma^{3}\rangle\sim \frac{1}{\re\langle T^{2}\rangle^{3}}\frac{\re\langle T\Sigma\rangle^{3}}{ \re\langle \Sigma^{2}\rangle^{3}}\re\langle \Sigma^{3}\rangle\propto \ts^{3}\re\langle \Sigma^{3}\rangle\ ,
  \end{align}
where in the last relation we have used
\eqref{MasslessSpin2TwoPointd3},  \eqref{PartiallyMasslessSpin2TwoPointd3}, and  \eqref{FormOfTSigmaMixing} to extract the time dependence of the two-point coefficients.  As noted in Sec.~\ref{sec:WavefunctionCoefficientSigma3}, the leading component of $\re\langle \Sigma^{3}\rangle$ is only $\mathcal{O}(\ts^{-2})$ and hence this contribution decays as $\ts$ at late times, and will not produce an interesting signal in graviton non-Gaussianities.

\subsubsection{Graviton bispectrum from $\langle \Sigma T^{2}\rangle$\label{sec:BispectrumFromST2}}
Next, we consider the contribution to $\langle\gamma^3\rangle$ from interactions involving a single partially massless spin-2 and two massless gravitons, induced by the $\langle \Sigma T^{2}\rangle$ wavefunction coefficient. The associated non-Gaussianity is rather strange: it grows as $\propto\ts^{-1}$, but it also has purely the same shape as the ambiguous boundary term.

The contribution from $\langle \Sigma T^{2}\rangle$ to $\langle \gamma^{3}\rangle$ is of the schematic form (see Fig.~\ref{fig:GravitonnPointFunctionsFromPMFields})
\be
\langle \gamma^{3}\rangle\sim \frac{1}{\re\langle T^{2}\rangle^{3}}\frac{\re\langle T\Sigma\rangle}{ \re\langle \Sigma^{2}\rangle}\re\langle \Sigma T^{2}\rangle\sim \ts\re\langle \Sigma T^{2}\rangle\ ,
\ee
as follows from an analysis similar to that of the preceding section. As noted in Sec.~\ref{sec:WavefunctionCoefficientSigmaT2}, the leading component of $\re\langle \Sigma T^{2}\rangle$ at late times grows as $\mathcal{O}(\ts^{-2})$, hence the $\re\langle \Sigma T^{2}\rangle$ term would seem to produce non-Gaussianity which diverges as $\ts^{-1}$.  Specifically, the following shapes are generated:
\begin{align}
\langle \gamma_{\bfk_{1}}^{P}(\ts)\gamma_{\bfk_{2}}^{P}(\ts)\gamma_{\bfk_{3}}^{P}(\ts)\rangle'_{\rm bdy.}&=-16\left (b_{4b}+12H^{2}b_{5}\right )\frac{\Lambda_{\rm mix}^{2}H^{6}}{\MP ^{8}\ts}\frac{k_{T}\prod_{i}(k_{T}-2k_{i})}{\prod_{i}k_{i}^{5}} \left (\sum	_{i}k_{i}^{2}\right )^{2}\nn
\langle \gamma_{\bfk_{1}}^{P}(\ts)\gamma_{\bfk_{2}}^{X}(\ts)\gamma_{\bfk_{3}}^{X}(\ts)\rangle'_{\rm bdy.}&=16\left (b_{4b}+12H^{2}b_{5}\right )\frac{2\Lambda_{\rm mix}^{2}H^{6}}{\MP ^{8}\ts}\frac{k_{T}\prod_{i}(k_{T}-2k_{i})}{k_{1}\prod_{i}k_{i}^{4}} \left (\sum_{i}k_{i}^{2}\right )\ .\label{NGFromST2}
\end{align}
As also noted in Sec.~\ref{sec:WavefunctionCoefficientSigmaT2}, these correspond to the local shapes which can be reproduced (or removed) by field redefinitions, so it is somewhat unclear what---if any---significance we should ascribe to the late-time divergence.

\subsubsection{Graviton bispectrum from $\langle T\Sigma^{2}\rangle$\label{sec:BispectrumFromTS2}}
In this section, we discuss the graviton bispectrum arising from 3-point interactions involving two partially massless fields and one massless graviton. This corresponds to the contribution of  the $\langle T\Sigma^{2}\rangle$ wavefunction coefficient to $\langle\gamma^3\rangle$.   
This is the only coefficient which sources the $\ts$-independent component of $\langle \gamma^{3}\rangle$, and is therefore the most phenomenologically-interesting situation.
  
The contribution from $\langle T\Sigma^{2}\rangle$ to $\langle \gamma^{3}\rangle$ is of the schematic form (see Fig.~\ref{fig:GravitonnPointFunctionsFromPMFields})
  \begin{align}
  \langle \gamma^{3}\rangle\sim \frac{1}{\re\langle T^{2}\rangle^{3}}\frac{\re\langle T\Sigma\rangle^{2}}{ \re\langle \Sigma^{2}\rangle^{2}}\re\langle T\Sigma^{2}\rangle\sim \ts^{2}\re\langle T \Sigma^{2}\rangle\ ,
  \end{align}
  as follows from an analysis similar to that of the preceding sections. The leading component of $\re \langle T\Sigma^{2}\rangle$ scales with conformal time as $\mathcal{O}(\ts^{-2})$, hence $\langle T\Sigma^{2}\rangle$ sources long-lived $\mathcal{O}(\ts^{0})$ graviton non-Gaussianities.  We find the following shapes:
   \begin{itemize}
   \item The shape produced by the $b_{3a}$ coefficient is:
    \begin{align}
  \langle \gamma_{\bfk_{1}}^{P}(\ts)\gamma_{\bfk_{2}}^{P}(\ts)\gamma_{\bfk_{3}}^{P}(\ts)\rangle'_{b_{3a}}&=\frac{16b_{3a}\Lambda_{\rm mix}^{4}H^{4}}{\MP ^{10}}\frac{\prod_{i}(k_{T}-2k_{i})}{k_{T}\prod_{i}k_{i}^{5}}\Big[\sum_{i\neq j}\big(k_{i}^{6}k_{j}+k_{i}^{5}k_{j}^{2}+6k_{i}^{4}k_{j}^{3}\big)\nn
  &\quad+\sum_{i\neq j\neq l}\big(k_{i}^{5}k_{j}k_{l}-k_{i}^{4}k_{j}^{2}k_{l}-2k_{i}^{3}k_{j}^{3}k_{l}-2k_{i}^{3}k_{j}^{2}k_{l}^{2}\big)\Big]\nn
    \langle \gamma_{\bfk_{1}}^{P}(\ts)\gamma_{\bfk_{2}}^{X}(\ts)\gamma_{\bfk_{3}}^{X}(\ts)\rangle'_{b_{3a}}&=\frac{64b_{3a}\Lambda_{\rm mix}^{4}H^{4}}{\MP ^{10}}\frac{\prod_{i}(k_{T}-2k_{i})}{k_{1}^{2}k_{T}\prod_{i}k_{i}^{3}}\Big[4 k_{1}^3+k_{1}^2
   (k_{2}+k_{3})\nn
   &\quad-k_{2}^2
   k_{3}-k_{2} k_{3}^2-k_{2}^3-k_{3}^3\Big].\label{NGFromTS23a}
  \end{align}	
   \item The shape produced by the $b_{3b}$ coefficient is:
    \begin{align}
  \langle \gamma_{\bfk_{1}}^{P}(\ts)\gamma_{\bfk_{2}}^{P}(\ts)\gamma_{\bfk_{3}}^{P}(\ts)\rangle'_{b_{3b}}&=\frac{16b_{3b}\Lambda_{\rm mix}^{4}H^{4}}{\MP ^{10}}\frac{\prod_{i}(k_{T}-2k_{i})}{k_{T}\prod_{i}k_{i}^{5}}\Big[\sum_{i\neq j}\big(-k_{i}^{6}k_{j}-k_{i}^{5}k_{j}^{2}-6k_{i}^{4}k_{j}^{3}\big)\nn
  &\quad+\sum_{i\neq j\neq l}\big(-5k_{i}^{5}k_{j}k_{l}-3k_{i}^{4}k_{j}^{2}k_{l}-8k_{i}^{3}k_{j}^{3}k_{l}-2k_{i}^{3}k_{j}^{2}k_{l}^{2}\big)\Big]\nn
    \langle \gamma_{\bfk_{1}}^{P}(\ts)\gamma_{\bfk_{2}}^{X}(\ts)\gamma_{\bfk_{3}}^{X}(\ts)\rangle'_{b_{3b}}&=\frac{32b_{3b}\Lambda_{\rm mix}^{4}H^{4}}{\MP ^{10}}\frac{\prod_{i}(k_{T}-2k_{i})}{k_{1}k_{T}\prod_{i}k_{i}^{4}}\Big[k_{1}^4 (3 k_{2}+3 k_{3})\nn
    &\quad+k_{1}^3 \left(3
   k_{2}^2+14 k_{2} k_{3}+3
   k_{3}^2\right)+k_{1}^2 \left(k_{2}^3+3 k_{2}^2
   k_{3}+3 k_{2} k_{3}^2+k_{3}^3\right)\nn
    &\quad+k_{1}
   \left(k_{2}^4+6 k_{2}^3 k_{3}+2 k_{2}^2
   k_{3}^2+6 k_{2} k_{3}^3+k_{3}^4\right)\nn
    &\quad+2
   k_{2}^4 k_{3}+2 k_{2}^3 k_{3}^2+2 k_{2}^2
   k_{3}^3+2 k_{2} k_{3}^4\Big].\label{NGFromTS23b}
  \end{align}
   \item The shape produced by the $b_{4a}$ coefficient is:
    \begin{align}
  \langle \gamma_{\bfk_{1}}^{P}(\ts)\gamma_{\bfk_{2}}^{P}(\ts)\gamma_{\bfk_{3}}^{P}(\ts)\rangle'_{b_{4a}}&=\frac{32b_{4a}\Lambda_{\rm mix}^{4}H^{6}}{\MP ^{10}}\frac{\prod_{i}(k_{T}-2k_{i})}{k_{T}\prod_{i}k_{i}^{5}}\Big[\sum_{i\neq j}\big(-2k_{i}^{5}k_{j}^{2}+2k_{i}^{4}k_{j}^{3}\big)\nn
  &\quad+\sum_{i\neq j\neq l}\big(\frac{1}{2}k_{i}^{5}k_{j}k_{l}-2k_{i}^{4}k_{j}^{2}k_{l}+9k_{i}^{3}k_{j}^{3}k_{l}-8k_{i}^{3}k_{j}^{2}k_{l}^{2}\big)\Big]\nn
    \langle \gamma_{\bfk_{1}}^{P}(\ts)\gamma_{\bfk_{2}}^{X}(\ts)\gamma_{\bfk_{3}}^{X}(\ts)\rangle'_{b_{4a}}&=\frac{32b_{4a}\Lambda_{\rm mix}^{4}H^{6}}{\MP ^{10}}\frac{\prod_{i}(k_{T}-2k_{i})}{k_{1}k_{T}\prod_{i}k_{i}^{4}}\Big[k_{1}^4 (-7 k_{2}-7 k_{3})\nn
    &\quad+k_{1}^3 \left(-3
   k_{2}^2-30 k_{2} k_{3}-3
   k_{3}^2\right)\nn
    &\quad+k_{1}^2 \left(3 k_{2}^3+13
   k_{2}^2 k_{3}+13 k_{2} k_{3}^2+3
   k_{3}^3\right)\nn
    &\quad+k_{1} \left(-k_{2}^4-10 k_{2}^3
   k_{3}+6 k_{2}^2 k_{3}^2-10 k_{2}
   k_{3}^3-k_{3}^4\right)\nn
    &\quad	-2 k_{2}^4 k_{3}+2
   k_{2}^3 k_{3}^2+2 k_{2}^2 k_{3}^3-2 k_{2}
   k_{3}^4\Big].\label{NGFromTS24a}
  \end{align}
   \item The shape produced by the $b_{4b}$ coefficient is:
    \begin{align}
  \langle \gamma_{\bfk_{1}}^{P}(\ts)\gamma_{\bfk_{2}}^{P}(\ts)\gamma_{\bfk_{3}}^{P}(\ts)\rangle'_{b_{4b}}&=\frac{8b_{4b}\Lambda_{\rm mix}^{4}H^{6}}{\MP ^{10}}\frac{\prod_{i}(k_{T}-2k_{i})}{k_{T}^{3}\prod_{i}k_{i}^{5}}\Big[\sum_{i\neq j}\big(k_{i}^{6}k_{j}-7k_{i}^{5}k_{j}^{2}+14k_{i}^{4}k_{j}^{3}\big)\nn
  &\quad+\sum_{i\neq j\neq l}\big(4k_{i}^{5}k_{j}k_{l}-5k_{i}^{4}k_{j}^{2}k_{l}+44k_{i}^{3}k_{j}^{3}k_{l}+6k_{i}^{3}k_{j}^{2}k_{l}^{2}\big)\Big]\nn
    \langle \gamma_{\bfk_{1}}^{P}(\ts)\gamma_{\bfk_{2}}^{X}(\ts)\gamma_{\bfk_{3}}^{X}(\ts)\rangle'_{b_{4b}}&=\frac{16b_{4b}\Lambda_{\rm mix}^{4}H^{6}}{\MP ^{10}}\frac{\prod_{i}(k_{T}-2k_{i})}{k_{1}k_{T}\prod_{i}k_{i}^{4}}\Big[k_{1}^4 (-17 k_{2}-17 k_{3})\nn
    &\quad+k_{1}^3 \left(-9
   k_{2}^2-110 k_{2} k_{3}-9
   k_{3}^2\right)\nn
    &\quad+k_{1}^2 \left(5 k_{2}^3-13
   k_{2}^2 k_{3}-13 k_{2} k_{3}^2+5
   k_{3}^3\right)\nn
    &\quad+k_{1} \left(-3 k_{2}^4-26 k_{2}^3
   k_{3}+10 k_{2}^2 k_{3}^2-26 k_{2} k_{3}^3-3
   k_{3}^4\right)\nn
    &\quad-6 k_{2}^4 k_{3}+2 k_{2}^3
   k_{3}^2+2 k_{2}^2 k_{3}^3-6 k_{2} k_{3}^4\Big].\label{NGFromTS24b}
  \end{align}
   \item The shape produced by the $b_{5}$ coefficient is:
    \begin{align}
  \langle \gamma_{\bfk_{1}}^{P}(\ts)\gamma_{\bfk_{2}}^{P}(\ts)\gamma_{\bfk_{3}}^{P}(\ts)\rangle'_{b_{5}}&=\frac{128b_{5}\Lambda_{\rm mix}^{4}H^{8}}{\MP ^{10}}\frac{\prod_{i}(k_{T}-2k_{i})}{k_{T}^{3}\prod_{i}k_{i}^{5}}\Big[\sum_{i\neq j}\big(-2k_{i}^{8}k_{j}-7k_{i}^{7}k_{j}^{2}-19k_{i}^{6}k_{j}^{3}\nn
    &\quad-36k_{i}^{5}k_{j}^{4}\big)+\sum_{i\neq j\neq l}\big(-\frac{11}{2}k_{i}^{7}k_{j}k_{l}-16k_{i}^{6}k_{j}^{2}k_{l}-5k_{i}^{5}k_{j}^{3}k_{l}\nn
  &\quad+2k_{i}^{4}k_{j}^{4}k_{l}+15k_{i}^{4}k_{j}^{3}k_{l}^{2}+55k_{i}^{3}k_{j}^{3}k_{l}^{3}\big)\Big]\nn
    \langle \gamma_{\bfk_{1}}^{P}(\ts)\gamma_{\bfk_{2}}^{X}(\ts)\gamma_{\bfk_{3}}^{X}(\ts)\rangle'_{b_{5}}&=\frac{128b_{5}\Lambda_{\rm mix}^{4}H^{8}}{\MP ^{10}}\frac{\prod_{i}(k_{T}-2k_{i})}{k_{1}k_{T}^{3}\prod_{i}k_{i}^{4}}\Big[k_{1}^6 (-5 k_{2}-5 k_{3})\nn
    &\quad+k_{1}^5 \left(-13
   k_{2}^2-84 k_{2} k_{3}-13
   k_{3}^2\right)\nn
    &\quad+k_{1}^4 \left(-10 k_{2}^3-111
   k_{2}^2 k_{3}-111 k_{2} k_{3}^2-10
   k_{3}^3\right)\nn
    &\quad+k_{1}^3 \left(-2 k_{2}^4-48
   k_{2}^3 k_{3}-388 k_{2}^2 k_{3}^2-48 k_{2}
   k_{3}^3-2 k_{3}^4\right)\nn
    &\quad+k_{1}^2
   \left(-k_{2}^5-19 k_{2}^4 k_{3}+16 k_{2}^3
   k_{3}^2+16 k_{2}^2 k_{3}^3-19 k_{2}
   k_{3}^4-k_{3}^5\right)\nn
    &\quad+k_{1} \left(-k_{2}^6+4
   k_{2}^5 k_{3}+17 k_{2}^4 k_{3}^2+24 k_{2}^3
   k_{3}^3+17 k_{2}^2 k_{3}^4+4 k_{2}
   k_{3}^5-k_{3}^6\right)\nn
    &\quad+7 k_{2}^6 k_{3}+23
   k_{2}^5 k_{3}^2+34 k_{2}^4 k_{3}^3+34
   k_{2}^3 k_{3}^4+23 k_{2}^2 k_{3}^5+7 k_{2}
   k_{3}^6\Big].\label{NGFromTS25}
  \end{align}
   \end{itemize}

We also calculate the contribution to $\langle \gamma^{3}\rangle$ from the purely boundary terms \eqref{PM2MasslessBdyTermCorrelators}:
\begin{align}
\langle \gamma_{\bfk_{1}}^{P}(\ts)\gamma_{\bfk_{2}}^{P}(\ts)\gamma_{\bfk_{3}}^{P}(\ts)\rangle'_{\rm bdy.}&=-\lambda\frac{64H^{4}\Lambda_{\rm mix}^{4}}{\MP ^{10}}\frac{k_{T}\prod_{i}(k_{T}-2k_{i})}{\prod_{i}k_{i}^{5}}\left (\sum_{i}k_{i}^{2}\right )\left (\sum_{i\neq j}k_{i}^{2}k_{j}\right )\nn
\langle \gamma_{\bfk_{1}}^{P}(\ts)\gamma_{\bfk_{2}}^{X}(\ts)\gamma_{\bfk_{3}}^{X}(\ts)\rangle'_{\rm bdy.}  &=\lambda\frac{128\Lambda_{\rm mix}^{4}H^{4}}{\MP ^{10}}\frac{k_{T}\prod_{i}(k_{T}-2k_{i})}{k_{1}\prod_{i}k_{i}^{4}}\left (\sum_{i\neq j}k_{i}^{2}k_{j}\right ) \, .\label{NGFromTS2Bdy}
\end{align}
The precise details of the shapes enumerated in this section are perhaps not particularly important at this juncture. What we would like to emphasize, however, is that they are {\it different} from the pure graviton shapes considered in Section~\ref{sec:BispectrumFromT3}. In this sense, a measurement of one of these shapes would be a sharp indication of the presence of an additional degree of freedom during inflation.

\subsection{Soft limits}
In the context of scalar non-Gaussianity, soft limits of correlation functions are a particularly sensitive probe of the presence of additional heavy particles~\cite{Chen:2009zp,Assassi:2012zq,Noumi:2012vr,Arkani-Hamed:2015bza,Lee:2016vti}. We are therefore motivated to see how the presence of additional spin-2 fields can show up in tensor 3-point functions.
In this section we discuss these soft limits---in particular the dominant contribution comes from
  bulk mixing diagrams that have been neglected up to this point, where the PM spin-2 converts into the massless graviton during inflation.
In order to connect to familiar formulae in soft limits, we present results expressed in terms of arbitrary polarization tensors, rather than expressing equations in terms of $\epsilon^{X}$ and $\epsilon^{P}$.

\subsubsection{Heretofore neglected bulk diagrams}

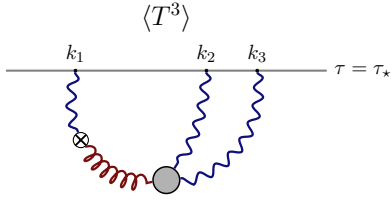
\begin{figure}
\begin{center}
\resizebox{5.5cm}{!}{
\begin{tikzpicture}

\begin{scope}[xshift=0cm]
\node at (0,1) {\Large $\langle T^{3}\rangle$};
\draw[gray, line width=1.1pt] (-3,0) -- (3,0) node [right,black] {$\tau=\ts$};
\node at (0,-2.05) [vertex,line width=.8pt,minimum size=.5cm](vertex) {};
\node[label=$k_{1}$] at (-1.7,0) [smalldot,minimum size=.05cm](dotL) {};
\node[label=$k_{2}$] at (.75,0) [smalldot,minimum size=.05cm](dotC) {};
\node[label=$k_{3}$] at (1.7,0) [smalldot,minimum size=.05cm](dotR) {};

	\node at (-1.6,-1.3) [circle,draw=black,inner sep=.1cm,line width= .2mm](mix) {};
	\node at (-1.6,-1.3) [cross,line width=1pt,minimum size=.2cm](mixcross) {};
\path[]
	(dotL) edge[bend right=15,graviton, line width=1.2 pt] node  {} (mix)
	(mix) edge[bend right=32,pmfield, line width=1.2 pt] node  {} (vertex)
	(dotC) edge[bend left=14,graviton, line width=1.2 pt] node  {} (vertex)
	(dotR) edge[bend left=43,graviton, line width=1.2 pt] node  {} (vertex);
\end{scope}
\end{tikzpicture}
}
\end{center}
\caption{The simplest bulk mixing diagram in which a PM field contributes to $\langle T^{3}\rangle$.}
\label{fig:Graviton3PtCoefficientFromPMMixingLabeled}
\end{figure}

  The mixing term produced by the interaction \eqref{MixingInteraction} also allows partially massless fields to imprint themselves on the $\langle T^{3}\rangle$ coefficient via thus-far neglected diagrams of the form shown in Fig.~\ref{fig:Graviton3PtCoefficientFromPMMixingLabeled}. These diagrams are expected to contribute to non-Gaussianity at the same order as the contributions which have been presented so far. Unfortunately, the calculation of such contributions is difficult, as they require the use of bulk-to-bulk propagators for the internal lines.   We only consider the single diagram in Fig.~\ref{fig:Graviton3PtCoefficientFromPMMixingLabeled} for what follows, and leave the calculation of more complicated bulk-mixing calculations for future work.

For computations such as that shown in Fig.~\ref{fig:Graviton3PtCoefficientFromPMMixingLabeled}, only the spatial, transverse, traceless parts of the bulk-to-bulk propagator for the PM field are required. These components take on the form
\begin{align}
{\cal G}^{TT}(\tau,\tau',k)_{ij}{}^{lm}\equiv {\cal G}(\tau,\tau',k)(\Pi^{\bfk}_{TT} )_{ij}{}^{lm}.
\end{align} 
The function ${\cal G}(\tau,\tau',k)$ can be derived as a Green's function for the equation of motion $\mathcal{E}_{\tau,k}{\cal G}(\tau,\tau',k)=-\delta(\tau-\tau')$, where $\mathcal{E}$ is the differential operator appearing in the EOM: $\frac{\delta S_{2}[\gamma]}{\delta \gamma(\tau,\bfx)^{ij}}\equiv \mathcal{E}_{\tau,\bfx}\gamma_{ij}(\tau,\bfx)$, subject to the following boundary conditions:
\begin{itemize}
\item Symmetry: ${\cal G}(\tau,\tau',k)={\cal G}(\tau',\tau,k)$.
\item The Dirichlet condition: $\lim_{\tau\to \ts} {\cal G}(\tau,\tau',k)=0$.
\item The vacuum condition: ${\cal G}(\tau,\tau',k) $ must behave as $\sim e^{+i\tau}$ at early times when $\tau<\tau'$.
\end{itemize}
\vspace{2mm}
along with a jump condition on the first derivative that follows from integrating the equation of motion.
The solution for ${\cal G}(\tau,\tau',k)$ is given by 
\be
{\cal G}(\tau,\tau',k)=\frac{2i\tau\tau' H^{2}}{\MP ^{2}k}\left(e^{ik(\tau'-\tau)}\theta(\tau-\tau')+e^{ik(\tau-\tau')}\theta(\tau'-\tau) - e^{ik(\tau-\ts)}e^{ik(\tau'-\ts)}
\right).
\label{PMBulktoBulk}
\ee
Given \eqref{PMBulktoBulk}, we can calculate the contribution shown in Fig.~\ref{fig:Graviton3PtCoefficientFromPMMixingLabeled} to $\langle T^{3}\rangle$ via (see Appendix~\ref{app:PathIntegralCalculationOfPsi} for details)
\be
\quad\int\rd^{3}\tilde{k}_{1}\rd^{3}\tilde{k}_{2}\rd^{3}\tilde{k}_{3} \langle T^{ij}_{\bfk_{1}}T^{lm}_{\bfk_{2}}T^{no}_{\bfk_{3}}\rangle_{\rm bulk-mix}\, \bar{\gamma}^{\bfk_{1}}_{ij}\bar{\gamma}^{\bfk_{2}}_{lm}\bar{\gamma}^{\bfk_{2}}_{no}\equiv-3!\int \rd\tau\rd\tau'\rd^{3}\tilde{p}\, \frac{\delta S_{\rm mix}}{\delta \sigma^{\bfp}_{ij}(\tau)}{\cal G}(\tau,\tau',p)_{ijlm}\frac{\delta S_{\rm int}}{\delta \sigma^{-\bfp}_{lm}(\tau')},\label{T3FromBulkMixing}
\ee
where $S_{\rm int}$ contains the $\mathcal{O}(\gamma^{2}\sigma)$ cubic interactions of Sec.~\ref{sec:PMM0M0CubicVerts} and $S_{\rm mix}$ is the linear $\mathcal{O}(\gamma \sigma)$ mixing term which we concretely write as
\begin{align}
S_{\rm mix}&=\int\rd^{4}x\, \frac{1}{\tau^{2}}\left (\frac{\Lambda_{\rm mix}}{H}\right )^{2}\, \gamma'_{ij}\sigma'^{ij}\ .\label{ExplicitMixingTerm}
\end{align}
This is just the explicit form of the bulk interaction which generates the mixing term in \eqref{FormOfTSigmaMixing}, though there could be more general possibilities.

\subsubsection{Soft-Limits from bulk mixing}

We now compute \eqref{T3FromBulkMixing} in the limit where one of the momenta is soft.  Until now we have been symmetrizing over the three momenta appearing in $\langle T^{3}\rangle$.  For the mixing diagram in Fig.~\ref{fig:Graviton3PtCoefficientFromPMMixingLabeled} it is convenient to instead only symmetrize over $k_{2}$ and $k_{3}$, and to keep $k_{1}$ as the leg where the mixing occurs, as indicated in Fig.~\ref{fig:Graviton3PtCoefficientFromPMMixingLabeled}.  When exploring soft limits, this is useful since it lets us isolate the effect of taking the leg where the mixing occurs to be soft, versus the effect of taking one  of the other two legs soft. Both calculations can be done analytically.

First, we take one of the non-mixing legs to be soft by sending $\bfk_{3}\to \bfq$, $\bfk_{2}\to \bfk$, and $\bfk_{1}\to -\bfk-\bfq$.  In the $\bfq\to 0$ limit, the leading terms are,\footnote{Explicit calculation demonstrates that both $\Tr  (\epsilon^{\bfk}\cdot\epsilon^{\bfq}\cdot\epsilon^{-\bfk-\bfq} )$ and $\Tr  (\epsilon^{\bfk}\cdot\epsilon^{-\bfk-\bfq} ) (\hat{\bfk}\cdot\epsilon^{\bfq}\cdot\hat{\bfk} )$ are $\mathcal{O}(1)$ while all other contractions are at least $\mathcal{O}(q/k)$. When wavefunction coefficients are converted into $\langle \gamma^{3}\rangle$ correlators we effectively replace $\bar{\gamma}\to\epsilon$ and we have used this fact to simplify expressions when deriving leading contributions.} 
\be
\lim _{\bfq\to 0} \re \langle T^{{ij}}_{-\bfk-\bfq}T^{{lm}}_{\bfk}T^{{no}}_{\bfq}\rangle_{\rm bulk-mix}'\bar{\gamma}^{-\bfk-\bfq}_{ij}\bar{\gamma}^{\bfk}_{lm}\bar{\gamma}^{\bfq}_{no}= \left (b_{4b}-12H^{2}b_{5}\right )\frac{9k^{3}}{2}\left (\frac{\Lambda_{\rm mix}}{\MP}\right )^{2}\Tr  (\bar{\gamma}^{\bfk}\cdot\bar{\gamma}^{\bfq}\cdot\bar{\gamma}^{-\bfk-\bfq} )\ .\label{BulkMixingNonMixLegSoft}
\ee
where we have used matrix product notation to suppress the indices.

Next, we can take the mixing leg to be soft by sending $\bfk_{1}\to \bfq$, $\bfk_{2}\to \bfk$, and $\bfk_{3}\to -\bfk-\bfq$. In this case, the leading terms are
\begin{align}
\lim _{\bfq\to 0} \re \langle T^{{ij}}_{\bfq}T^{{lm}}_{\bfk}T^{{no}}_{-\bfk-\bfq}&\rangle_{\rm bulk-mix}'\bar{\gamma}^{\bfq}_{ij}\bar{\gamma}^{\bfk}_{lm}\bar{\gamma}^{-\bfk-\bfq}_{no}\nn
 &= \frac{15q^{2}k}{2}\ln(q/k)\left (b_{4b}-12H^{2}b_{5}\right )\left (\frac{\Lambda_{\rm mix}}{\MP}\right )^{2}\Tr(\bar{\gamma}^{\bfk}\cdot\bar{\gamma}^{-\bfk-\bfq})(\hat{\bfk}\cdot\bar{\gamma}^{\bfq}\cdot\hat{\bfk})\nn
 &\quad+3q^{2}k\ln(q/k)\left (5b_{4b}-48H^{2}b_{5}\right )\left (\frac{\Lambda_{\rm mix}}{\MP}\right )^{2}\Tr  (\bar{\gamma}^{\bfk}\cdot\bar{\gamma}^{\bfq}\cdot\bar{\gamma}^{-\bfk-\bfq} )\ .\label{BulkMixingMixLegSoft}
\end{align}

We can then convert these wavefunctional coefficients to 
squeezed bispectra.  The wavefunctional coefficient~\eqref{BulkMixingNonMixLegSoft} gives a contribution\footnote{We use the abbreviated notation
$\langle\gamma_{\bfk_{1}}\gamma_{\bfk_{2}}\gamma_{\bfk_{3}}\rangle'\equiv\langle \gamma^{\bfk_{1}}_{ij}(\ts)\gamma^{\bfk_{2}}_{lm}(\ts)\gamma^{\bfk_{3}}_{no}(\ts)\rangle'\epsilon^{ij}_{\bfk_{1}}\epsilon^{lm}_{\bfk_{2}}\epsilon^{no}_{\bfk_{3}}$.
}
\be
\lim_{\bf q\to 0}\langle \gamma_{\bfq}\gamma_{\bfk}\gamma_{-\bfk-\bfq}\rangle'\supset \frac{6}{q^{3}k^{3}}\left (b_{4b}-12H^{2}b_{5}\right )\left (\frac{H}{\MP}\right )^{6}\left (\frac{\Lambda_{\rm mix}}{\MP}\right )^{2}\Tr  (\epsilon^{\bfk}\cdot\epsilon^{\bfq}\cdot\epsilon^{-\bfk-\bfq} )\ ,\label{BulkMixingNonMixLegSoftNG}
\ee
while \eqref{BulkMixingMixLegSoft} contributes a logarithmic piece in the squeezed limit:
\begin{align}
\langle \gamma_{\bfq}\gamma_{\bfk}\gamma_{-\bfk-\bfq}\rangle'&\supset-\frac{5\ln(q/k)}{qk^{5}}\left (b_{4b}-12H^{2}b_{5}\right )\left (\frac{H}{\MP}\right )^{6}\left (\frac{\Lambda_{\rm mix}}{\MP}\right )^{2}\Tr(\epsilon^{\bfk}\cdot\epsilon^{-\bfk-\bfq})(\hat{\bfk}\cdot\epsilon^{\bfq}\cdot\hat{\bfk})\nn
 &\quad-\frac{2\ln(q/k)}{qk^{5}}\left (5b_{4b}-48H^{2}b_{5}\right )\left (\frac{H}{\MP}\right )^{6}\left (\frac{\Lambda_{\rm mix}}{\MP}\right )^{2} \Tr  (\epsilon^{\bfk}\cdot\epsilon^{\bfq}\cdot\epsilon^{-\bfk-\bfq} )\,.\label{BulkMixingMixLegSoftNG}
\end{align}
The result \eqref{BulkMixingMixLegSoftNG} scales with a higher power of $\bfq$ than both \eqref{BulkMixingNonMixLegSoftNG} and subleading contributions to \eqref{BulkMixingNonMixLegSoftNG} which have been left unwritten.  We have separated out the sub-leading terms \eqref{BulkMixingMixLegSoftNG} in order to highlight the logarithmic scaling in the soft limit, which is reminiscent of findings in ``cosmological collider" studies \cite{Arkani-Hamed:2015bza,Lee:2016vti}.  In those cases, however, the logarithm appears as an oscillatory factor $\sim\cos\ln(q/k)$.

\subsubsection{Other soft limits}

In this section, we compute the graviton bispectrum soft-limit induced by the $\langle T^{3}\rangle$ and $\langle T\Sigma^{2}\rangle$ coefficients, which were the unique terms that sourced the time-independent component of $\langle \gamma^{3}\rangle$.

\paragraph{$\langle T^{3}\rangle$  Soft limits:} There is a single soft limit that can be taken:
\begin{align}
\lim _{\bfq\to 0} \re \langle T^{{ij}}_{\bfq}T^{{lm}}_{\bfk}T^{{no}}_{-\bfk-\bfq}\rangle'\bar{\gamma}^{\bfq}_{ij}&\bar{\gamma}^{\bfk}_{lm}\bar{\gamma}^{-\bfk-\bfq}_{no}\nn
&= -\frac{3k^{3}}{2}\left (\frac{a_{3}}{H^{2}}-5a_{4}-12H^{2}a_{5}\right )\Tr(\bar{\gamma}^{\bfk}\cdot\bar{\gamma}^{-\bfk-\bfq})(\hat{\bfk}\cdot\bar{\gamma}^{\bfq}\cdot\hat{\bfk})\nn
&\quad+2k^{3}\left (a_{4}+42H^{2}a_{5}\right )\Tr  (\bar{\gamma}^{\bfk}\cdot\bar{\gamma}^{\bfq}\cdot\bar{\gamma}^{-\bfk-\bfq} )\ .
\end{align}
The soft limit from the boundary term shape \eqref{Massless3BoundaryTerm} is:
\be
\lim _{\bfq\to 0} \re \langle T^{{ij}}_{\bfq}T^{{lm}}_{\bfk}T^{{no}}_{-\bfk-\bfq}\rangle'_{\rm bdy}\bar{\gamma}^{\bfk_{1}}_{ij}\bar{\gamma}^{\bfk_{2}}_{lm}\bar{\gamma}^{\bfk_{3}}_{no}
=\lambda\frac{12k^{3}}{H^{2}}\Tr  (\bar{\gamma}^{\bfk}\cdot\bar{\gamma}^{\bfq}\cdot\bar{\gamma}^{-\bfk-\bfq} )\ .
\ee
The above terms give the following contribution to the squeezed bispectrum:
\begin{align}
\lim _{\bfq\to0}\langle \gamma_{\bfq}\gamma_{\bfk}\gamma_{-\bfk-\bfq}\rangle'&\supset\frac{24}{q^{3}k^{3}}\left (\frac{H}{\MP}\right )^{6}\left (\frac{a_{3}}{H^{2}}-5a_{4}-12H^{2}a_{5}\right )\Tr(\epsilon^{\bfk}\cdot\epsilon^{-\bfk-\bfq})(\hat{\bfk}\cdot\epsilon^{\bfq}\cdot\hat{\bfk})\nn
&\quad -\frac{32}{q^{3}k^{3}}\left (\frac{H}{\MP}\right )^{6}\left (\frac{6\lambda}{H^{2}}+a_{4}+42H^{2}a_{5}\right )\Tr  (\epsilon^{\bfk}\cdot\epsilon^{\bfq}\cdot\epsilon^{-\bfk-\bfq} )\ .\label{TTTNG}
\end{align}
 The structure on the first line is what was found in \cite{Maldacena:2002vr} and the structure in the second line is produced by the ambiguous boundary term operator \eqref{Massless3BoundaryTerm}, as shown.  As discussed in Sec.~\ref{sec:BispectrumFromT3}, the Einstein--Hilbert shape corresponds to $a_{3}=\MP^{2}/4$ and $a_{4}=a_{5}=\lambda=0$, in which case we find the standard result
\be
\lim _{\bfq\to0}\langle \gamma_{\bfq}\gamma_{\bfk}\gamma_{-\bfk-\bfq}\rangle'_{\rm EH}\approx 6\left (\frac{H}{\MP}\right )^{4}\frac{1}{q^3k^3}\Tr(\epsilon^{\bfk}\cdot\epsilon^{-\bfk-\bfq})(\hat{\bfk}\cdot\epsilon^{\bfq}\cdot\hat{\bfk}) \ ,\label{EHSoftLimit}
\ee
which agrees with~\cite{Maldacena:2002vr}, after accounting for notational differences.\footnote{Namely, factors of two arise when converting from our $\gamma_{ij}$ to the $\gamma^{s}$ of \cite{Maldacena:2002vr} due to the relations $\gamma_{ij}=\sum_{s}\epsilon_{ij}^{s}\gamma^{s}$ and $\gamma^{s}=\frac{1}{2}\epsilon^{ij,s}\gamma_{ij}$ which follows from the standard polarization tensor conventions given in Appendix \ref{app:PolarizationTensors}.}

\paragraph{$\langle T\Sigma^{2}\rangle$ Soft limits:} There are two distinct soft limits which can be taken:
\begin{itemize}
\item First, we can take the $T$ leg to be soft:
\begin{align}
\lim _{\bfq\to 0} \re \langle T_{\bfq}^{ij}\Sigma_{\bfk}^{lm}&\Sigma_{-\bfk-\bfq}^{no}\rangle'\bar{\gamma}^{\bfq}_{ij}\bar{\sigma}^{\bfk}_{lm}\bar{\sigma}^{-\bfk-\bfq}_{no}\nn
&= \frac{k}{4\ts^{2}}\left (\frac{2\left (b_{3a}-b_{3b}\right )}{H^{2}}+4b_{4a}+5b_{4b}-24H^{2}b_{5}\right )\Tr(\bar{\sigma}^{\bfk}\cdot\bar{\sigma}^{-\bfk-\bfq})(\hat{\bfk}\cdot\bar{\gamma}^{\bfq}\cdot\hat{\bfk})\nn
&\quad +\frac{2k}{\ts^{2}}\left (b_{4a}+b_{4b}+2H^{2}b_{5}\right ) \Tr  (\bar{\sigma}^{\bfk}\cdot\bar{\gamma}^{\bfq}\cdot\bar{\sigma}^{-\bfk-\bfq} )\ .
\end{align}
\item Second, we can take one of the $\Sigma$ legs to be soft:
\begin{align}
\lim _{\bfq\to 0} \re \langle T_{-\bfk-\bfq}^{ij}\Sigma_{\bfk}^{lm}&\Sigma_{\bfq}^{no}\rangle'\bar{\gamma}^{-\bfk-\bfq}_{ij}\bar{\sigma}^{\bfk}_{lm}\bar{\sigma}^{\bfq}_{no}\nn
&=  \frac{3k}{4\ts^{2}}\left (-\frac{b_{3a}}{H^{2}}+b_{4a}+b_{4b}+20H^{2}b_{5}\right )\Tr(\bar{\sigma}^{\bfk}\cdot\bar{\gamma}^{-\bfk-\bfq})(\hat{\bfk}\cdot\bar{\sigma}^{\bfq}\cdot\hat{\bfk})\nn
&\quad +\frac{k}{4\ts^{2}}\left (\frac{2\left (-2b_{3a}+b_{3b}\right )}{H^{2}}+2b_{4a}+b_{4b}+72H^{2}b_{5}\right )\Tr  (\bar{\sigma}^{\bfk}\cdot\bar{\sigma}^{\bfq}\cdot\bar{\gamma}^{-\bfk-\bfq} )\ .
\end{align}
\end{itemize}

The same analysis can be performed on the boundary term \eqref{PM2MasslessBoundaryTerm}:
\begin{itemize}
\item First, we can take the $T$ leg to be soft:
\be
\lim _{\bfq\to 0} \re \langle T_{\bfq}^{ij}\Sigma_{\bfk}^{lm}\Sigma_{-\bfk-\bfq}^{no}\rangle_{\rm bdy}'\bar{\gamma}^{\bfq}_{ij}\bar{\sigma}^{\bfk}_{lm}\bar{\sigma}^{-\bfk-\bfq}_{no}
= \lambda\frac{4k}{\ts^{2}}\Tr  (\bar{\sigma}^{\bfk}\cdot\bar{\gamma}^{\bfq}\cdot\bar{\sigma}^{-\bfk-\bfq} )\ .
\ee
\item Second, we can take one of the $\Sigma$ legs to be soft:
\be
\lim _{\bfq\to 0} \re \langle T_{-\bfk-\bfq}^{ij}\Sigma_{\bfk}^{lm}\Sigma_{\bfq}^{no}\rangle_{\rm bdy}'\bar{\gamma}^{-\bfk-\bfq}_{ij}\bar{\sigma}^{\bfk}_{lm}\bar{\sigma}^{\bfq}_{no}
=\lambda\frac{2k}{\ts^{2}}\Tr  (\bar{\sigma}^{\bfk}\cdot\bar{\sigma}^{\bfq}\cdot\bar{\gamma}^{-\bfk-\bfq} )\ .
\ee
\end{itemize}
The above terms give the following contribution to the squeezed bispectrum:
\begin{align}
&\lim _{\bfq\to0}\langle \gamma_{\bfq}\gamma_{\bfk}\gamma_{-\bfk-\bfq}\rangle'\nn
&~~~~~~\supset-\frac{64}{q^{3}k^{3}}\left (\frac{H}{\MP}\right )^{6}\left (\frac{\Lambda_{\rm mix}}{\MP}\right )^{5}\left (\frac{2\left (b_{3a}-b_{3b}\right )}{H^{2}}+4b_{4a}+5b_{4b}-24H^{2}b_{5}\right )\Tr(\epsilon^{\bfk}\cdot\epsilon^{-\bfk-\bfq})(\hat{\bfk}\cdot\epsilon^{\bfq}\cdot\hat{\bfk})\nn
&~~~~~~\quad -\frac{512}{q^{3}k^{3}}\left (\frac{H}{\MP}\right )^{6}\left (\frac{\Lambda_{\rm mix}}{\MP}\right )^{5}\left (\frac{2\lambda}{H^{2}}+b_{4a}+b_{4b}+2H^{2}b_{5}\right )\Tr  (\epsilon^{\bfk}\cdot\epsilon^{\bfq}\cdot\epsilon^{-\bfk-\bfq} )\ .\label{Gamma3FromTSSSoft}
\end{align}
The first line of \eqref{Gamma3FromTSSSoft} is of the same form as the leading GR soft-limit result \eqref{EHSoftLimit}, though with a different coefficient, while the second line is produced by the ambiguous boundary term operator \eqref{PM2MasslessBoundaryTerm}, as shown.  Note that the dominant contributions to $\langle \gamma^{3}\rangle$ arise from contributions where $T$ is soft.   We cannot conclude from \eqref{Gamma3FromTSSSoft} that this violates the usual soft limit result \cite{Bordin:2016ruc}, since \eqref{Gamma3FromTSSSoft} is missing contributions from bulk-mixing diagrams.

\subsection{Comments on the graviton trispectrum \label{sec:Trispectrum}}

\begin{figure}
\begin{center}
\resizebox{16cm}{!}{
\begin{tikzpicture}

\begin{scope}[xshift=0cm]
\node[font=\Large] at (0,.5) {$\langle T^{4}\rangle$};
\draw[gray, line width=1.1pt] (-3,0) -- (3,0) node [right,black] {$\tau=\ts$};
\node at (-1.1,-1.9) [vertex,line width=.8pt,minimum size=.5cm](vertexL) {};
\node at (1.1,-1.9) [vertex,line width=.8pt,minimum size=.5cm](vertexR) {};
\node at (-2,0) [smalldot,minimum size=.05cm](dotLL) {};
\node at (-1.2,0) [smalldot,minimum size=.05cm](dotL) {};
\node at (2,0) [smalldot,minimum size=.05cm](dotRR) {};
\node at (1.2,0) [smalldot,minimum size=.05cm](dotR) {};
\path[]
	(dotLL) edge[bend right=27,graviton, line width=1.2pt] node  {} (vertexL)
	(dotL) edge[bend right=10,graviton, line width=1.2pt] node  {} (vertexL)
	(dotRR) edge[bend left=21,graviton, line width=1.2pt] node  {} (vertexR)
	(dotR) edge[bend left=10,graviton, line width=1.2pt] node  {} (vertexR)
	(vertexL) edge[bend right=10,graviton, line width=1.2pt] node  {} (vertexR);
\end{scope}

\begin{scope}[xshift=8cm]
\node[font=\Large] at (0,.5) {$\langle T^{4}\rangle$};
\draw[gray, line width=1.1pt] (-3,0) -- (3,0) node [right,black] {$\tau=\ts$};
\node at (-1.1,-1.9) [vertex,line width=.8pt,minimum size=.5cm](vertexL) {};
\node at (1.1,-1.9) [vertex,line width=.8pt,minimum size=.5cm](vertexR) {};
\node at (-2,0) [smalldot,minimum size=.05cm](dotLL) {};
\node at (-1.2,0) [smalldot,minimum size=.05cm](dotL) {};
\node at (2,0) [smalldot,minimum size=.05cm](dotRR) {};
\node at (1.2,0) [smalldot,minimum size=.05cm](dotR) {};
\path[]
	(dotLL) edge[bend right=27,graviton, line width=1.2pt] node  {} (vertexL)
	(dotL) edge[bend right=10,graviton, line width=1.2pt] node  {} (vertexL)
	(dotRR) edge[bend left=21,graviton, line width=1.2pt] node  {} (vertexR)
	(dotR) edge[bend left=10,graviton, line width=1.2pt] node  {} (vertexR)
	(vertexL) edge[bend right=10,pmfield, line width=1.2pt] node  {} (vertexR);
\end{scope}

\begin{scope}[xshift=16cm]
\node[font=\Large] at (0,.5) {$\langle T^{4}\rangle$};
\draw[gray, line width=1.1pt] (-3,0) -- (3,0) node [right,black] {$\tau=\ts$};
\node at (0,-1.9) [vertex,line width=.8pt,minimum size=.5cm](vertex) {};
\node at (-2,0) [smalldot,minimum size=.05cm](dotLL) {};
\node at (-1,0) [smalldot,minimum size=.05cm](dotL) {};
\node at (2,0) [smalldot,minimum size=.05cm](dotRR) {};
\node at (1,0) [smalldot,minimum size=.05cm](dotR) {};
\path[]
	(dotLL) edge[bend right=27,graviton, line width=1.2pt] node  {} (vertex)
	(dotL) edge[bend right=10,graviton, line width=1.2pt] node  {} (vertex)
	(dotRR) edge[bend left=21,graviton, line width=1.2pt] node  {} (vertex)
	(dotR) edge[bend left=10,graviton, line width=1.2pt] node  {} (vertex);
\end{scope}

\end{tikzpicture}
}
\end{center}
\caption{Contributions to $\langle T^{4}\rangle$ from gravitons and partially massless fields.}
\label{fig:Graviton4PtCoefficients}
\end{figure}
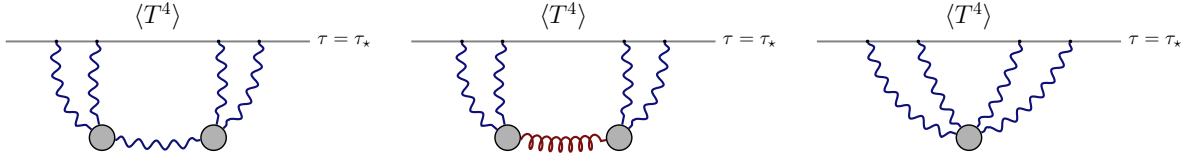

Finally, we make some brief comments on the contribution of the coefficients we have calculated to the graviton trispectrum $\langle \gamma^{4}\rangle$. 
While the graviton trispectrum is expected to be even less observationally relevant than the bispectrum, for our purposes it has the advantage that partially massless fields can imprint themselves on $\langle \gamma^{4}\rangle$ even in pure de Sitter, without a mixing term of the type considered in Sec.~\ref{sec:Mixing}.  A primary finding is that if the time-dependence of the $\langle \Sigma T^{2}\rangle$ coefficients found in Sec.~\ref{sec:WavefunctionCoefficientSigmaT2} is taken seriously, then these provide a contribution to $\langle \gamma^{4}\rangle$ which diverges as $\propto\ts^{-2}$.
   
\begin{figure}
\begin{center}
\resizebox{11cm}{!}{
\begin{tikzpicture}

	\begin{scope}[yshift=0cm,xshift=0cm]

	\node[font=\Large] at (0,0) {$\langle \gamma^{4}\rangle\supset$};

	\node at (1.5,0) [vertex,line width=.8pt,minimum size=.5cm](vertexL) {};
	\node at (3.5,0) [vertex,line width=.8pt,minimum size=.5cm](vertexR) {};

\coordinate (dotUpL) at ([shift=({120:1.5})]vertexL) {};

\coordinate (dotDownL) at ([shift=({240:1.5})]vertexL) {};

\coordinate (dotUpR) at ([shift=({60:1.5})]vertexR) {};

\coordinate (dotDownR) at ([shift=({-60:1.5})]vertexR) {};

\path[]
	(dotUpL) edge[graviton,line width=1.2 pt] node  {} (vertexL)
	(dotDownL) edge[graviton,line width=1.2 pt] node  {} (vertexL)
	(dotUpR) edge[graviton,line width=1.2 pt] node  {} (vertexR)
	(dotDownR) edge[graviton,line width=1.2 pt] node  {} (vertexR)
	(vertexR) edge[graviton,line width=1.2 pt] node  {} (vertexL);

	\begin{scope}[yshift=0cm,xshift=5cm]

	\node[font=\Large] at (0,0) {$+$};

	\node at (1.5,0) [vertex,line width=.8pt,minimum size=.5cm](vertexL) {};
	\node at (3.5,0) [vertex,line width=.8pt,minimum size=.5cm](vertexR) {};

\coordinate (dotUpL) at ([shift=({120:1.5})]vertexL) {};

\coordinate (dotDownL) at ([shift=({240:1.5})]vertexL) {};

\coordinate (dotUpR) at ([shift=({60:1.5})]vertexR) {};

\coordinate (dotDownR) at ([shift=({-60:1.5})]vertexR) {};

\path[]
	(dotUpL) edge[graviton,line width=1.2 pt] node  {} (vertexL)
	(dotDownL) edge[graviton,line width=1.2 pt] node  {} (vertexL)
	(dotUpR) edge[graviton,line width=1.2 pt] node  {} (vertexR)
	(dotDownR) edge[graviton,line width=1.2 pt] node  {} (vertexR)
	(vertexR) edge[pmfield,line width=1.2 pt] node  {} (vertexL);
	
	\end{scope}
	
	\end{scope}

	\begin{scope}[yshift=0cm,xshift=10cm]

	\node[font=\Large] at (0,0) {$+$};

	\node at (2,0) [vertex,line width=.8pt,minimum size=.5cm](vertex) {};

	\coordinate (dotTR) at ([shift=({45:1.5})]vertex) {};
	\coordinate (dotTL) at ([shift=({135:1.5})]vertex) {};
	\coordinate (dotBL) at ([shift=({225:1.5})]vertex) {};
	\coordinate (dotBR) at ([shift=({315:1.5})]vertex) {};

\path[]
	(dotTR) edge[graviton,line width=1.2 pt] node  {} (vertex)
	(dotTL) edge[graviton,line width=1.2 pt] node  {} (vertex)
	(dotBL) edge[graviton,line width=1.2 pt] node  {} (vertex)
	(dotBR) edge[graviton,line width=1.2 pt] node  {} (vertex);
	
	\end{scope}

\end{tikzpicture}
}
\end{center}
\caption{Contributions to the graviton trispectrum $\langle\gamma^{4}\rangle$ from gravitons and partially massless fields.  The three-point vertices come from the  $\langle T^{3}\rangle$ and $\langle \Sigma T^{2}\rangle$ coefficients we have already calculated.  The four-point vertex depends on the $\langle T^{4}\rangle$ coefficient which we have not calculated and whose associated diagrams are shown in Fig.~\ref{fig:Graviton4PtCoefficients}.}
\label{fig:Graviton4PtFunction}
\end{figure}
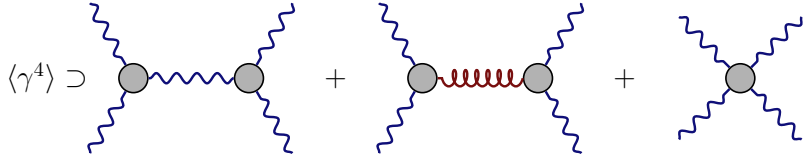

The leading-order contributions are due to interactions between one partially massless spin-2 particle and two gravitons.  This interaction both produces the $\langle \Sigma T^{2}\rangle$ coefficient considered in Sec.~\ref{sec:WavefunctionCoefficientSigmaT2} and contributes to $\langle T^{4}\rangle$ through the middle diagram in Fig.~\ref{fig:Graviton4PtCoefficients}.  These coefficients, along with $\langle T^{3}\rangle$, determine $\langle \gamma^{4}\rangle$ at leading order through the diagrams in Fig.~\ref{fig:Graviton4PtFunction}.

In order to understand the effects of partially massless spin-2 exchange, it is useful to compare and 
contrast the contributions of $\langle T^{3}\rangle$ and $\langle \Sigma T^{2}\rangle$ to $\langle \gamma^{4}\rangle$:
\begin{itemize}
\item The leftmost diagram in Fig.~\ref{fig:Graviton4PtFunction} is schematically given by 
\begin{align}
\langle \gamma^{4}\rangle\sim \frac{1}{\re \langle T^{2}\rangle^{4}} \re\langle T^{3}\rangle \frac{1}{\re\langle T^{2}\rangle} \re\langle T^{3}\rangle 
\end{align}
using the scaling of these wavefunction coefficients, we see that this is a $\ts$-independent expression scaling as $\langle \gamma^{4}\rangle\sim k^{-9}$.
\item The middle diagram in Fig.~\ref{fig:Graviton4PtFunction} is schematically given by 
\begin{align}
\langle \gamma^{4}\rangle\sim \frac{1}{\re \langle T^{2}\rangle^{4}} \re\langle \Sigma T^{2}\rangle \frac{1}{\re\langle \Sigma^{2}\rangle} \re\langle \Sigma T^{2}\rangle \label{TrispectrumSketch}
\end{align}
and from \eqref{PartiallyMasslessSpin2TwoPointd3} and \eqref{SigmaT2WavefunctionCoefficient}, both $\re\langle \Sigma^{2}\rangle$ and $\re\langle \Sigma T^{2}\rangle$ scale as $\propto k/\ts^{2}$.  Hence, the above will produce a result $\langle \gamma^{4}\rangle\sim k^{-11}\ts^{-2} $ which diverges strongly as $\ts\to 0$.
\end{itemize}

The calculation in \eqref{TrispectrumSketch} is of course not complete, as the contribution to $\langle T^4\rangle$ from the middle diagram in Fig.~\ref{fig:Graviton4PtCoefficients} has not yet been accounted for.  It would be interesting to perform this full calculation and check whether or not the $\sim \ts^{-2}$ time dependence in $\langle \gamma^{4}\rangle$ persists.

\section{Discussion\label{sec:Discussion}}
We have studied the imprints of massive spin-2 fields on inflationary tensor non-Gaussianities.
As compared to the single massless graviton present in typical inflationary models, new shapes can appear due to the presence of new possible on-shell cubic vertices involving the massive spin-2 fields. This can also be understood from the dual perspective: massive spin-2 fields correspond to non-conserved spin-2 currents as conformal representations, for which there are more possible 3-point structures. This provides evidence for the presence of new particles during inflation: measurement of a shape not allowed for a massless graviton would point toward other spinning degrees of freedom.

We have derived the general bulk-to-boundary propagators and on-shell cubic vertices required to compute arbitrary tensor non-Gaussianities involving spin-2 fields of any mass. In order to obtain closed-form results, we have focused on the case where the additional spin-2 field is partially massless.  These interactions also provide a mechanism for increasing the size of tensor non-Gaussianity while only negligibly affecting the tensor power spectrum. 

The tensor bi-spectrum is expected to be very small, and even the tensor spectrum itself has not been detected yet, so any observation is necessarily futuristic.  However, it is possible that information about the tensor bi-spectrum may be gained from the future LISA gravitational wave experiment \cite{Bartolo:2018qqn}.
Measurement of a non-Einstein--Hilbert 3-point correlation function for the graviton would be exciting for a number of reasons. Observation of one of the shapes we enumerate in this paper (or their general mass versions) would provide evidence for the presence of an additional heavy field during inflation, but it would also strongly suggest that there should be additional heavy states to discover. In~\cite{Camanho:2014apa}, it was shown that large higher-derivative cubic interactions for a massless graviton in flat space lead to asymptotic causality problems. These problems can be cured by introducing a tower of higher-spin states---as in string theory. If we assume that something similar happens on de Sitter space (though this is not proven), we should expect that large higher-derivative non-Gaussianity for the graviton would also imply the presence of a tower of higher-spin states~\cite{Maldacena:2011nz,Camanho:2014apa}. The shapes that we are interested in here are not pure gravity shapes, but rather involve other spin-2 fields: however, it was shown in~\cite{Hinterbichler:2017qyt,Bonifacio:2017nnt}---again in flat space---that theories involving massive spin-2 fields with large cubic couplings have the same superluminality issues, unless their couplings are of the Einstein--Hilbert form. Therefore, a natural expectation is that non-Einstein--Hilbert non-Gaussian shapes in general require a tower of higher-spin fields to be consistent.

It would be very interesting to understand more fully to what extent the signatures we have computed here are degenerate with either modifications of the initial vacuum or to slow-roll suppressed corrections to graviton correlation functions. Our expectation is that at least the non-analytic exchange contribution cannot be mimicked by any local interaction or non-singular vacuum state, but it would be interesting to understand these issues systematically.  We leave such a study for the future.

In this paper we have focused on the ``bulk" version of the wavefunctional calculation, by directly computing the on-shell action in de Sitter space. However, our final results are controlled by the isometries of de Sitter space, which act as conformal transformations at late times. Conformal symmetry is known to fix the 3-point correlation function of any operators, up to a finite number of constants. Therefore, it should be possible to re-derive our results from the ``boundary" perspective using conformal symmetry. Though this seems to be a formidable task, given that the constraints of conformal symmetry for spinning operators in momentum space are quite complex, the final results we obtain are rather simple: they are rational functions of momenta, so it may be more tractable than it naively seems. 
It would be very interesting to carry out such an analysis, as it may provide a systematic way to compute to higher-order correlation functions involving massive and massless spinning operators in cosmology and in AdS/CFT applications.

\paragraph{Acknowledgements:} We thank James Bonifacio, Emanuela Dimastrogiovanni, Valerie Domcke, Lam Hui, Juan Maldacena, Alberto Nicolis, Guilherme Pimentel, Rachel Rosen, John Stout, Zimo Sun, and Sam Wong for helpful discussions. The Mathematica package {\tt xAct}~\cite{xAct} was used extensively in the course of this work. GG is partly supported by the research programme VIDI with Project No. 680-47-535, which is financed by Netherlands organization for scientific research (NWO).  KH acknowledges support from DOE grant DE-SC0019143. AJ is supported in part by NASA grant NNX16AB27G. The work of MT is supported in part by US Department of Energy (HEP) Award DE-SC0013528, and by NASA ATP grant NNH17ZDA001N.

\appendix

\section{The wavefunction\label{app:TheWavefunction}}
One of the central objects of our interest is the late-time cosmological wavefunction. For the convenience of the reader,
in this Appendix we briefly review the wavefunctional approach to cosmological correctors. This is a straightforward extension of flat space Schr\"odinger field theory. For some other applications, see~\cite{Anninos:2011kh,Anninos:2014lwa,Arkani-Hamed:2017fdk,Benincasa:2018ssx}

\subsection{Introduction to $\Psi$}

Consider a theory of a scalar field $\varphi$, for concreteness, whose action is of the form
\begin{align}
S_{2}&=\int\rd^{d+1}x\, \sqrt{-g}\,\left (-\frac{1}{2}(\nabla\varphi)^{2}-\frac{m^{2}}{2}+{\cal L}_{\rm int}\right )\,,
\end{align}
where ${\cal L}_{\rm int}$ stands for any interaction terms we may be interested in.
The generalization to other types of fields is straightforward.

The wavefunction of the universe is simply the amplitude for finding $\varphi(x^{\mu})$ in a particular spatial configuration $\bar{\varphi}(\bfx)$ at time $\ts$: $\Psi[\bar{\varphi},\ts]=\langle \bar{\varphi}|\Psi(\ts)\rangle$. This is the field-basis representation of a state in the Hilbert space. Given $\Psi[\bar{\varphi},\ts]$, arbitrary equal time expectation values can be calculated via the usual quantum mechanics formula
\begin{align}
\langle \varphi(\ts,\bfx_{1})\ldots \varphi(\ts,\bfx_{n})\rangle&=\int\mathcal{D}\bar{\varphi}\, |\Psi[\bar{\varphi},\ts]|^{2}\, \bar{\varphi}(\bfx_{1})\ldots \bar{\varphi}(\bfx_{n})\ .
\end{align}
There are two ways in which $\Psi[\bar{\varphi},\ts]$ may be practically calculated.  First, from $S$ one can build the hamiltonian $\mathcal{H}[\varphi,\Pi_{\varphi}]$, where $\Pi_{\varphi}$ is the momentum conjugate to $\varphi$, and solve the functional Schr\"odinger equation
\begin{align}
i\partial_{\tau}\Psi[\bar{\varphi},\tau]&=\mathcal{H}\left [\varphi,-i\frac{\delta}{\delta\varphi}\right ]\Psi[\bar{\varphi},\tau]\ . \label{FunctionalSchrodingerEquation}
\end{align}
This quickly becomes infeasible, so a more tractable approach is to use the path integral
\be
 \Psi[\bar{\varphi},\ts]=\int^{\varphi(\ts)=\bar{\varphi}}\mathcal{D}\varphi\, e^{iS}\ ,
\ee
where we have suppressed the dependence on the initial state; typically we project onto the vacuum state of the theory by employing an $i\epsilon$ prescription as we send $\tau\to -\infty$.

 \subsection{The path integral calculation of $\Psi$\label{app:PathIntegralCalculationOfPsi}}
In order to compute the late-time wavefunction in de Sitter space, we utilize the path integral method
 and restrict all of our calculations to the tree-level, classical approximation in which we can approximate the path integral by its saddle point so that
 \begin{align}
 \Psi[\bar{\varphi},\ts]\approx \exp\left (iS_{\rm cl}[\varphi_{\rm cl}]\right )\,,
 \end{align}
 where $\varphi_{\rm cl}$ is a solution to the classical equations of motion and we have 
omitted the overall normalization factor (it is immaterial).  The particular classical solution, $\varphi_{\rm cl}(\tau,\bfx)$, that we are interested in is the one satisfying the boundary conditions
 \begin{align}
 \varphi_{\rm cl}(\tau,\bfx)\xrightarrow{\tau\to \ts}\bar{\varphi}(x)\, , \qquad\qquad \varphi_{\rm cl}(\tau,\bfx)\xrightarrow{\tau\to-\infty}\varphi_{\rm vac}\, .
 \end{align}
 The choice of $\varphi_{\rm vac}$ will be discussed momentarily.
 
 The on-shell action, $S_{\rm cl}[\varphi_{\rm cl}]$, is constructed perturbatively as a series in $\bar{\varphi}(\bfx)$.  To this end, it is useful to construct the bulk-to-boundary propagator ${\cal K}(\tau,\bfx;\bfy)$ and bulk-to-bulk propagator ${\cal G}(\tau,\bfx;\tau',\bfy)$ (borrowing terminology from AdS/CFT) which obey:
 \begin{itemize}
 \item $\mathcal{E}_{\tau,\bfx}{\cal K}(\tau,\bfx;\bfy)=0$ and $\lim _{\tau\to \ts}{\cal K}(\tau,\bfx;\bfy)=\delta^{d}(\bfx-\bfy)$.
 \item $\mathcal{E}_{\tau,\bfx}{\cal G}(\tau,\bfx;\tau',\bfy)=-\delta(\tau-\tau')\delta^{d}(\bfx-\bfy)$ and $\lim _{\tau\to \ts}{\cal G}(\tau,\bfx;\tau',\bfy)=0$,
 \end{itemize}
 where $\mathcal{E}_{\tau,\bfx}$ is the differential operator appearing in the free EOM:
\be
 \frac{\delta S_{2}}{\delta\varphi(\tau,\bfx)}\equiv \mathcal{E}_{\tau,\bfx}\varphi(\tau,\bfx)\ .
\ee
 The Bunch--Davies vacuum condition corresponds to choosing ${\cal K}$ to only oscillate as $\sim e^{+i\tau}$ at early times.  Similarly, ${\cal G}$ must behave as $\sim e^{+i\tau}$ for $\tau< \tau'$  and $\sim e^{+i\tau'}$ for $\tau'<\tau$.
 
 Using these ingredients, an implicit expression for the classical solution is given by
\be
 \varphi_{\rm cl}(\tau,\bfx)=\int\rd^{d}y\, {\cal K}(\tau,\bfx;\bfy) \bar{\varphi}(\bfy)+\int\rd\tau'\rd^{d}y\,{\cal G}(\tau,\bfx;\tau',\bfy)\frac{\delta S_{\rm int}}{\delta\varphi(\tau',\bfy)}\ .\label{PhiClassicalSoln}
\ee
 This can be solved iteratively to find $\varphi_{\rm cl}$ to any desired order in $\bar{\varphi}$.
 
 The on-shell action $S_{\rm cl}[\bar{\varphi}_{\rm cl}]$ is efficiently calculated by first writing the quadratic action as a boundary contribution plus the equation of motion:
\be
 S=\frac{1}{2}\int_{\tau=\ts}\rd^{d}x\, \sqrt{\bar{g}}n^{\mu}\varphi\nabla_{\mu}\varphi+\int\rd^{d+1}x\,  \frac{1}{2}\varphi\mathcal{E}\varphi+S_{\rm int}\ ,
\ee
 where $n^{\mu}$ is the outward pointing normal vector to the $\tau=\ts$ surface\footnote{Fall-off conditions are chosen to ensure that $\varphi$ vanishes on all other (spatial) boundaries.} and $\bar{g}_{ij}$ is the induced metric on this hypersurface.  Plugging in the expression \eqref{PhiClassicalSoln} for $\varphi_{\rm cl}$ into the above and doing some algebra, the end result is\footnote{Deriving this requires the relation
 \be
  \lim _{\tau'\to \ts}\sqrt{\bar{g}}\, n^{\mu}\nabla_{\mu}^{(\tau',\bfy)}{\cal G}(\tau,\bfx;\tau',\bfy)={\cal K}(\tau,\bfx;\bfy)\,,
 \ee
 which follows from the integral identity $\int_{\mathcal{M}}\, {\cal K}\square {\cal G}-{\cal G}\square {\cal K}=\int_{\partial\mathcal{M}}\, n^{2}\left ({\cal K}n\cdot\nabla {\cal G}-{\cal G}n\cdot\nabla {\cal K}\right )$ where $n^{\mu}$ the outward pointing normal vector.  This property is responsible for important cancellations in the derivation of \eqref{OnShellAction}.
 }
 \begin{align}
 S_{\rm cl}[\varphi_{\rm cl}]&=\int_{\tau=\ts}\rd^{d}x\,\rd^{d}y\, \sqrt{\bar{g}}\,\frac{1}{2}\bar{\varphi}(\bfx) n^{\mu}\nabla_{\mu}^{(\tau,\bfx)}\left ({\cal K}(\tau,\bfx;\bfy)\right )\bar{\varphi}(\bfy)+\int\rd^{d+1}x\, \sqrt{-g}\mathcal{L}_{\rm int}[\varphi_{\rm cl}]\nn
 &\quad-\frac{1}{2}\int\rd^{d+1}x\rd^{d+1}y\, \frac{\delta S_{\rm int}}{\delta\varphi(\tau,\bfx)}{\cal G}(\tau,\bfx;\tau',\bfy)\frac{\delta S_{\rm int}}{\delta\varphi(\tau',\bfy)}\Big|_{\varphi=\varphi_{\rm cl}}\ .\label{OnShellAction}
 \end{align}
Here (and throughout this paper), all temporal integrals  range from $-\infty$ to $\ts$, while spatial integrals range from $-\infty$ to $\infty$.
 In the present paper, we will primarily focus on three-point correlation functions, in which case the final line of \eqref{OnShellAction} does not contribute.

 \subsection{Correlators from $\Psi$}
 Once we have the wavefunction, there is yet another step required to extract from it correlation functions.
 The wavefunction, $\Psi[\bar{\varphi},\ts]$, can be written as an expansion in $\bar{\varphi}$.  We are interested in momentum space correlators, in which case we write
 \begin{align}
 \Psi[\bar{\varphi}_{\bfk},\ts]=\exp\Bigg(&-\frac{1}{2}\int\rd^{d}\tilde{k}_{1}\rd^{d}\tilde{k}_{2}\,\langle\mathcal{O}_{\bfk_{1}}\mathcal{O}_{\bfk_{2}}\rangle\bar{\varphi}_{\bfk_{1}}\bar{\varphi}_{\bfk_{2}}\nn
&~~~~-\frac{1}{3!}\int\rd^{d}\tilde{k}_{1}\rd^{d}\tilde{k}_{2}\rd^{d}\tilde{k}_{3}\,\langle\mathcal{O}_{\bfk_{1}}\mathcal{O}_{\bfk_{2}}\mathcal{O}_{\bfk_{3}}\rangle\bar{\varphi}_{\bfk_{1}}\bar{\varphi}_{\bfk_{2}}\bar{\varphi}_{\bfk_{3}}+\cdots\Bigg)\ ,\label{GenericWavefunctionNotation}
 \end{align}
 where the $\langle \mathcal{O}_{\bfk_{1}}\ldots\mathcal{O}_{\bfk_{n}}\rangle$'s are simply functions of the indicated arguments. These coefficients contain the information that specifies correlation functions of the fields $\varphi$.  
 Due to homogeneity and isotropy of $\tau= {\rm const.}$ slices, these wavefunction coefficients are proportional to delta functions which enforce $d$-momentum conservation.  We use a prime to indicate that a delta function and the associated $2\pi$ factors have been dropped from a quantity, {\it i.e.},
\be
\langle \mathcal{O}_{\bfk_{1}}\ldots\mathcal{O}_{\bfk_{n}}\rangle\equiv \tilde{\delta}^{d}\left ({\textstyle \sum_{i=1}^{n}\bfk_{i}}\right )\langle \mathcal{O}_{\bfk_{1}}\ldots\mathcal{O}_{\bfk_{n}}\rangle'\ .
\ee

Expectation values are related to wavefunction coefficients via standard perturbative Gaussian integral formulas, such as
\begin{align}
\langle \varphi_{\bfk_{1}}(\ts)\varphi_{\bfk_{2}}(\ts)\rangle'&=\frac{1}{2 \re \langle\mathcal{O}_{\bfk_{1}}\mathcal{O}_{-\bfk_{1}}\rangle'}\nn
\langle \varphi_{\bfk_{1}}(\ts)\varphi_{\bfk_{2}}(\ts)\varphi_{\bfk_{3}}(\ts)\rangle'&=-\frac{1}{4}\prod_{i=1}^{3}\left (\frac{1}{\re \langle\mathcal{O}_{\bfk_{i}}\mathcal{O}_{-\bfk_{i}}\rangle'}\right ) \left (\re \langle\mathcal{O}_{\bfk_{1}}\mathcal{O}_{\bfk_{2}}\mathcal{O}_{\bfk_{3}}\rangle'\right )\nn
\langle \varphi_{\bfk_{1}}(\ts)\varphi_{\bfk_{2}}(\ts)\varphi_{\bfk_{3}}(\ts)\varphi_{\bfk_{4}}(\ts)\rangle'&=-\frac{1}{8}\prod_{i=1}^{4}\left (\frac{1}{\re \langle\mathcal{O}_{\bfk_{i}}\mathcal{O}_{-\bfk_{i}}\rangle'}\right ) \Big[\re \langle\mathcal{O}_{\bfk_{1}}\mathcal{O}_{\bfk_{2}}\mathcal{O}_{\bfk_{3}}\mathcal{O}_{\bfk_{4}}\rangle'\nn
&\quad -\frac{\re\langle\mathcal{O}_{\bfk_{1}}\mathcal{O}_{\bfk_{2}}\mathcal{O}_{-\bfk_{12}}\rangle'\re\langle\mathcal{O}_{\bfk_{12}}\mathcal{O}_{\bfk_{3}}\mathcal{O}_{\bfk_{4}}\rangle'}{\re \langle\mathcal{O}_{\bfk_{12}}\mathcal{O}_{\bfk_{12}}\rangle'}+{\rm 2 \ permutations}\Big].\label{CorrelatorExpressions}
\end{align}
where $\bfk_{ij}\equiv\bfk_{i}+\bfk_{j}$.  When generalized to tensor fields, inverse wavefunction factors become matrix inverses.  Note that the real components of the $\langle \mathcal{O}^{n}\rangle$ wavefunction coefficients arise from the imaginary component of the action because $\Psi \sim e^{iS}$. Though the action is defined in terms of real fields, it  develops imaginary terms because it is being evaluated on complex field configurations. This is essentially due to the early time vacuum condition, which is an imaginary constraint.

\section{Projectors\label{app:Projectors}}

In several places it is useful to be able to project spatial tensors onto their irreducible components. We therefore introduce a basis of projection tensors which accomplishes this task. It is most straightforwardly built out of the transverse tensor
\be
\pi_{ij}^\bfk = \delta_{ij}-\frac{k_ik_j}{k^2}.
\ee
Using this, we form the tensors (dropping the $\bfk$ label on $\pi_{ij}$ to simplify notation)
\begin{subequations}
\begin{align}
(\Pi^\bfk_{TT})^{ij}{}_{lm} &= \pi^{(i}_{(l} \pi^{j)}_{k)}-\frac{1}{d-1}\pi^{ij}\pi_{lm}\\
(\Pi^\bfk_{V})^{ij}{}_{lm} &= \frac{2}{k^2}k^{(i}k_{(l}\pi^{j)}_{m)}\\
(\Pi^\bfk_{S})^{ij}{}_{lm} &= \frac{1}{d}\delta_{ij}\delta^{lm}\\
(\Pi^\bfk_Q)^{ij}{}_{lm} &= \frac{d}{(d-1)}\left (\frac{k_{i}k_{j}}{k^{2}}-\frac{1}{d}\delta_{ij}\right )\left (\frac{k^{l}k^{m}}{k^{2}}-\frac{1}{d}\delta^{lm}\right )\ .
\end{align}
\label{Projectors}
\end{subequations}
These tensors obey the usual orthonormality and completeness relations:
\begin{align}
\nonumber
&\Pi^\bfk_{TT}+\Pi^\bfk_{V}+\Pi^\bfk_{{S}}+\Pi^\bfk_{Q} = \mathds{1} ,\\
&(\Pi^\bfk_{TT})^{2}=\Pi^\bfk_{TT} \ ,\qquad (\Pi^\bfk_{V})^{2}=\Pi^\bfk_{V} \ ,\qquad (\Pi^\bfk_{{S}})^{2}=\Pi^\bfk_{{S}} \ ,\qquad (\Pi^\bfk_{Q})^{2}=\Pi^\bfk_{Q} \nn
&\Pi^\bfk_{TT}\cdot \Pi^\bfk_{V}=\Pi^\bfk_{TT}\cdot \Pi^\bfk_{{S}}=\Pi^\bfk_{TT}\cdot \Pi^\bfk_{Q}=\Pi^\bfk_{V}\cdot \Pi^\bfk_{{S}}=\Pi^\bfk_{V}\cdot \Pi^\bfk_{Q}=\Pi^\bfk_{{S}}\cdot \Pi^\bfk_{Q}=0\, , \label{ProjectorsPropertiesBasic}
\end{align}
in condensed notation.\footnote{For instance, the identity matrix $\mathds{1}$ is explicitly given by $(\mathds{1})^{ij}{}_{lm}=\delta^{(i}_{(l}\delta^{j)}_{m)}$ and $(\Pi^\bfk_{TT}\cdot\Pi^\bfk_{TT})^{ij}{}_{lm}=(\Pi^\bfk_{TT})^{ij}{}_{rs}(\Pi^\bfk_{TT})^{rs}{}_{lm}$.}

 \section{On-shell, gauge invariant cubic interactions\label{app:OnShellCubicTerms}}
 In this Appendix, we describe the derivation of the on-shell cubic vertices involving massless or partially massless fields, which require the imposition of on-shell gauge invariance. A similar problem was treated in~\cite{Joung:2011ww,Joung:2012rv} making use of embedding space techniques, but for our purposes it will be more convenient to derive expressions directly in the physical $(d+1)$-dimensional de Sitter space.

The bases of operators introduced in Sec.~\ref{sec:CubicVertices} (see \eqref{ThreehOperatorBasis}, \eqref{TwohOperatorBasis} and \eqref{OnehOperatorBasis}) are written for spin-2 fields with generic masses.  When a $h_{(i)\mu\nu}$ field is massless or partially massless, gauge invariance places further constraints on the operator combinations.  In this Appendix, we discuss how gauge invariance is enforced in the on-shell action. We produce the counting of operators found in \cite{Joung:2012rv} for cases where $d$ is arbitrary and find additional invariant operators for special choices of $d$, namely $d=3$.

\subsection{Off-shell vs. on-shell gauge invariance}

Given a theory involving a spin-2 field, $h_{\mu\nu}$, which enjoys a gauge symmetry, we can expand both the action and gauge transformation in powers of $h$:
\begin{align}
S[h]&=S_{2}[h]+S_{3}[h]+\ldots\nn
h_{\mu\nu}&\mapsto h'_{\mu\nu}=h_{\mu\nu}+\Delta_{\xi}^{(0)} h_{\mu\nu}+\Delta_{\xi}^{(1)} h_{\mu\nu}+\ldots
\end{align}
where $S_{n}[h]$ and $\Delta_{\xi}^{(n)}h_{\mu\nu}$ are $\mathcal{O}(h^{n})$ and $\xi$ represents the dependence on the gauge parameter.  The \textit{off-shell} statement of gauge invariance is that $S[h]=S[h']$ which can be written order-by-order in the fields as
\begin{align}
\delta^{(0)}_{\xi}S_{2}[h]&=0\nn
\delta^{(0)}_{\xi}S_{3}[h]+\delta^{(1)}_{\xi}S_{2}[h]&=0\nn
\delta^{(0)}_{\xi}S_{4}[h]+\delta^{(1)}_{\xi}S_{3}[h]+\delta^{(2)}_{\xi}S_{2}[h]&=0\nn
&~~\vdots\label{GaugeInvarianceConstraintOffShell}
\end{align}
where we have defined
\be
\delta^{(n)}_{\xi}\equiv \int\rd^{d+1}x\, \Delta^{(n)}_{\xi}h_{\mu\nu}(x)\frac{\delta}{\delta h_{\mu\nu}(x)}\ .
\ee
All of the conditions \eqref{GaugeInvarianceConstraintOffShell} mean equality up to total derivatives in the action.

However, to compute 3-point correlation functions, we only
require the \textit{on-shell} action. In this case, the constraints simplify.  In particular, when the fields are put on-shell, the second condition in \eqref{GaugeInvarianceConstraintOffShell} becomes
\begin{align}
\delta^{(0)}_{\xi}S_{3}[h]&\cong 0\ ,\label{GaugeInvarianceConstraintOnShell}
\end{align}
where $\cong$ indicates equality up to terms which vanish on-shell. This happens because $\delta S_2/\delta h$ is the linear equation of motion, which is zero on-shell to this order in the fields.
  The on-shell gauge invariance constraints are in some ways simpler than the off-shell ones since the knowledge of the first-order gauge transformation $\delta^{(1)}_{\xi}$ is not required.  However, the requirement of keeping all fields on-shell also brings in complications which are discussed in the following section using a concrete example.

\subsubsection{Example: the Einstein--Hilbert interaction in $d=3$\label{app:EHOnShellCubicTerms}}

There are two main subtleties in solving \eqref{GaugeInvarianceConstraintOnShell}: the form of the gauge parameter is restricted, since it must preserve $\nabla^{\mu}h_{\mu\nu}=h=0$,\footnote{We include the gauge conditions $h = \nabla^\mu h_{\mu\nu} = 0$ as on-shell conditions.}  and there exist terms which are total derivatives only when the EOM are used (and hence are not annihilated by the naive variational derivative). We demonstrate both of these subtleties and how to deal with them in the concrete case of a massless spin-2 in $d=3$, where we derive the on-shell cubic vertex corresponding to the Einstein--Hilbert action \eqref{EinsteinHilbertActiondS}.

\paragraph{On-Shell Gauge Transformations:}

Generic gauge transformations do not preserve the conditions $\nabla^{\mu}h_{\mu\nu}=h=0$.  Under the massless gauge transformation $h_{\mu\nu}\mapsto h'_{\mu\nu}= h_{\mu\nu}+\nabla_{(\mu}\xi_{\nu)}$ \eqref{MasslesSymmetry} we instead have in $d=3$:
\begin{align}
\nabla^{\nu}h_{\mu\nu}&\mapsto \nabla^{\nu}h'_{\mu\nu}=\nabla^{\nu}h_{\mu\nu}+\left (\square +3H^{2}\right )\xi_{\mu}\ , \\
h&\mapsto h'=h+\nabla_{\mu}\xi^{\mu}\ .
\end{align}
Therefore, after imposing these conditions, we can only demand on-shell gauge invariance under residual gauge transformations that obey the equations
\be
\left (\square+3H^{2}\right )\xi^{\mu}=\nabla_{\mu}\xi^{\mu}=0\, .\label{GROnShellGaugeTransConditions}
\ee

\paragraph{On-Shell Total Derivatives:}

The condition $\delta^{(0)}_{\xi}S_{3}[h]\cong 0$ means that $\delta^{(0)}_{\xi}S_{3}[h]$ is a sum of terms which are total derivatives after accounting for equations of motion.  We can parametrize the most general possible total derivative as
\be
\delta^{(0)}_{\xi}S_{3}[h]\cong\int\rd^{4}x\, \sqrt{-g}\, \left [\nabla_{\mu}J^{\mu}+\xi^{\nu}\left (\nabla_{\nu}O+\left (\square+3H^{2}\right )O_{\nu}\right )+h^{\mu\nu}\left (\nabla_{\mu}\mathcal{O}_{\nu}+\left (\square-2H^{2}\right )\mathcal{O}_{\mu\nu}\right )\right ]\label{GROnShellGaugeVariationExpresion}
\ee
for some operators $J^{\mu},O,O_{\nu},\mathcal{O}_{\nu}$, and $\mathcal{O}_{\mu\nu}$ built from appropriate powers of $\xi_{\mu}$ and $h_{\mu\nu}$.  If we were solving for the off-shell gauge invariance constraints, only the explicit total derivative term, $\nabla_{\mu}J^{\mu}$, would appear, but due to the on-shell conditions \eqref{MasslesshEOM}, \eqref{GROnShellGaugeTransConditions}, and $\nabla^{\mu}h_{\mu\nu}=0$, the additional terms above are also total derivatives.

The appearance of the extra terms in \eqref{GROnShellGaugeVariationExpresion} makes the process of solving for the gauge-invariant interaction somewhat different than in the off-shell scenario.  A practical method for solving \eqref{GROnShellGaugeVariationExpresion} is to integrate all derivatives off of the gauge parameter $\xi_{\mu}$.  This is accomplished by acting on both sides with the operator $\int\rd^{4}x\, \xi_{\mu}(x)\frac{\delta}{\delta \xi_{\mu}(x)}$, yielding\footnote{Note that the LHS is more simply written as $\int\rd^{4}x\, \xi_{\mu}(x)\frac{\delta}{\delta \xi_{\mu}(x)}\big (\delta^{(0)}_{\xi}S_{3}[h]\big )=\int\rd^{4}x\, \xi_{\mu}(x)\frac{\delta}{\delta \xi_{\mu}(x)}S_{3}[h_{\mu\nu}+\nabla_{(\mu}\xi_{\nu)}]$.} 
\be
\int\rd^{4}x\, \xi_{\mu}(x)\frac{\delta}{\delta \xi_{\mu}(x)}\left (\delta^{(0)}_{\xi}S_{3}[h]\right )\cong\int\rd^{4}x\, \sqrt{-g}\,  \xi^{\nu}\left (\nabla_{\nu}O'+\left (\square+3H^{2}\right )O'_{\nu}\right )\ .\label{GROnShellGaugeVariationExpresionExplicit}
\ee
for some operators $O'$, $O'_\mu$.
In contrast, the analogous off-shell gauge invariance constraint at this order would read
\be
\int\rd^{4}x\, \xi_{\mu}(x)\frac{\delta}{\delta \xi_{\mu}(x)}\left (\delta^{(0)}_{\xi}S_{3}[h]+\delta^{(1)}_{\xi}S_{2}[h]\right )=0,\label{GROffShellGaugeVariationExpresionExplicit}
\ee
with nothing appearing on the RHS.

\paragraph{Finding the Einstein--Hilbert Interaction:}  Accounting for the previous points, we can now solve for the $d=3$ Einstein--Hilbert interaction.  From derivative counting, the on-shell action must take on the form \eqref{OnehOperatorBasis} with $a_{4}=a_{5}=0$:
\begin{align}
S_{3}[h]&=\int\rd^{4}x\sqrt{-g}\,\left ( a_{1}h^{\mu }{}_{\nu}h^{\nu}{}_{\sigma}h^{\sigma}{}_{\mu }+a_{2}h^{\rho \sigma }\nabla_{\rho }h^{\mu \nu }\nabla_{\sigma }h_{\mu \nu }  +a_{3}h^{\mu \rho }\nabla_{\nu }h^{\sigma }{}_{\rho}\nabla_{\sigma }h^{\nu }{}_{\mu}\right )\ .\label{GRExampleBasisS3}
\end{align}
Calculating the LHS of~\eqref{GROnShellGaugeVariationExpresionExplicit}  for this action yields
\begin{align*}
&\int\rd^{4}x\,\xi_{\mu}(x)\frac{\delta}{\delta \xi_{\mu}(x)}S_{3}[h_{\mu\nu}+\nabla_{(\mu}\xi_{\nu)}]\nn
&~~=\int\rd^{4}x\, \sqrt{-g}\, \Big(2a_{3}\xi^\mu\nabla_{\rho }\nabla_{\mu }h_{\nu  \sigma }\nabla^{\sigma }h^{\nu  \rho }+\left (4a_{2}-2a_{3}\right )\xi^{\mu }\nabla_{\rho }\nabla_{\nu  }h_{\mu  \sigma }\nabla^{\sigma }h^{\nu  \sigma }-2a_{2}\xi^{\mu } \nabla_{\sigma }\nabla_{\mu }h_{\nu  \rho }\nabla^{\sigma } h^{\nu  \rho }\nn
&\quad +2a_{3}\xi^{\mu } \nabla_{\sigma }\nabla_{\rho }h_{\mu  \nu }\nabla^{\sigma } h^{\nu  \rho }-4(a_{2}-2a_{3})H^{2}\xi ^{\mu }h^{\nu  \rho }\nabla_{\mu }h_{\nu  \rho }+6\left (a_{1}-2\left (3a_{2}+a_{3}\right )H^{2}\right )h^{\nu  \rho }\xi^{\mu }\nabla_{\rho }h_{\mu  \nu }\Big),
\end{align*}
and we need to find the operators $O$ and $O_{\nu}$ for which 
\be
\int\rd^{4}x\left (\xi_{\mu}(x)\frac{\delta}{\delta \xi_{\mu}(x)}S_{3}[h_{\mu\nu}+\nabla_{(\mu}\xi_{\nu)}]-\sqrt{-g}\,\xi^{\nu}\left (\nabla_{\nu}O+\left (\square+3H^{2}\right )O_{\nu}\right ) \right )\cong 0\ .\label{GRExampleEquationToBeSolved}
\ee

The relation \eqref{GRExampleEquationToBeSolved} is then solved at each order in derivatives, writing $O=\sum_{n}O^{(n)}$ where $O^{(n)}$ is $\mathcal{O}(\nabla^{n})$ and similar for $O_{\nu}$. 
Starting with the $\mathcal{O}(\nabla^{3})$ terms, we can systematically include all on-shell independent terms in  $O$ and $O_{\nu}$ in \eqref{GRExampleEquationToBeSolved} which contribute at this order with arbitrary coefficients.  These are explicitly given by
\begin{align}
O^{(2)}&=A _{1}\nabla_{\rho }h_{\nu  \sigma }\nabla^{\sigma } h^{\nu  \rho } +A _{2}\nabla_{\sigma } h_{\nu  \rho }\nabla^{\sigma } h^{\nu  \rho }\, , \\
O^{(1)}_{\nu }&=A _{3}h^{\rho  \sigma }\nabla_{\nu }h_{\rho  \sigma }+A _{4}h^{\rho  \sigma }\nabla_{\sigma }h_{\nu  \rho }\,,
\end{align}
  Solving \eqref{GRExampleEquationToBeSolved} at this order places constraints both on the $O$ and $O_{\nu}$ coefficients and the  $a_{i}$'s:
  \begin{align}
   a_{2}=\frac{a_{3}}{2}\, , \qquad A _{1}=a_{3}\, , \qquad A _{2}=-A _{3}-\frac{a_{3}}{2}\,,\qquad A _{4}=a_3 .
   \end{align} 
The remaining terms in \eqref{GRExampleEquationToBeSolved} are $\mathcal{O}(\nabla)$ and read
\begin{equation*}
\int\rd^{4}x\, \sqrt{-g}\, \left ( 2H^{2}\left (a_{3}-5A _{3}\right )\xi^{\mu}h^{\nu\rho}\nabla_{\mu}h_{\nu\rho}-6\left (a_{1}-3a_{3}H^{2}\right )h^{\nu\rho}\xi^{\mu}\nabla_{\rho}h_{\mu\nu}-\xi^{\nu}\nabla_{\nu}O^{(0)}\right )\cong 0\ .
\end{equation*}
Choosing $O^{(0)}=(a_{3}-5A _{3})H^{2}h^{\mu\nu}h_{\mu\nu}$ cancels the first term above, while the second cannot be canceled so we are forced to set $a_{3}=a_{1}/(3H^2)$. 

The interaction has therefore been fixed up to an overall factor:
\begin{align}
S_{3}[h]&=a_{1}\int\rd^{4}x\,\sqrt{-g}\,\left ( h^{\mu }{}_{\nu}h^{\nu}{}_{\sigma}h^{\sigma}{}_{\mu }+\frac{1}{6H^{2}}h^{\rho \sigma }\nabla_{\rho }h^{\mu \nu }\nabla_{\sigma }h_{\mu \nu }  +\frac{1}{3H^{2}}h^{\mu \rho }\nabla_{\nu }h^{\sigma }{}_{\rho}\nabla_{\sigma }h^{\nu }{}_{\mu}\right )\ .
\end{align}
Expanding the Einstein--Hilbert action \eqref{EinsteinHilbertActiondS} to cubic order, integrating by parts, and using the on-shell conditions, it can be explicitly shown that the cubic interactions agree if we set $a_{1}=\frac{3}{4}\MP ^{2}H^{2}$. 

In the following section we describe the generalization of this procedure to general dimension and to include partially massless fields.

\subsection{Generic $d$ residual gauge transformations}

The analysis leading to the constraints on the $a_{i}$, $b_{i}$, and $c_{i}$'s of Sec.~\ref{sec:CubicVertices} is a straightforward generalization of the procedure used in the preceding section.    In generic dimensions, massless and partially massless fields have the residual gauge transformations
\begin{align}
\delta h_{\mu\nu}^{(\gamma)}&=2\nabla_{(\mu}\xi_{\nu)} & &{\rm with} ~~~\left (\square+d\, H^{2}\right ) \xi_{\nu}=0 \ ,\quad \nabla^{\nu}\xi_{\nu}=0,\nn
\delta h_{\mu\nu}^{(\sigma)}&=\left (\nabla_{\mu}\nabla_{\nu}+H^{2}g_{\mu\nu}\right )\chi & &{\rm with}~~~\left (\square+(d+1)H^{2}\right )\chi=0,\label{OnShellGaugeTransformations}
\end{align}
where $h^{(\gamma)}$ is massless and $h^{(\sigma)}$ is the PM field.  These preserve $\nabla^{\mu}h^{(\gamma)}_{\mu\nu}=h^{(\gamma)}=0$ and $\nabla^{\mu}h^{(\sigma)}_{\mu\nu}=h^{(\sigma)}=0$.  Starting from the generic operator bases \eqref{ThreehOperatorBasis}, \eqref{TwohOperatorBasis}, and \eqref{OnehOperatorBasis}, we systematically impose the constraints of gauge-invariance following the steps of Appendix \ref{app:EHOnShellCubicTerms}. This results in the constraints quoted in Sec.~\ref{sec:CubicVertices}.

\section{Explicit $d=3$ transverse-traceless polarization tensors\label{app:PolarizationTensors}}

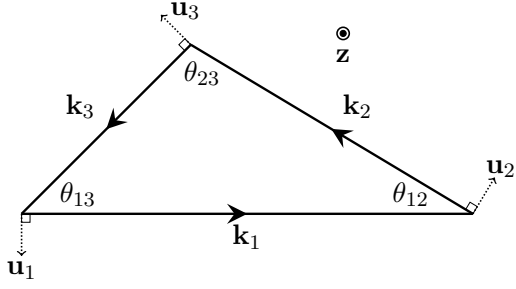
\begin{figure}
\begin{center}
 \scalebox{.75}{\begin{tikzpicture}[ decoration={markings,
mark=at position 0.5 with {\arrow[very thick,scale=2]{stealth}}}
]
\coordinate (L) at (-4,0);
\coordinate (R) at (4,0);
\coordinate (T) at (-1,3);

\Perp[0]{(L)}{(R)}{U1}[-.75cm];
\Perp[0]{(R)}{(T)}{U2}[-.75cm];
\Perp[0]{(T)}{(L)}{U3}[-.75cm];

\tkzMarkRightAngle[draw=black,size=.15](L,T,U3);
\tkzMarkRightAngle[draw=black,size=.15](U2,R,T);
\tkzMarkRightAngle[draw=black,size=.15](R,L,U1);

\path 	node[scale=1.3] at (-1.1,3.6) {$\bfu_3$}
		node[scale=1.3] at (4.5,.8) {$\bfu_2$}
		node[scale=1.3] at (-4,-1) {$\bfu_1$};

\path	(L)	edge[very thick,postaction={decorate}] node [midway,below,scale=1.3]{$\bfk_{1}$} node {} (R)
		(R)	edge[very thick,postaction={decorate}] node [midway,above right,scale=1.3]{$\bfk_{2}$} node {} (T)
		(T)	edge[very thick,postaction={decorate}] node [midway,above left,scale=1.3]{$\bfk_{3}$} node {} (L);
		
\path 	node[scale=1.3] at (2.9,.35) {$\theta_{12}$}
		node[scale=1.3] at (-.8,2.5) {$\theta_{23}$}
		node[scale=1.3] at (-3,.35) {$\theta_{13}$}
		node[circle,thick, draw=black,fill=white,scale=.6] at (1.7,3.2) {}
		node[circle, draw=black,fill=black,scale=.3] at (1.7,3.2) {}
		node[scale=1.3] at (1.7,2.8) {$\bfz$};
\end{tikzpicture}}
\end{center}
   \caption{\small Diagram defining the vectors which make up the polarization tensors in \eqref{appendixSpinTwoGuiMaldacenaPolarizationTensors}.}
   \label{fig:PolarizationDefinitionTriangle}
\end{figure}

In this Appendix, we discuss the explicit transverse, traceless polarization tensors which are used when evaluating tensor non-Gaussianity in $d=3$.   We write our basis as
\begin{align}
\epsilon^{\bfk_{a}P}_{ij }&=\left (z_{i}z_{j}-u_{i}^{a}u_{j}^{a}\right )\ , \quad
\epsilon^{\bfk_{a}X}_{ij }=\left (u^{a}_{i}z_{j}+z_{i}u^{a}_{j}\right )\label{appendixSpinTwoGuiMaldacenaPolarizationTensors}\ ,
\end{align}
following \cite{Maldacena:2011nz}, though changing the overall normalization. In \eqref{appendixSpinTwoGuiMaldacenaPolarizationTensors},
 $z_{i}$ is a unit vector orthogonal to the plane spanned by the three $\bfk_{a}$'s and $\bfu_{a}$ is a unit vector in the plane of this triangle which is orthogonal to $\bfk_{a}$; see Fig.~\ref{fig:PolarizationDefinitionTriangle}. The action of parity is reflection across the plane in which the momenta lie. The $\bfu_{a}$ are invariant under this transformation, while $\bfz$ flips sign.  Hence, $\epsilon^{P}$ is even under parity, while $\epsilon^{X}$ is odd.

The normalization in \eqref{appendixSpinTwoGuiMaldacenaPolarizationTensors} is chosen such that
 \begin{align}
\sum_{s}\epsilon^{*,\,s}_{\bfk_{a}ij}\epsilon_{\bfk_{a}lm}^{s}&=2(\Pi^{\bfk_{a}}_{TT}){}_{ijlm} \ , \quad \epsilon^{sij}_{\bfk_{a}}\epsilon^{*,\,s'}_{\bfk_{a}ij}=2\delta^{ss'}\ , 
  \end{align}
  where $s\in \{X,P\}$ and $\Pi_{TT}$ is the transverse, traceless projector defined in \eqref{Projectors}. The factors of 2 are conventional in the literature.  For instance, the dimensionless tensor power spectrum $\Delta_{\gamma}^{2}$ is naturally defined as  
  \begin{align}
 \Delta_{\gamma}^{2}\equiv \frac{k^{3}}{\pi^{2}} \sum_{s} \langle \gamma^{\bfk,s}(\ts)\gamma^{-\bfk,s}(\ts)\rangle'= \frac{k^{3}}{\pi^{2}}\frac{1}{4}\sum_{s} \epsilon^{\bfk ij}_{s}\langle \gamma^{\bfk}_{ij}(\ts)\gamma^{-\bfk}_{lm}(\ts)\rangle'\epsilon^{-\bfk lm}_{s}
 \end{align}
 and using \eqref{MasslessSpin2TwoPointd3}, which reads $\langle \gamma^{\bfk}_{ij}(\ts)\gamma^{-\bfk}_{lm}(\ts)\rangle'=\frac{2H^{2}}{\MP ^{2}k^{3}}(\Pi^{\bfk}_{TT} ){}_{ijlm}$ and $ (\Pi_{TT}^{\bfk} ){}_{ij}{}^{ij}=2$, it follows that $\Delta_{\gamma}^{2}=\frac{2}{\pi^{2}}\frac{H^{2}}{\MP ^{2}}$  which is the standard expression.

\subsection{Dimensionally-dependent identities}
A motivation for introducing an explicit basis of polarizations is that
in $d=3$,
there are many accidental identities involving contractions between transverse, traceless polarization tensors and 3-momenta, which can make it difficult to identify the independent structures appearing in contractions such as  $\langle T^{ ij }_{\bfk_{1}}T^{ lm }_{\bfk_{2}}T^{ no }_{\bfk_{3}}\rangle' \bar{\gamma}^{\bfk_{1}}_{ij}\bar{\gamma}^{\bfk_{2}}_{lm}\bar{\gamma}^{\bfk_{3}}_{no}$. As an illustration of this difficulty we present some highly non-trivial identities relating various contractions of momenta and polarizations. After using the transverse property and 3-momentum conservation to canonicalize all contractions between $\bfk_{a}$'s and $\epsilon$'s by eliminating all contractions of the form $k^{i}_{a}\epsilon^{\bfk_{b}}_{ij}$ with $a<b$, cyclically understood, we find the following identities via direct calculation.  First, we find
\begin{align}
\epsilon^{ i }_{\bfk_{ 1 } j }\epsilon^{ j }_{\bfk_{ 2 } l }\epsilon^{ l }_{\bfk_{ 3 } i }&=\frac{-4}{k_{T}\prod_{a}(k_{T}-2k_{a})}\Big[\left (k_{1}^{i}k_{1}^{j}\epsilon_{ij}^{\bfk_{3}}\right )\left (k_{2}^{l}\epsilon_{lm}^{\bfk_{1}}\epsilon_{n}^{\bfk_{2}m} k^{n}_{3}\right )+{\rm 2 \ perms.}\Big]\ .
\end{align}
Second, we have the more involved identity
\begin{align}
\epsilon^{ i }_{\bfk_{ 1 } j }\epsilon^{ j }_{\bfk_{ 2 } l }\epsilon^{ l }_{\bfk_{ 3 } i }&=F_{1}(k_{1},k_{2},k_{3})\,k^{l}_{1}k^{m}_{1}\epsilon_{lm}^{\bfk_{3}} \epsilon^{\bfk_{1}ij}\epsilon_{ij}^{\bfk_{2}}+{\rm 2 \ perms.}\nn
&\quad+F_{2}(k_{1},k_{2},k_{3})\,k_{2}^{i}\epsilon_{ij}^{\bfk_{1}}\epsilon_{jl}^{\bfk_{1}}\epsilon^{\bfk_{2}l}_{m} k_{3}^{m}+{\rm 2 \ perms.}
\end{align}
where the $F_{i}$ are symmetric under any permutation of $k_{i}$'s for any $F_{2}$ obeying\footnote{For instance, $F_{2}(k_{1},k_{2},k_{3})=\frac{-1}{3k_{3}^{2}}$ or $F_{2}(k_{1},k_{2},k_{3})=-\frac{1}{3}\frac{k_{3}^{n}}{k_{1}^{n+2}+k_{2}^{n+2}+k_{3}^{n+2}}$ satisfy this relation.}
\begin{align}
0&=1+k_{ 1 }^{2}F_{2}(k_{ 2 },k_{  3  },k_{ 1 })+k_{ 2 }^{2}F_{2}(k_{ 3 },k_{  1  },k_{ 2 })+k_{ 3 }^{2}F_{2}(k_{ 1 },k_{  2  },k_{ 3 })
\end{align}
and $F_{1}$ is determined in terms of $F_{2}$ via
\begin{align}
F_{1}(k_{1},k_{2},k_{3})&=\frac{-1}{2 \left(k_{1}^2+k_{2}^2-k_{3}^2\right)}\Big[2+\left (k_{1}^{2}-k_{2}^{2}\right )\left (F_{2}(k_{2},k_{3},k_{1})-F_{2}(k_{3},k_{1},k_{2})\right )\nn
&\quad+k_{3}^{2}\left (2F_{2}(k_{1},k_{2},k_{3})+F_{2}(k_{2},k_{3},k_{1})+F_{2}(k_{3},k_{1},k_{2})\right )\Big]\ .
\end{align}

Rather than using these identities to simplify contractions like $\langle T^{ ij }_{\bfk_{1}}T^{ lm }_{\bfk_{2}}T^{ no }_{\bfk_{3}}\rangle' \bar{\gamma}^{\bfk_{1}}_{ij}\bar{\gamma}^{\bfk_{2}}_{lm}\bar{\gamma}^{\bfk_{3}}_{no}$ for arbitrary $\bar{\gamma}$, we instead find it more convenient to simply evaluate the expressions using the explicit $\epsilon^{X}$ and $\epsilon^{P}$ tensors for $\bar{\gamma}$.

\section{Integrations by parts and the wavefunction\label{app:IBPAndTheWavefunction}}
Perturbative computation of the wavefunction in terms of the on-shell action appears to have an inherent ambiguity: the addition of boundary terms supported on the time slice $\tau = \ts$.\footnote{More specifically, the only boundary terms which can have  interesting effects on the wavefunction are those involving derivatives normal to the boundary.  Operators which only depend on tangential derivatives ({\it i.e.}~the analogues of holographic counterterms \cite{Skenderis:2002wp}) can only affect the phase of the wavefunction, due to the reality condition on fields at the $\tau=\ts$ surface.} Actions related by integrations-by-parts differ precisely by such boundary terms, so it behooves us to understand their effect on the wavefunctional. In this Appendix, we discuss the effect that integrations-by-parts have on wavefunction coefficients. These operations do not leave the wavefunction invariant.  Instead, they change $\Psi$ in a very specific way: integration-by-parts has the same effect as local field redefinitions.  We illustrate these concepts using the example of a massless scalar field. See~\cite{Arroja:2011yj,Burrage:2011hd,Seery:2010kh} for discussions of boundary terms in the canonical in-in formalism and \cite{Seery:2006tq} for a discussion from a dS/CFT and wavefunctional perspective.
 
 \subsection{Example: massless scalar field}
 
Consider the action
\be
S[\varphi]=\int\rd^{d+1}x\,\sqrt{-g}\,\left( -\frac{1}{2}\left (\nabla\varphi\right )^{2}+a_{1}\varphi\left (\nabla\varphi\right )^{2}+a_{2}\nabla^{\mu}\varphi\nabla_{\mu}\nabla_{\nu}\varphi\nabla^{\nu}\varphi+a_{3}\varphi^{2}\square\varphi\right)\ .\label{MasslessScalarActionExample}
\ee
Each of the $a_{i}$ operators in~\eqref{MasslessScalarActionExample} can be integrated-by-parts to be proportional to $\square\phi$.  Hence, when evaluated on the classical linear solution obeying $\square\varphi_{\rm cl}=0$, the entire effect of these operators is captured by a boundary term:
\be
 S[\varphi_{\rm cl}]=\int\rd^{d+1}x\,\sqrt{-g}\left( -\frac{1}{2}\left (\nabla\varphi_{\rm cl}\right )^{2}\right)-\int_{\tau=\ts}\rd^{d}x\,\sqrt{\bar{g}}\, n_{\mu}\left( \frac{a_{1}}{2}\varphi_{\rm cl}^{2}\nabla^{\mu}\varphi_{\rm cl}+\frac{a_{2}}{2}\left (\nabla\varphi_{\rm cl}\right )^{2}\nabla^{\mu}\varphi_{\rm cl}\right)\ .\label{ScalarExOnShellAction}
\ee
The boundary terms above affect $\Psi$ non-trivially since $\varphi_{\rm cl}$ is non-vanishing on the $\tau=\ts$ surface.  However, their contributions to the wavefunction have the same effect as redefining $\bar{\varphi}$ in $\Psi[\bar{\varphi},\ts]$.  

We illustrate this by explicitly calculating the wavefunction corresponding to \eqref{MasslessScalarActionExample} in the $d = 3$, super-horizon limit.  The linear solution is
\be
\varphi_{\rm cl}^{\bfk}(\tau)=\bar{\varphi}^{\bfk}\frac{(1-ik\tau)}{(1-ik\ts)}e^{ik(\tau-\ts)}\, . \label{MasslessScalarLinearSolution}
\ee
Substituting this solution into~\eqref{ScalarExOnShellAction}, the result is
\be
\Psi[\bar{\varphi},\ts] \approx\exp\left(-\frac{1}{2}\int\rd^{3}\tilde{k}\, \frac{k^{3}}{H^{2}}\bar{\varphi}^{\bfk}\bar{\varphi}^{-\bfk}+\frac{a_{1}}{3!H^{2}}\int\rd^{3}\tilde{k}_{1}\rd^{3}\tilde{k}_{2}\rd^{3}\tilde{k}_{3}\,\tilde{\delta}^{3}(\textstyle{\sum_{i}\bfk_{i}}) \left (\sum_{i}k_{i}^{3}\right )\bar{\varphi}^{\bfk_{1}}\bar{\varphi}^{\bfk_{2}}\bar{\varphi}^{\bfk_{3}}\right)\, .\label{ScalarExWavefunction}
\ee
The boundary contribution to \eqref{ScalarExWavefunction} is non-zero, but it can also be removed by a local field redefinition, due to the $\sim k^{3}$ form of the quadratic terms in \eqref{ScalarExWavefunction}. Specifically, making the following local, position space field redefinition precisely cancels the contribution of boundary terms to the wavefunction:
\be
\bar{\varphi}(\bfx)\mapsto \bar{\varphi}(\bfx)+\frac{a_{1}}{2}\bar{\varphi}(\bfx)^{2}\ .
\ee
A systematic construction of all possible cubic boundary terms involving up to five derivatives confirms that total derivatives only generate the $\sim \left (\sum_{i}k_{i}^{3}\right )\bar{\varphi}^{\bfk_{1}}\bar{\varphi}^{\bfk_{2}}\bar{\varphi}^{\bfk_{3}}$ structure. 
We have found that the same pattern holds for spin-2 wavefunctions: integration-by-parts ambiguities in the wavefunction are of precisely the same form as field redefinition ambiguities.

\subsection{A general cubic argument: field redefinitions correspond to boundary terms}
In fact, it is possible to show more generally that
local field redefinitions of the type $\varphi(x^{\mu})\mapsto \varphi(x^{\mu})+ a\, \varphi(x^{\mu})^{2}$, with constant $a$, have the same effect on the cubic order wavefunction as adding boundary terms to the action, so that the ambiguities associated with boundary terms are the same contact term ambiguities associated with making a choice of field variables.

Consider a generic cubic action 
\begin{align}
S[\varphi]=S^{(2)}[\varphi]+S^{(3)}[\varphi] \ ,
\end{align}
where $\varphi$ stands for one or more fields of arbitrary type.   Making the above field redefinition yields
\begin{align}
S[\varphi]\mapsto S[\varphi]&=S^{(2)}[\varphi+ a \, \varphi^{2}]+S^{(3)}[\varphi]+\mathcal{O}(\varphi^{4})\nn
&=S^{(2)}[\varphi]+S^{(3)}[\varphi]+\int\rd^{d+1}x\,  a \, \varphi^{2}\,\frac{\delta S^{(2)}[\varphi]}{\delta\varphi}+ a\, S^{(3)}_{\rm bdy}[\varphi]+\mathcal{O}(\varphi^{4})\ ,\label{ActionVariationFieldRedef}
\end{align}
where $S^{(3)}_{\rm bdy}[\varphi]$ contains the boundary terms which arise from the integrations-by-parts needed to form the variational derivative.
When evaluated on-shell, the variational derivative term in \eqref{ActionVariationFieldRedef} vanishes, resulting in:
\begin{align}
S_{\rm cl}[\varphi_{\rm cl}]=S^{(2)}_{\rm cl}[\varphi_{\rm cl}]+S^{(3)}[\varphi_{\rm cl}]+ a\,  S^{(3)}_{\rm bdy,cl}[\varphi_{\rm cl}]+\mathcal{O}(\varphi_{\rm cl}^{4})\ .
\end{align}

At the $\tau=\ts$ surface, the similar local field redefinition $\varphi(\bfx)\mapsto  \varphi(\bfx)+ a\,  \varphi(\bfx)^{2}$ is induced, which causes the position space wavefunction to change from
\begin{align}
 \Psi[\bar{\varphi}(\bfx),\ts]=\exp\bigg(&-\frac{1}{2}\int\rd^{d}x\rd^{d}y\,\langle\mathcal{O}(\bfx)\mathcal{O}(\bfy) \rangle\bar{\varphi}(\bfx)\bar{\varphi}(\bfy)\nn
 &\quad-\frac{1}{3!}\int\rd^{d}x\rd^{d}y\rd^{d}z\,\langle\mathcal{O}(\bfx)\mathcal{O}(\bfy)\mathcal{O}(\bfz)\rangle\bar{\varphi}(\bfx)\bar{\varphi}(\bfy)\bar{\varphi}(\bfz)+\ldots\bigg),
\end{align}
to
\begin{align}
 \Psi[\bar{\varphi}(\bfx),\ts]=\exp\bigg(&-\frac{1}{2}\int\rd^{d}x\rd^{d}y\,\langle\mathcal{O}(\bfx)\mathcal{O}(\bfy) \rangle\bar{\varphi}(\bfx)\bar{\varphi}(\bfy)\nn
 &-\frac{ a }{2}\int\rd^{d}x\rd^{d}y\,\langle\mathcal{O}(\bfx)\mathcal{O}(\bfy) \rangle\left (\bar{\varphi}(\bfx)^{2}\bar{\varphi}(\bfy)+\bar{\varphi}(\bfx)\bar{\varphi}(\bfy)^{2}\right )\nn
 &-\frac{1}{3!}\int\rd^{d}x\rd^{d}y\rd^{d}z\,\langle\mathcal{O}(\bfx)\mathcal{O}(\bfy)\mathcal{O}(\bfz)\rangle\bar{\varphi}(\bfx)\bar{\varphi}(\bfy)\bar{\varphi}(\bfz)+\ldots\bigg).
\end{align}
Finally, from the relation $\Psi[\bar{\varphi}]\approx e^{iS_{\rm cl}[\varphi_{\rm cl}]}$, it follows that we must have
\begin{align}
iS^{(3)}_{\rm bdy,cl}[\varphi_{\rm cl}]&=-\frac{ 1 }{2}\int\rd^{d}x\rd^{d}y\,\langle\mathcal{O}(\bfx)\mathcal{O}(\bfy) \rangle\left (\bar{\varphi}(\bfx)^{2}\bar{\varphi}(\bfy)+\bar{\varphi}(\bfx)\bar{\varphi}(\bfy)^{2}\right )\, ,\label{FieldRedefBdyTermEquilvanceExample}
\end{align}
explicitly demonstrating that the result of a local field redefinition is to add a boundary term to the on-shell action.  The terms on the RHS of \eqref{FieldRedefBdyTermEquilvanceExample} only affect three-point correlators by contact terms, where two of the three operators are brought to coincident points, {\it i.e.}~their contribution is proportional to a delta function.  Such semi-local contributions, in the language of \cite{Bzowski:2013sza}, are often ignored in position space, but can be relevant for cosmological correlators.

\section{CFT two-point functions\label{app:CFTTwoPointFunctions}}

In this Appendix, we review the structure of momentum-space CFT two-point functions for arbitrary symmetric, traceless primary operators.  For fields on de Sitter, both the quadratic wavefunction coefficients and the two-point equal-time correlators take on the form of CFT correlators.

 An efficient method for deriving the momentum space two-point functions of generic spin, symmetric, traceless primary fields $\varphi_{i_{1}\ldots i_{s}}$ is to first contract all indices with auxiliary null-vectors $\bfy$ and $\bfz$ to create the index-free correlator
\begin{align}
\langle \varphi^{s}_{\bfk}\varphi_{-\bfk}^{s}\rangle'&\equiv y_{i_{1}}\ldots y_{i_{s}}z_{j_{1}}\ldots z_{j_{s}}\langle \varphi_{\bfk}^{ i_{1}\ldots i_{s}}\varphi^{ j_{1}\ldots j_{s}}_{-\bfk}\rangle'\ .
\end{align}
The constraints of scale and special conformal invariance can then be efficiently imposed on $\langle \varphi_{s}^{\bfk}\varphi^{-\bfk}_{s}\rangle'$ which is forced to take the form \cite{Arkani-Hamed:2015bza}
\begin{align}
\langle \varphi_{s}^{\bfk}\varphi^{-\bfk}_{s}\rangle'&\propto k^{2\Delta-d-2s}\left (\bfy\cdot\bfk\right )^{s}\left (\bfz\cdot\bfk\right )^{s}{}_{2}F_{1}\left [-s,-1+\Delta,1-\frac{d}{2}-s+\Delta,\frac{k^{2}}{2}\frac{(\bfy\cdot\bfz)}{(\bfy\cdot\bfk)(\bfz\cdot\bfk)}\right ]\label{GenericCFTTwoPointFunction}
\end{align}
with ${}_{2}F_{1}$ a hypergeometric function.\footnote{The ${}_2F_1$ convention we use is that of \href{http://functions.wolfram.com/HypergeometricFunctions/Hypergeometric2F1}{\tt functions.wolfram.com/HypergeometricFunctions/Hypergeometric2F1}.}  The $\bfy$ and $\bfz$ factors can then be stripped from the above to produce a traceless correlator by acting $s$-times with the operator \cite{Costa:2011mg}
\begin{align}
D_{i}^{(y)}&\equiv\left (\frac{d}{2}-1-y^{j}\frac{\partial}{\partial y^{j}}\right )\frac{\partial}{\partial y^{i}}-\frac{1}{2}y_{i}\frac{\partial^{2}}{\partial y^{i}\partial y_{i}}\label{DStripOperator}
\end{align}
and $s$-times with the similar $D_{i}^{(z)}$ operator. 

 The above construction also reproduces the quadratic wavefunction coefficients and two-point functions derived in Sec.~\ref{sec:TwoPointFunctions}.  In the present paper we are primarily concerned with the transverse, traceless parts of correlators, but the other components can be restored using the preceding results.

\renewcommand{\em}{}
\bibliographystyle{utphys}
\bibliography{PartiallyMasslessNonGaussianity-arxiv}

\end{document}